\documentclass[times,12pt]{elsarticle}

\usepackage{moreverb} 
\usepackage[colorlinks,bookmarksopen,bookmarksnumbered,citecolor=red,urlcolor=red]{hyperref}
\usepackage{amsfonts,amsmath,amssymb}
\usepackage{psfrag,pstricks,pst-node}
\usepackage{mathrsfs}
\usepackage{psfrag}
\usepackage{lineno}

\usepackage{booktabs}
\usepackage{float}
\usepackage{caption}
\usepackage{subcaption}
\usepackage{appendix}
\DeclareCaptionLabelFormat{cont}{#1~#2\alph{ContinuedFloat}}
\usepackage{multirow}
\usepackage{longtable}
\usepackage{seqsplit}

\usepackage{natbib}
\setcitestyle{authoryear,open={(},close={)}}
\bibpunct[, ]{(}{)}{;}{a}{,}{,}

\usepackage{enumerate}
\usepackage{fancyhdr}
\usepackage[margin=3cm]{geometry}
\usepackage[normalem]{ulem}
\usepackage{bm}
\usepackage{algorithm}
\usepackage{algorithmic}

\newcommand\BibTeX{{\rmfamily B\kern-.05em \textsc{i\kern-.025em b}\kern-.08em
T\kern-.1667em\lower.7ex\hbox{E}\kern-.125emX}}
\usepackage[makeroom]{cancel}
\usepackage{setspace}
\usepackage{soul}

\makeatletter
\def\ps@pprintTitle{%
 \let\@oddhead\@empty
 \let\@evenhead\@empty
 \let\@oddfoot\@empty
 \let\@evenfoot\@empty
}
\makeatother

\begin{document}
%\linenumbers

\begin{frontmatter}
\renewcommand{\thefootnote}{\fnsymbol{footnotemark}}
\fancypagestyle{plain}{%
\fancyhf{} % clear all header and footer fields
\fancyhead[RO,RE]{\thepage} %RO=right odd, RE=right even
\renewcommand{\headrulewidth}{0pt}
\renewcommand{\footrulewidth}{0pt}
}

\title{Machine Learning Aided Modeling of Granular Materials: A Review}

    \author[lab1]{Mengqi Wang}   
    \author[lab2]{Krishna Kumar}
    
    \author[lab1]{Y.~T. Feng\corref{cor1}}
    \cortext[cor1]{}
    \ead{y.feng@swansea.ac.uk}

    \author[lab3]{Tongming Qu}
    
    \author[lab4]{Min Wang\corref{cor1}}
    \cortext[cor1]{Corresponding author}
    \ead{minw@lanl.gov}

    \address[lab1]{Zienkiewicz Centre for Computational Engineering, Faculty of Science and Engineering, Swansea University, \\Swansea, Wales, SA1 8EP, UK}
    \address[lab2]{Department of Civil, Architecture and Environmental Engineering, University of Texas at Austin, Austin, Texas, 78701, USA.}
    \address[lab3]{Department of Civil and Environmental Engineering, Hong Kong University of Science and Technology, Clearwater Bay, Kowloon, Hong Kong SAR, China.}
    \address[lab4]{Fluid Dynamics and Solid Mechanics Group, Theoretical Division, Los Alamos National Laboratory, Los Alamos, New Mexico 87545, USA}

\begin{abstract}
Artificial intelligence (AI) has become a buzz word since Google’s AlphaGo beat a world champion in 2017. In the past five years, machine learning as a subset of the broader category of AI has obtained considerable attention in the research community of granular materials. This work offers a detailed review of the recent advances in machine learning-aided studies of granular materials from the particle-particle interaction at the grain level to the macroscopic simulations of granular flow. This work will start with the application of machine learning in the microscopic particle-particle interaction and associated contact models. Then, different neural networks for learning the constitutive behaviour of granular materials will be reviewed and compared. Finally, the macroscopic simulations of practical engineering or boundary value problems based on the combination of neural networks and numerical methods are discussed. We hope readers will have a clear idea of the development of machine learning-aided modelling of granular materials via this comprehensive review work.
\end{abstract}

\end{frontmatter}

\section*{Highlights} 
\begin{itemize}
\setlength{\itemsep}{0pt}
\setlength{\parsep}{0pt}
\setlength{\parskip}{0pt}
\item Application of existing ML algorithms to the particle-particle interaction models
\item Constitutive study using different ML algorithms and their comparisons
\item ML-aided macroscopic (deformation or flow) simulations of granular materials
\end{itemize}

\section{Introduction}
The granular material, as a macroscopic continuum, showcases complicated features, involving anisotropy~\citep{petalas2020sanisand,ueda2019constitutive}, strain localization~\citep{ueda2019constitutive,voyiadjis2005evolving}, non-coaxiality~\citep{tian2017modelling}, and path-and-states dependence~\citep{das2019influence,alipour2018sand,hu2018cyclic}, under external loading due to their micro-discrete nature. Traditional numerical techniques, such as the finite element method (FEM), finite differential method (FDM), material point method (MPM), discrete element method (DEM), and smoothed particle hydrodynamics (SPH), have been widely employed to investigate the micro/discrete and macro/continuous duality~\citep{wood2017geotechnical} of granular media. Wherein the FEM, SPH, and MPM solve the mechanical responses of granular materials from the macroscopic scale, while the DEM focuses on the mechanical behaviour of granular media at the microscale.

In the mesh-based numerical method, such as FEM~\citep{zienkiewicz2005finite}, the research domain is discretized into finite element meshes that incorporate Gaussian points or grids, and the local mechanical response of the material is characterized by continuum-theory-based phenomenological models embedded in each Gaussian point. The classic constitutive models used in these methods include linear elasticity (e.g. Hooke’s law), nonlinear elasticity (e.g. Duncan-Chang~\citep{duncan1970nonlinear}), the elastic-perfectly-plastic model (e.g. Mohr-Coulomb model~\citep{ti2009review}, Drucker-Prager model, and hardening soil (HS) model~\citep{brinkgreve2005selection}), and the critical-state-based model (e.g. modified cam-clay MCC~\citep{roscoe1968generalized} UH~\citep{yao2009uh,yao2004critical}, and hypoplastic model~\citep{wu1996hypoplastic,mavsin2005hypoplastic}), etc.

Different from the mesh-based method, the MPM, a hybrid Eulerian-Lagrangian meshless approach~\citep{harlow1964particle,zhang2011material}, governs the macroscopic deformation of the target body via a set of discretized material points which carry the material’s local physical features (e.g. mass, density, and velocity). In MPM, the information stored on each material point is first projected to the node of the Eulerian mesh on which the motion equation is solved. Then the updated nodal kinematic features are interpolated back to the corresponding material points to update the deformation of the material, making the MPM especially superior in solving the large deformation and flow problems of granular materials.

Unlike the continuum-based methods, the DEM, proposed in Cundall and Strack’s works~\citep{cundall1971computer,cundall1974rational,strack1978distinct,cundall1979discrete} in the 1970s, represents the overall behaviour of particle systems by explicitly modelling interactions between particles, where adhesive/frictional contact plays a crucial role. In the early framework of the DEM, the discrete particle is normally simplified as discs for two-dimensional (2D) problems. With the advancement of non-spherical models~\citep{williams1992superquadrics,bowman2001particle,fu2012polyarc,lai2020fourier,wang2021novel} and corresponding contact theories~\citep{zhou2018geometric,feng2021energy,feng2021strain} over the last four decades, the DEM has been firmly established as one powerful numerical tool to model the deformation~\citep{ma2006statistics,tavarez2007discrete} or flow~\citep{you2003development,deen2007review} of the collection of grains from the microscopic scale in engineering and scientific problems.

However, these methods have their intrinsic deficiencies. For mesh-based methods, the majority of constitutive models they leveraged tend to homogenize or smear out the underlying discrete features of granular assemblies~\citep{guo2014coupled}, which limits their ability to describe how materials behave following the onset of particle accurately collectively motion-induced catastrophic instabilities~\citep{,andrade2009multiscale} (e.g. landslides and liquefaction). Furthermore, to perfectly fit one certain experimental phenomenon, constitutive models are increasingly sophisticated and inevitably introduce assumptions~\citep{yao2009uh,yin2011modeling} or arguments without physical meanings but need much effort to calibrate~\citep{zhang2021modelling,qu2021adaptive,yin2019practice}, which limits their further applications in other tests. It is also worth mentioning that the constitutive model developed for granular flow is more complicated than the above-mentioned ones. Typical work can be found in~\citep{zhang2000stress}, where the effect of particle rotation at the grain level results in the Jaumann derivative in the evolution equation of the macroscopic contact stress. Additionally, when addressing large deformation problems, the local mesh may suffer from severe distortion and thus deteriorate modelling results. While the mesh distortion problem can be avoided in MPM, it requires the information exchange process between the background (Eulerian) grid and material points, resulting in data breaches and significantly increasing computational complexity.

For DEM, although the discrete nature of granular particles can be taken into account, it still confronts some challenges. One is the prohibitive computational expense primarily attributed to the contact detection procedure in DEM simulations of particle collections~\citep{williams1999discrete}, and this issue is further exacerbated in nonspherical particle assemblies where contact detection involves an iteration-based optimization method~\citep{lai2020fourier,lim2013granular}. Related to this, to minimize the computational cost, irregular shape particles are typically simplified as spheres (3D) or circles (2D)~\citep{bagheri2015characterization,lai2019reconstructing} but may lead to the spuriousness of the bulk mechanical properties in particle assemblies.

In addition to the aforementioned numerical methods used for the modelling of granular media, the requirement of amounts of grain-scale computation has also fostered the emergence of multiscale approaches, such as the FEM-DEM~\citep{andrade2009multiscale,andrade2011multiscale,nitka2011two,guo2014coupled,guo20163d} and MPM-DEM~\citep{liang2019multiscale} approaches, which bridges the micro-features to the computation of macroscopic response of grain materials. Although these methods can compensate for defects of continuum methods to some extent, they are also subjected to the problem of prohibitive computing costs in large-scale simulations.

More recently, the machine learning (ML) method resurged in various fields~\citep{lee2018background,wang2019various,li2021applications,amroune2021machine} seems a promising scheme to circumvent the aforementioned shortcomings of traditional numerical methods in simulating the behaviour of granular materials because of its following advantages: 1) the ML or neural network-based surrogate model can directly extract mechanical features of granular materials from raw data without any assumption; 2) due to its remarkable high-dimension mapping capability~\citep{hornik1989multilayer,cybenko1989approximation}, ML models can approximate desired constitutive laws at high precision by sufficient training data rather than sophisticated formula and well-calibrated physically meaningless parameters; 3) ML models are highly computationally efficient and can instantaneously update the materials’ micro or macro-states according to the received deformation information once their training parameters are determined. Resorting to these advantages of the ML method, it is possible to develop the ML-based stress-strain model to replace the traditional constitutive model used in continuum-level computational approaches. Meanwhile, employing ML algorithms to establish surrogate contact/interaction models for both micro-grains or macroscopic material points also shows promise in alleviating the computational burden for mesh-based and meshless numerical methods (e.g. MPM and DEM).

This paper primarily aims to provide one comprehensive review of the latest advances in ML-assisted granular material modelling at both microscopic and macroscopic levels from the following aspects: 1) the application of ML models in microscopic grain scale computation; 2) the development of ML-based constitutive models of granular materials; 3) the development of ML-aided macroscopic simulation of granular materials, including both the macroscopic kinematic features-based ML model and the application of the ML method in mesh-based numerical methods (e.g. the FEM-ML framework).

The remaining paper is structured as follows: Section~\ref{section2} compares and summarises the architecture and features of some typically used neural networks in ML-aided modelling of granular materials. A brief review of grain information-based ML simulations, including the ML-based inter-particle contact feature and kinematic feature modelling of grain systems, is provided in Section~\ref{section3} from both their advantages and deficiencies of each aspect. Section~\ref{section4} showcases the latest development of ML-based constitutive studies of granular materials from aspects of the used training data sources and training methodologies of different ML frameworks. Furthermore, the prediction capability of varying ML models for stress-strain response under different loading paths is also intuitively assessed in this section. Section~\ref{section5} detailed reviews several commonly used macroscopic simulation schemes of granular materials and their coupling approaches with the ML methods, followed by a specific biaxial modeling example with the FEM-ML scheme. The discussion and concluding remarks are made in the last two sections.

\section{Typical neural networks in ML-aided modeling of granular materials} \label{section2}
A thorough comprehension of different neural networks is indispensable for effectively applying them to specific tasks. In this section, seven frequently used ML models in the simulation of granular materials are introduced, including the multi-layer perceptron (MLP), the basic recurrent neural network (RNN) as well as its family members, i.e. the long-short-term memory (LSTM) and gated recurrent unit (GRU), temporary convolutional neural networks (TCNN), convolutional neural network (CNN), and graph neural networks (GNN). According to their unique properties, they are grouped into three categories: 1) single-step-based networks, 2) multi-step-based or time-sequence networks, and 3) spatial information-based neural networks.

\subsection{The single-time step-based neural network} \label{section2.1}

A notable characteristic of single-step-based neural networks like the MLP is that their prediction relies solely on the input of the current time step, necessitating a strict surjective relationship between the input and output of the network.

\begin{figure}[H]
\centering 
\includegraphics[width =0.55 \linewidth,angle=0,clip=true]{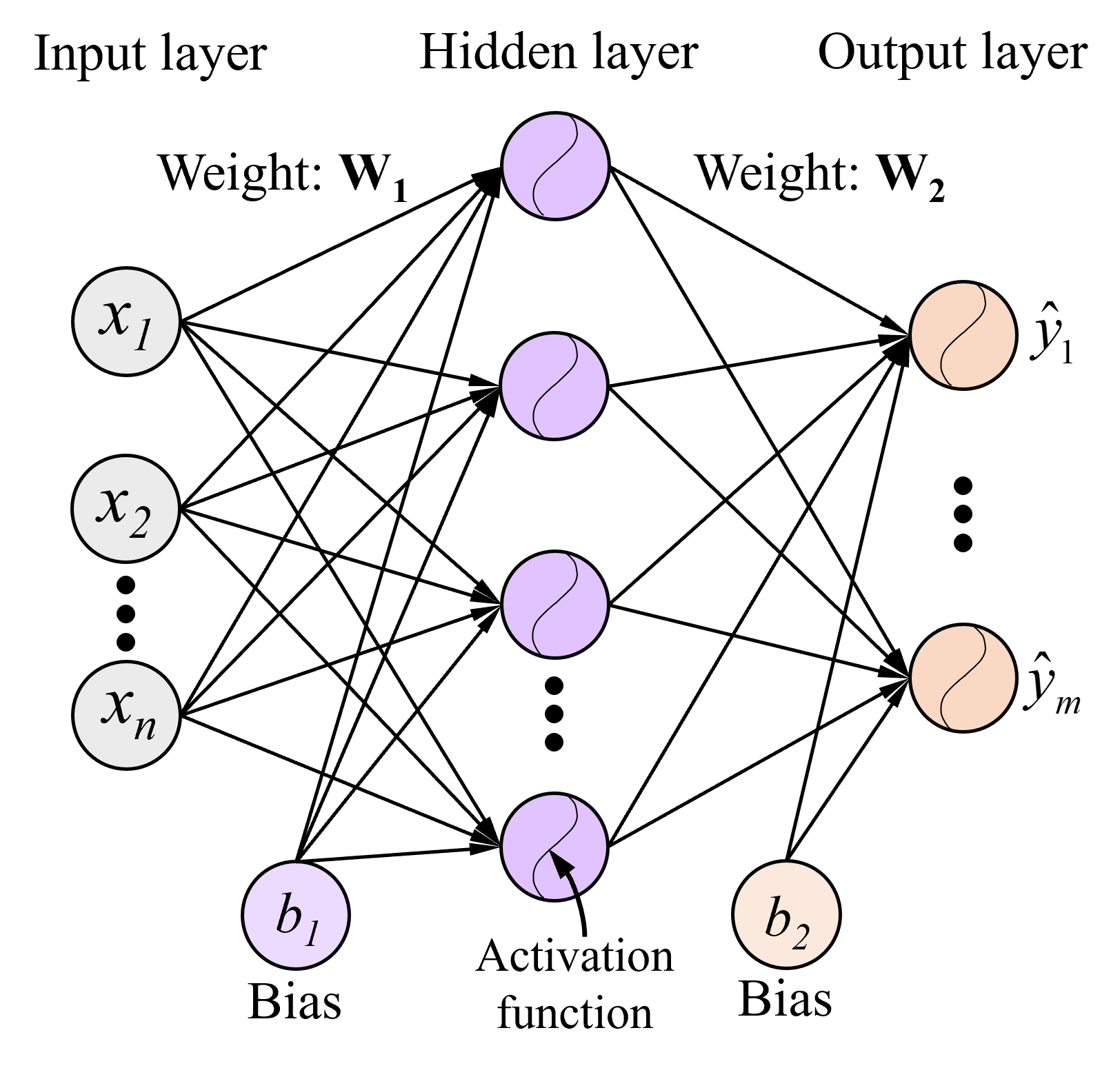}
\caption{{\color{red}The architecture of the MLP with one hidden layer}}
\label{BPNN}
\end{figure}

Fig.~\ref{BPNN} presents the architecture of the MLP with an input layer, hidden layer, and output layer, in between connected by trainable parameters (weight matrixes \(\mathbf{W_1, W_2}\), and bias vectors \(\mathbf{b_1, b_2}\)). The training process of the MLP consists of two parts: the feedforward and backpropagation procedures. To obtain the prediction values \(O\), i.e. \((\hat{y}_1,..., \hat{y}_m)\), the feedforward process of input data \(\mathbf{x}(x_1, x_2,..., x_n)\) can be expressed as:

\begin{equation}
\mathbf{H}=g\left(\mathbf{W}_1 \mathbf{x}+\boldsymbol{b}_1\right)
\end{equation}
\begin{equation}
\mathbf{O}=g\left(\mathbf{W}_2 \mathbf{H}+\boldsymbol{b}_2\right)
\end{equation}
where \(\mathbf{H}\) means the output of the hidden layer; \(f\) and \(g\) are activation functions used in hidden and output layers, respectively. After finishing the feedforward process, the error between the predicted and target values is calculated and returned to each layer to update the trainable parameters via the backpropagation algorithm. The whole process is iterated until the computed errors reach a minimum value or remain constant. The simple architecture and strong capability of mapping extremely intricate functions~\citep{rumelhart1986learning,lecun2015deep,lee2018background} make the MLP widely applied in ML-assisted numerical techniques for both regression and classification tasks~\citep{lai2022machine,hwang2022machine}.

However, the MLP also has one apparent shortcoming. In history-dependent problems, the MLP has no architecture to record history variables to distinguish history states, and thus highly relies on the artificially added history information.

\subsection{The time-sequence neural network} \label{section2.2}

The most notable feature of time-sequence ML models involves incorporating historical information to forecast the current time step. This means the output of the neural network depends not only on the current input but also on past information, making them well-suited for tackling time series forecasting tasks.

\subsubsection{Recurrent neural networks} \label{section2.2.1}

One prominent example of a time-sequence ML model is the basic RNN, which achieves this function by substituting each node of the hidden layer in the MLP with an RNN neuron~\citep{elman1990finding,werbos1990backpropagation}. Such history-dependent property makes it originally prevalent in language modeling~\citep{sutskever2011generating,graves2013generating} and machine translation~\citep{sutskever2014sequence,bahdanau2014neural}.

\begin{figure}[H]
\centering 
\includegraphics[width =0.97 \linewidth,angle=0,clip=true]{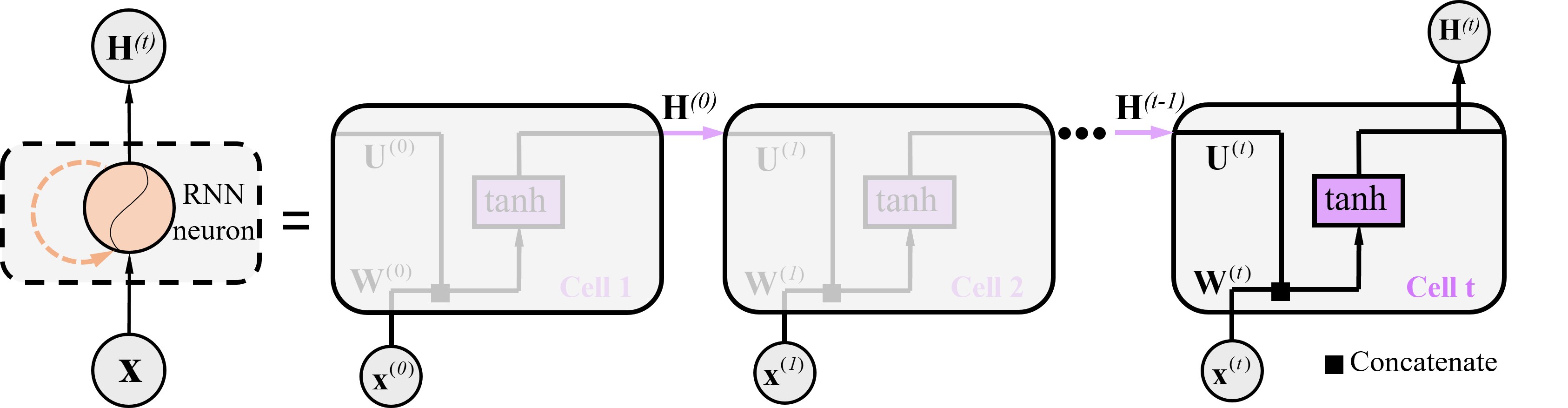}
\caption{The recurrent neuron in the basic RNN}
\label{RNN}
\end{figure}

As shown in Fig.~\ref{RNN}, the input vector \( \mathbf{x} (\mathbf{x}^{(_0)}, \ldots, \mathbf{x}^{(t)}) \) is fed into each RNN neuron featured with the recurrent connection (the orange dash line), which can be envisioned as a succession of unrolled cells across time steps, and output the hidden state \( \mathbf{H}^{(t)} \) of the current time step. In each cell, the hidden state of the last time step and the input of the current time step are considered together to generate the hidden state of the next time step until the hidden state of time step \( t \) is obtained, enabling the history information to be integrated into the output layer to predict current time step.

The feedforward process in one RNN neuron can be expressed as:
\begin{equation}
\mathbf{H}^{(t)}=\tanh \left(\mathbf{W}^{(t)} \mathbf{x}^{(t)}+\mathbf{U}^{(t)} \mathbf{H}^{(t-1)}+\boldsymbol{b}\right)
\end{equation}
\begin{equation}
\mathbf{O}^{(t)}=g\left(\mathbf{W}_2 \mathbf{H}^{(t)}+\boldsymbol{b}_{\mathbf{2}}\right)
\end{equation}
where \( \tanh \) is the activation in each RNN cell; the weight matrices connecting the input and RNN layer, as well as the adjacent cells, are denoted by \( \mathbf{W}^{(t)} \) and \( \mathbf{U}^{(t-1)} \), respectively; \( \boldsymbol{b} \) denotes the bias between the input layer and the RNN layer.

However, the basic RNN architectures normally suffer from gradient vanishing or exploding problems along the timeline, resulting in their weak capability to capture long-term dependencies. Therefore, more elaborate RNN architectures, such as the long short-term memory (LSTM)~\citep{hochreiter1997long} and the gated recurrent (GRU)~\citep{cho2014properties}, are innovated and more commonly used.

\begin{figure}[H]
  \centering
  \begin{subfigure}[b]{0.47\linewidth}
    \includegraphics[width=\linewidth]{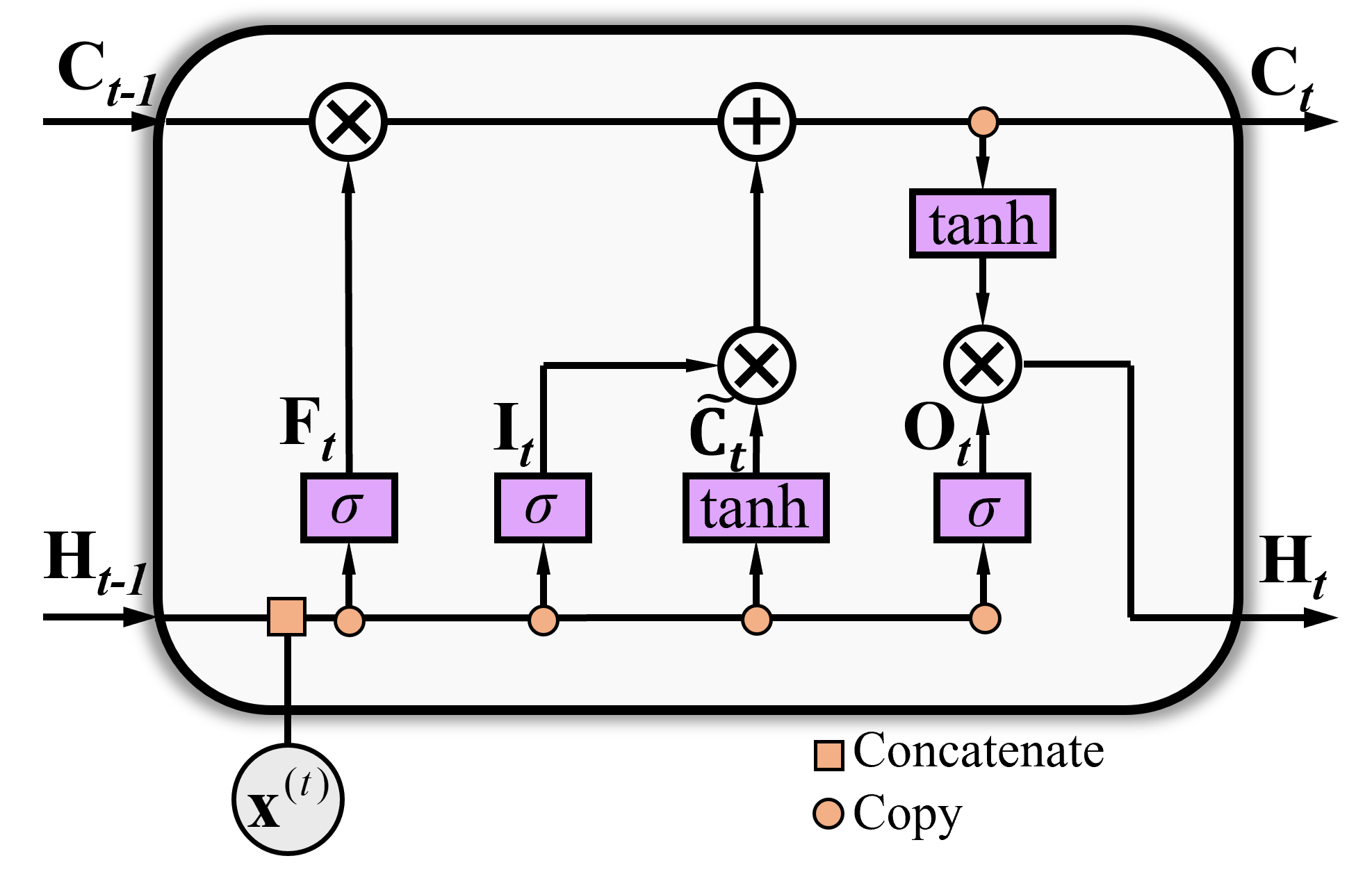}
    \caption{The memory cell of LSTM}
    \label{LSTM}
  \end{subfigure}
  \hfill
  \begin{subfigure}[b]{0.47\linewidth}
    \includegraphics[width=\linewidth]{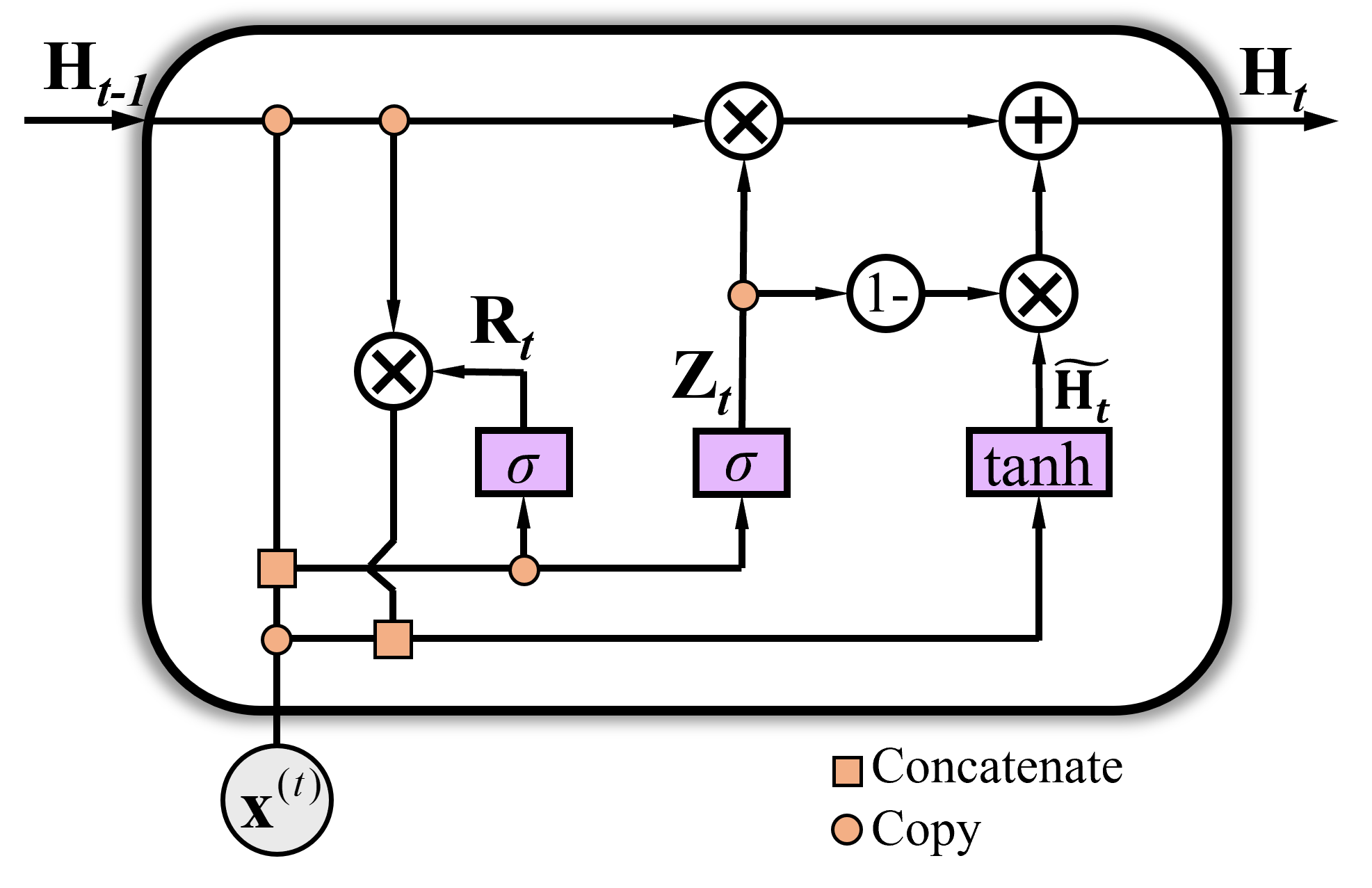}
    \caption{ the memory cell of GRU}
    \label{GRU}
  \end{subfigure}
  \caption{The different memory cells in RNNs}
  \label{LSTM_GRU}
\end{figure}

As illustrated in Fig.~\ref{LSTM}, which shows the memory cell of LSTM, the key distinction between the standard RNN and LSTM is the introduction of internal state \( \mathbf{C}_t \) and the novel inclusion of multiplicative gates i.e. the input gate \( \mathbf{I}_t \), input node \( \tilde{\mathbf{C}}_t \), forget gate \( \mathbf{F}_t \), and output gate \( \mathbf{O}_t \), in each RNN neuron, which can be calculated as:
\begin{equation}
\mathbf{F}_t=\sigma\left(\mathbf{W}_F \mathbf{x}^{(t)}+\mathbf{U}_F \mathbf{H}^{(t-1)}+\mathbf{b}_F\right)
\end{equation}
\begin{equation}
\mathbf{I}_t=\sigma\left(\mathbf{W}_I \mathbf{x}^{(t)}+\mathbf{U}_I \mathbf{H}^{(t-1)}+\mathbf{b}_I\right)
\end{equation}
\begin{equation}
\mathbf{O}_t=\sigma\left(\mathbf{W}_O \mathbf{x}^{(t)}+\mathbf{U}_O \mathbf{H}^{(t-1)}+\mathbf{b}_O\right)
\end{equation}
\begin{equation}
\tilde{\mathbf{C}}_t=\tanh \left(\mathbf{W}_C \mathbf{x}^{(t)}+\mathbf{U}_C \mathbf{H}^{(t-1)}+\mathbf{b}_C\right)
\end{equation}
where \( \mathbf{I}_t \) and \( \tilde{\mathbf{C}}_t \) determine how much new information is adopted and \( \mathbf{F}_t \) governs how much old internal state \( \mathbf{C}_{t-1} \) is inherited to update the new internal state \( \mathbf{C}_t \):
\begin{equation}
\mathbf{C}_t=\mathbf{F}_t \odot \mathbf{C}_{t-1}+\mathbf{I}_t \odot \tilde{\mathbf{C}}_t
\end{equation}
and then, the current hidden state \( \mathbf{H}^{(t)} \) which relies on the obtained internal state \( \mathbf{C}_t \) and output gate \( \mathbf{O}_t \) is updated by:
\begin{equation}
\mathbf{H}^{(\mathrm{t})}=\mathbf{O}_t \odot \tanh \left(\mathbf{C}_t\right)
\end{equation}
where $\odot$ denotes the elementwise product operator; $\sigma$ here represents the sigmoid activation function.

On the other hand, the gated recurrent unit (GRU) neural network shown in Fig.~\ref{GRU} provides a streamlined version of LSTM with a comparable performance but speeds up computation by only introducing two gates: update gate \( \mathbf{Z}_t \) and reset gate \( \mathbf{R}_t \) to the standard RNN cell. The output of these two gates can be expressed as:
\begin{equation}
\mathbf{R}_t=\sigma\left(\mathbf{W}_R \mathbf{x}^{(t)}+\mathbf{U}_R \mathbf{H}^{(t-1)}+\boldsymbol{b}_R\right)
\end{equation}
\begin{equation}
\mathbf{Z}_t=\sigma\left(\mathbf{W}_Z \mathbf{x}^{(t)}+\mathbf{U}_Z \mathbf{H}^{(t-1)}+\boldsymbol{b}_Z\right)
\end{equation}
where \( \mathbf{R}_t \) determines how much previous hidden state is kept in the current candidate hidden state \( \tilde{\mathbf{H}}^{(t)} \):
\begin{equation}
\widetilde{\mathbf{H}}^{(t)}=\tanh \left[\mathbf{W} \mathbf{x}^{(t)}+\mathbf{U}\left(\mathbf{R}_t \odot \mathbf{H}^{(t-1)}\right)\right]
\end{equation}
and \( \mathbf{Z}_t \) controls the extent to which the new hidden state \( \mathbf{H}^{(t)} \) matches the old state \( \mathbf{H}^{(t-1)} \) and how similar it is to the new candidate state \( \tilde{\mathbf{H}}^{(t)} \), which can be expressed as:
\begin{equation}
\mathbf{H}^{(t)} = \mathbf{Z}_t \odot \mathbf{H}^{(t-1)} + (1 - \mathbf{Z}_t) \odot \tilde{\mathbf{H}}^{(t)}
\end{equation}

Incorporating more sophisticated gate architectures in the standard RNN can effectively alleviate the gradient vanishing or explosion problem, but such structures also lead to the rapid growth of training parameters and excessive memory usage to store the result of each cell with the increase of the time step. Furthermore, the nature that the prediction of the current hidden state must wait for the completion of its predecessors in RNNs limits their parallelism.

\subsubsection{The temporal convolutional neural network} \label{section2.2.2}

The temporal convolutional neural network (TCNN)~\citep{bai2018empirical}, a variant of the convolutional neural network (CNN) which will be introduced in detail later in Section~\ref{section2.3}, is another ML model that can be used in time-sequence forecasting problems. Different from the RNN which uses the memory cells to integrate history information, the TCNN extracts the past information by scanning the input data along the time-step direction using filters with the shape of \textit{kernel size} \( \times \) \textit{Depth}, in which trainable parameters are stored.

\begin{figure}[H]
\centering 
\includegraphics[width =0.9 \linewidth,angle=0,clip=true]{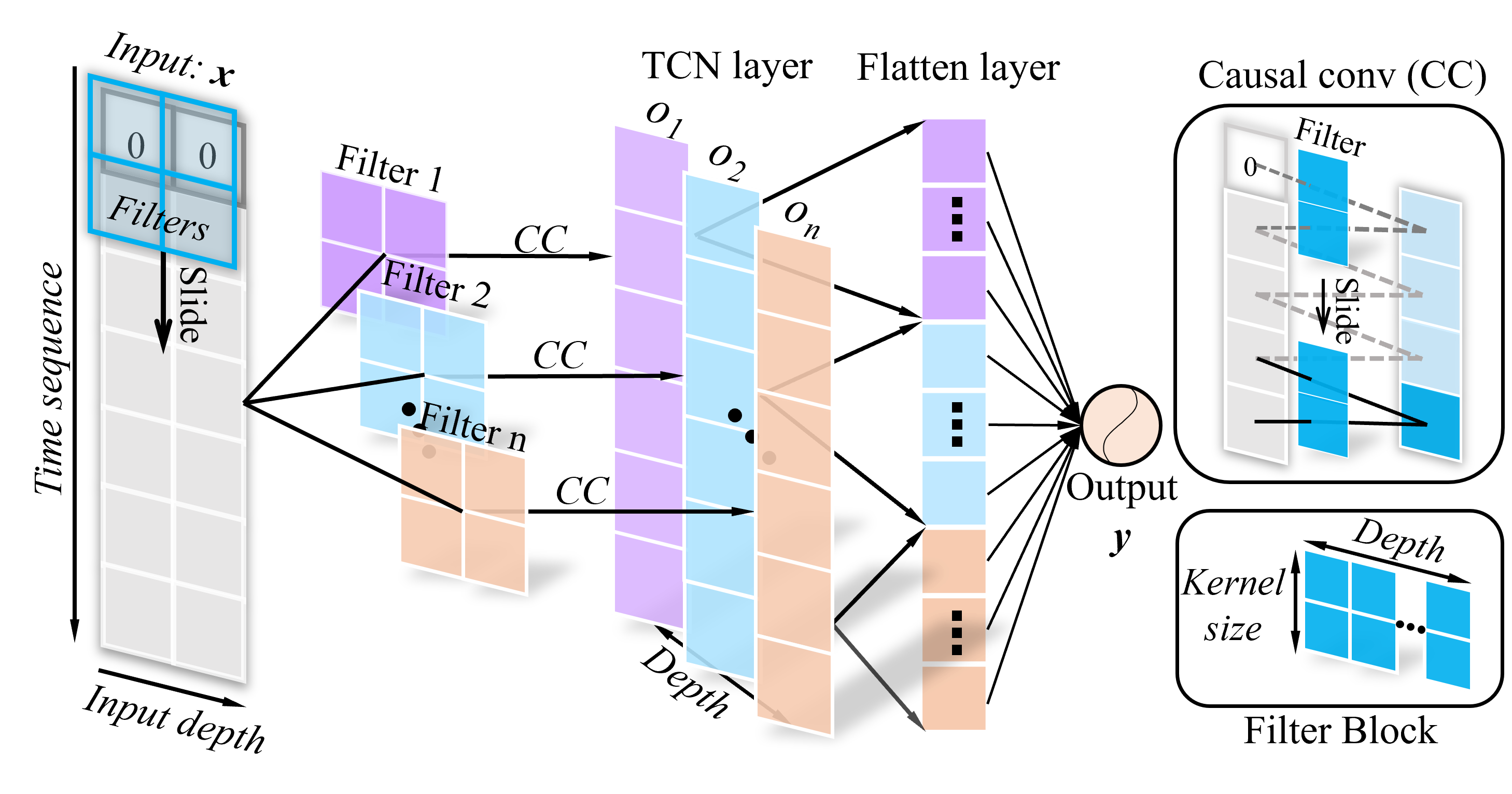}
\caption{{\color{red}The feedforward process in the temporal convolution neural network.}}
\label{TCN}
\end{figure}

Fig.~\ref{TCN} provides a detailed feedforward process in the TCNN. The input \( x \) is composed of a 2D array with one representing time steps and the other representing depth (i.e. the number of columns of the input data). To guarantee only the past and current data are involved and an equal length between the input and output columns of each convolution layer, the causal convolution (CC) in which the forefront of input data is padded with zero along the direction of time sequence. Where the number of the padded zero line is equal to $kernel size-1$. In the CC procedure, \( n \) filters will simultaneously make dot product calculations with the corresponding elements of input data along the direction of the time step with fixed stride \( s \) and obtain \( n \) corresponding output series (\( o_1, o_2, \ldots, o_n \)). The obtained output vectors are then flattened to one column and fed into the output neuron to predict the value of the current time step.

As the TCNN filters are independent, it is easy to parallelize the training process across GPU cores~\citep{bai2018empirical}. Furthermore, the number of training parameters is solely dependent on the number of filters, which means when the time steps of the training data increase, there are no extra training parameters introduced. In addition, the transfer learning process is easy to conduct between two TCNN models as long as the number and shape of filters remain constant.

However, the TCNN also has some notable disadvantages. Unlike the LSTM or GRU in which the needed history information is adjustable to predict the current time step, the TCNN cannot filter out the redundant historical information when transferring a model from a domain where only little memory is needed (i.e., small \textit{kernel size}) to a domain where much longer memory (i.e., large \textit{kernel size}) is required. In addition, the development of one TCNN normally needs several CC layers, which may need a large memory to store the trained parameters, resulting in a lower training efficiency.

\subsection{The geometry information-based neural networks} \label{section2.3}

The prediction of the granular mechanical response is traditionally considered a time-sequential problem and mainly treats granular information as vectors to develop the ML surrogate model, which ignores the rich spatial features of granular assembly. The recent development of ML technology also enables researchers to directly extract physical information of particles from figures or graphics to predict the behaviour of granular media. The two most commonly used geometry information-based neural networks are the CNN and the graph neural network (GNN), respectively.

\subsubsection{The convolutional neural network} \label{section2.3.1}

The advent of CNN~\citep{fukushima1980neocognitron,weng1993learning}, has revolutionized fields like image recognition~\citep{ciregan2012multi,ciresan2011flexible} and video analysis~\citep{baccouche2011sequential,ji20123d}. Leveraging CNNs, it becomes feasible to predict granular properties by encoding the structural characteristics of particle assemblies into a pixel matrix. Similar to the TCNN which extracts the temporal features with filters, the CNN collects spatial features of figures by filters.

\begin{figure}[H]
\centering 
\includegraphics[width =0.99 \linewidth,angle=0,clip=true]{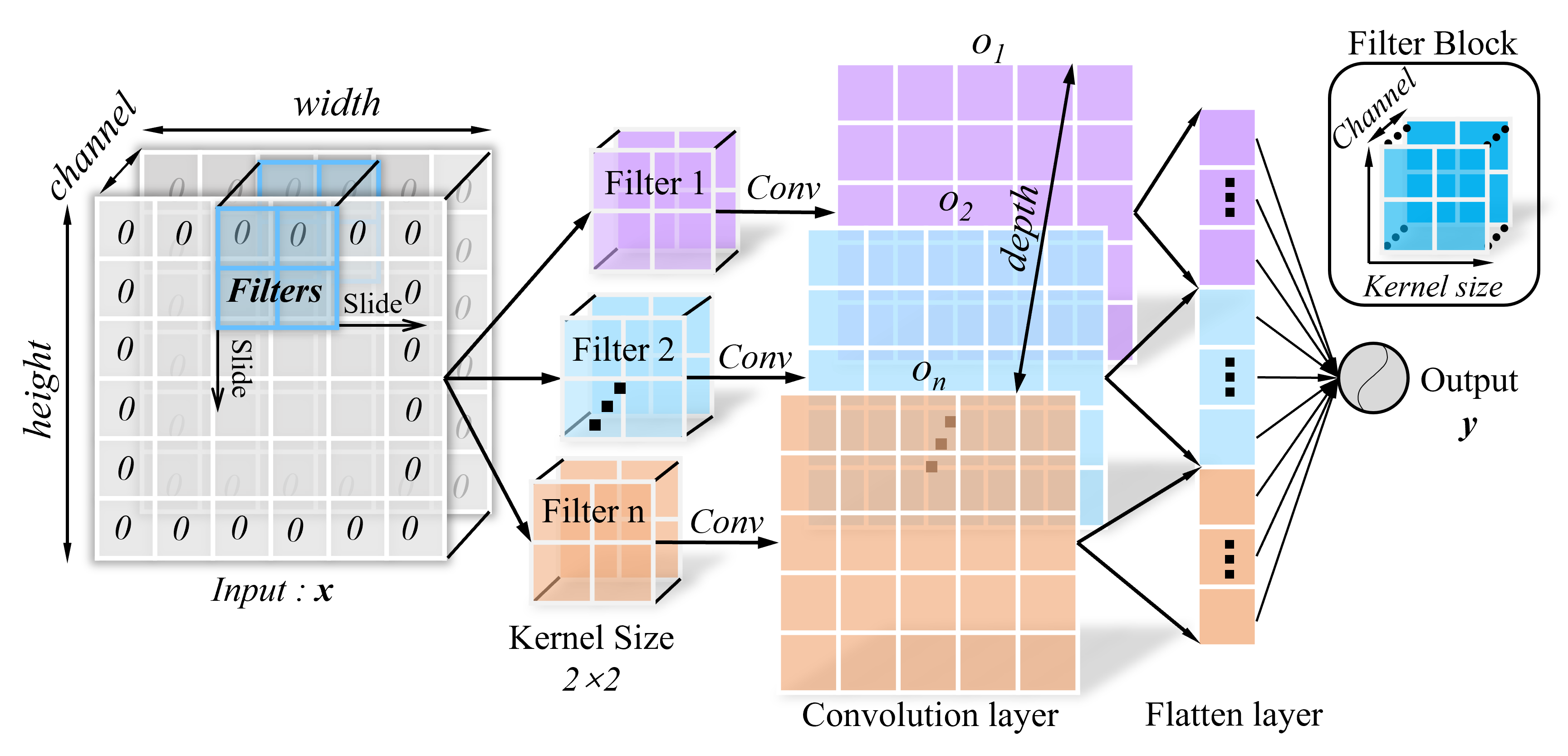}
\caption{{\color{red}The feedforward process of the convolutional neural network.}}
\label{CNN}
\end{figure}

Fig.~\ref{CNN} illustrates the whole feedforward process in CNN. Unlike the TCNN, which requires input data in a sequential time-based format, CNN regards the image with abundant spatial information as the input data; thus, the time dimension is not required. Instead, the input is represented by its height (\(h\)), width (\(w\)), and channel (\(c\)) dimensions, where \(c\) represents the number of input images. Correspondingly, the dimension of the filter is changed to 3D but with the same height and width as the kernel size \(k\). Furthermore, unlike TCNN, where the filter can only move along the direction of the time step, the filter in CNN scans the input data from both the height and width directions at a fixed stride. It is worth noting that to keep the constant shape between the input and output, zero-padding around the input pixel is still required, similar to the approach used in TCNN.

Compared to TCNN, CNNs have some unique features. The first is its interpretability, as CNNs can be visualized to show which part is significant for the prediction. The second notable feature is its location invariance. Filters are trained to detect various spatial or shape features, such as edges or corners, enabling CNNs to recognize objects in images regardless of their position. 

However, CNNs also have some limitations. For example, CNNs are not suitable for non-grid input data. Furthermore, their performance can be impacted when input images are occluded or contain significant noise. In addition, training one CNN to recognize meaningful patterns normally needs a large amount of labelled data, which is a great challenge when data is expensive to acquire.

\subsubsection{Graph neural network} \label{section2.3.2}

In addition to CNN, the GNN~\citep{scarselli2008graph, micheli2009neural,wu2022graph} are also capable of learning the structural information and topological features of graphs, which makes them a powerful tool for capturing the dynamic features of grain assemblies~\citep{aoyama2023optimal, choi2024graph}.

\begin{figure}[H]
\centering 
\includegraphics[width =0.9 \linewidth,angle=0,clip=true]{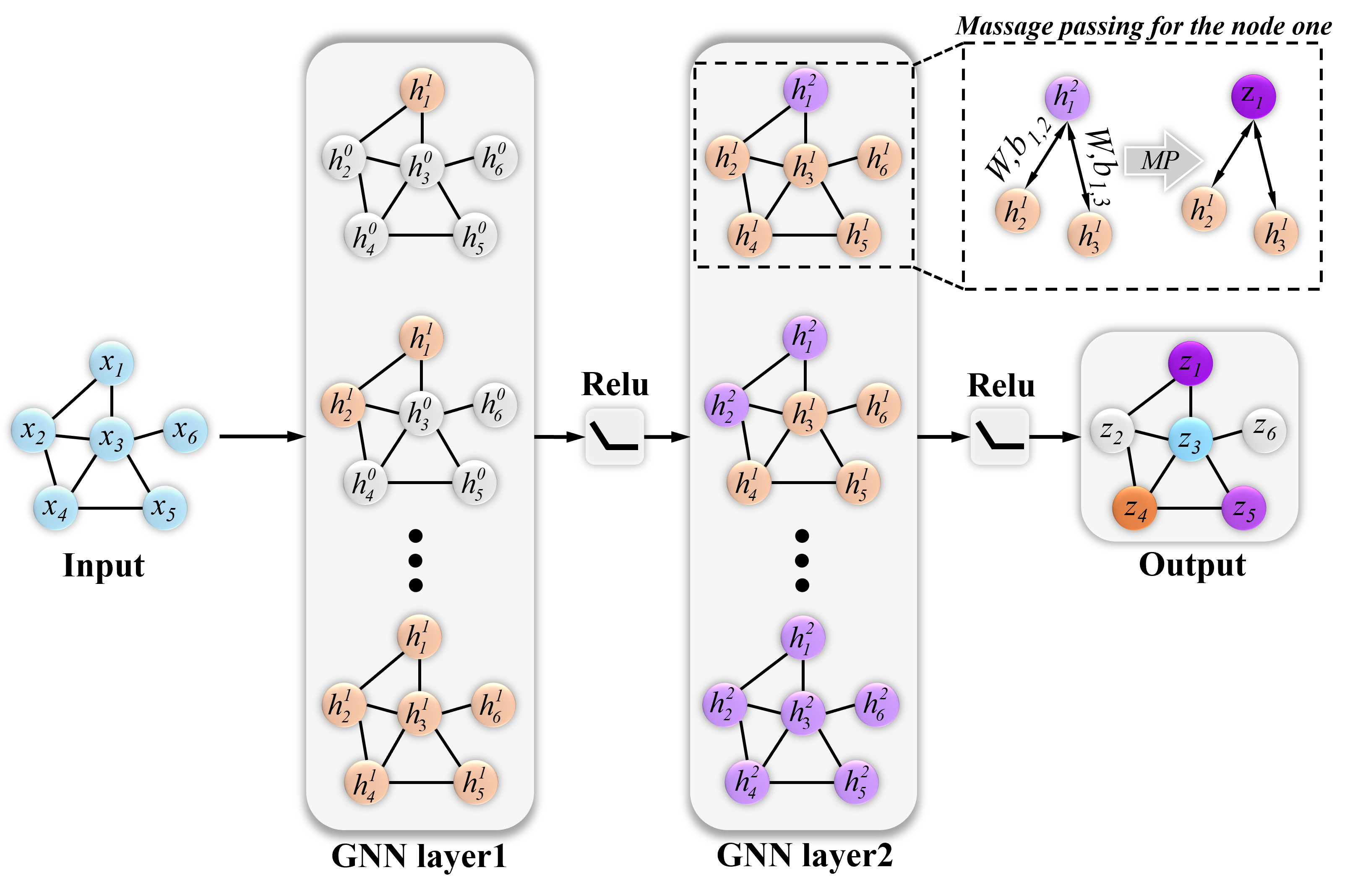}
\caption{The architecture of the graph neural network.}
\label{GNN}
\end{figure}

An example architecture of the GNN is illustrated in Fig.~\ref{GNN}. The GNN takes one graph consisting of six nodes as the input. In the input layer, also called the \(0^{th}\) layer, the node feature of each vertex is represented by \(x_v\), and all node information \(\boldsymbol{h}_{v}^{0}\left(h_{1}{ }^{0}, h_{2}{ }^{0}, \ldots, h_{6}{ }^{0}\right)\) is transported to the GNN layer1. In the GNN layer1, the node feature is alternately updated and the newly-generated output \(\boldsymbol{h}_v^{1} (h_1^{1}, h_2^{1}, \ldots, h_6^{1})\) is passed to the next GNN layer. This process is repeated until the output layer acquires the prediction result \(\boldsymbol{z}_v\) where \(\boldsymbol{z}_v = \boldsymbol{h}_v^{K}\), and \(K\) denotes the output layer.

In GNN, the feature of each node \( h_v^k \) is updated by the message passing (MP) process, which aggregates the information received from their adjacent nodes \( u \). There are normally three different ways to integrate the message from the neighbouring nodes, i.e. mean, sum, and max. Regarding the mean method as one example, the MP procedure of the node \( v \) in the \( k^{th} \) GNN layer can be expressed as:
\begin{equation}
    h_v^k = \boldsymbol{f} \left( \mathbf{W}^k \sum_{u \in N(v)} \frac{h_u^{k-1}}{N(v)} + \mathbf{b}^k h_v^{k-1}, \forall k \in \{1, ..., K\} \right)
\end{equation}
where \( h_u^{k-1} \) represents the output features of nodes that connect to the vertex \( v \) from the \( (k-1)^{th} \) layer. \( N(v) \) denotes the number of adjacent nodes of vertex \( v \). The item \( \sum_{u \in N(v)} \frac{h_u^{k-1}}{N(v)} \) aims to aggregate the received message of node \( v \) from the previous layer in an average way. \( \mathbf{W}^k \) and \( \mathbf{b}^k \) are trainable parameters stored on each edge in the \( k^{th} \) GNN layer. \( f \) represents the activation function (e.g. ReLU).

Compared with the MLP, the RNNs, CNNs, and the GNN have some distinct advantages. Different from RNNs and CNNs, whose input data needs to be arranged in one certain order, the GNN acquires knowledge by updating the message stored in each node and thus disregarding the order of input data. In addition, the interaction relationship between different objects can be captured by the edges of graphs, but networks, like the MLP and RNNs, cannot explicitly reflect this dependency relationship. Therefore, the GNN is quite suitable for analyzing kinematic features of structural scenarios, e.g. granular systems.

However, it is worth noting that the GNN also confronts many challenges. One obvious limitation is its vulnerability to the modification of graph structure. When nodes and edges are added or removed, the GNN cannot adaptively adjust the network structure. Furthermore, similar to CNNs, the GNN is also not robust to noise. Introducing a slight noise into the graph via node perturbation or addition/deletion of edges can cause an adversarial impact on the output of GNNs. In addition, in graphs consisting of numerous nodes, a large amount of training parameters are required to represent relationships between adjacent objects, which results in a low training efficiency of GNN.

\subsection{Summary}
\label{section2.4}

The networks used in the ML-aided granular material simulation have their unique architectures, advantages, and shortcomings, which are summarized in Table~\ref{table1}. It is found that all these neural networks excel at solving non-linear or high-dimensional problems. Meanwhile, to enhance their versatility in addressing various complex tasks, neural network architectures tend to become more sophisticated, for example, the evolution of basic RNN to the LSTM and GRU. The growing structural complexity of neural networks does enhance their ability to approximate or predict more intricate functions but meanwhile increases training parameters, potentially reducing the training efficiency.

\begin{table}[H]
\centering
\caption{Features of seven typically used neural networks in the ML-aided granular materials simulation}
\label{table1}
\scalebox{0.8}{
\begin{tabular}{c|l|l}
\hline
\begin{tabular}[c]{@{}c@{}}Neural\\ networks\end{tabular} & \multicolumn{1}{c|}{Advantages}                                                                                                                                                                                                                                                                         & \multicolumn{1}{c}{Disadvantages}                                                                                                                                                                                                                                  \\ \hline
MLP                                                      & \begin{tabular}[c]{@{}l@{}}1). Simple architecture;\\ 2). High training efficiency;\end{tabular}                                                                                                                                                                                                        & \begin{tabular}[c]{@{}l@{}}1). Requiring artificially added history variables\\   in time-sequence problems; \\ 2). Gradient vanishing or explosion;\\ 3). Sensitive to noise and irregularities\end{tabular}                                                    \\ \hline
RNN                                                       & \begin{tabular}[c]{@{}l@{}}1). No extra history variables are needed\\   in sequential prediction; \\ 2) Strong non-linear mapping capability.\end{tabular}                                                                                                                                             & \begin{tabular}[c]{@{}l@{}}1). More complex network architecture than MLP;\\ 2). More training parameters than MLP; \\ 3). Gradient vanishing or explosion; \\ 4). Weak ability to record long-history information; \\ 5). Weak parallelism;\end{tabular} \\ \hline
LSTM                                                      & \begin{tabular}[c]{@{}l@{}}1). No extra history variables are needed \\   in sequential prediction;\\ 2). Strong non-linear mapping capability;\\ 3). Eliminate the gradient vanishing or explosion\\   by the gate structure;\\ 4). Strong ability to capture long-term\\   dependencies.\end{tabular} & \begin{tabular}[c]{@{}l@{}}1). Increased structural complexity than RNN; \\ 2). More training parameters than RNN; \\ 3). Weak parallelism\end{tabular}                                                                                                           \\ \hline
GRU                                                       & \begin{tabular}[c]{@{}l@{}}1). Similar to LSTM; \\ 2). Simpler gate structures than LSTM\end{tabular}                                                                                                                                                                                                   & Similar to LSTM                                                                                                                                                                                                                                                    \\ \hline
TCNN                                                      & \begin{tabular}[c]{@{}l@{}}1). No extra history variables are needed\\   in sequential prediction; \\ 2). Strong non-linear mapping capability; \\ 3). Excellent parallelism; \\ 4). Good portability of trained parameters; \\ 5). Suitable for longer history information.\end{tabular}               & \begin{tabular}[c]{@{}l@{}}1). Numerous training parameters; \\ 2). Weak ability to filter out the  redundant history information\end{tabular}                                                                                                                 \\ \hline
CNN                                                       & \begin{tabular}[c]{@{}l@{}}1). Strong high-dimensional mapping capability; \\ 2). Excellent parallelism; \\ 3). Good portability of trained parameters; \\ 4). Good visualization; \\ 5). Location invariance of input; \\ 6). Strong ability to get rich spatial features.\end{tabular}                  & \begin{tabular}[c]{@{}l@{}}1). Numerous training parameters; \\ 2). Not available in non-grid input data; \\ 3). Weaker resistance to noise; \\ 4). Requirement for large amounts of labelled data\end{tabular}                                                    \\ \hline
GNN                                                       & \begin{tabular}[c]{@{}l@{}}1). Available in non-grid input data; \\ 2). Strong ability to get rich spatial features; \\ 3). Strong ability to reflect the \\   relationship of adjacent objects.\end{tabular}                                                                                           & \begin{tabular}[c]{@{}l@{}}1). Numerous training parameters;  \\ 2). Vulnerability to the modification of graph structure; \\ 3). Less resistance to noise\end{tabular}                                                                                        \\ \hline
\end{tabular}}
\end{table}

Furthermore, different neural networks possess distinct advantages and limitations. For instance, the MLP is particularly useful in approximating one-to-one or many-to-one mappings, while RNNs (including LSTM and GRU) and TCNNs are more adept at time series forecasting tasks, owing to their unique architectures. While CNNs and GNNs perform well at extracting spatial features from the input data. Instead, the focus should be on identifying the most suitable neural network for the designated task based on its specific strengths and weaknesses. 

In addition to the seven typical neural networks used in ML-assist granular materials modelling, other neural networks, such as the radial basis function neural network (RBFNN), Bi-LSTM, residual CNN, etc., are also developed based on the aforementioned typical neural networks. These networks inherit the strengths and limitations of their predecessors, but will not be discussed in detail here.

\section{The microscopic grain information-based ML models}\label{section3}

In the field of granular mechanics, DEM has been very popular for modelling the mechanical behaviour of various granular materials and related engineering problems, such as landslides~\citep{liu2020investigation}, shear deformation of soil and sand~\citep{lai2017characterization,thakur2020triaxial}, failure of tunnel surface~\citep{yin2020effect} and fluidized bed~\citep{lu2021machine}. While these methods can reflect the discrete nature of granular media from the grain scale to a certain degree, they typically suffer from intensive computational costs. The advent of deep learning methods offers a potential way to alleviate the computational burdens by integrating ML models with these conventional (microscopic) grain/particle-based techniques. In this section, a concise review of relevant studies in this area is presented. Before that, a brief introduction to the basic framework of the DEM is given.

\subsection{The basic framework of the discrete element method}\label{section3.1}

In contrast to the continuum approach, the DEM represents granular material as an assembly of distinct particle entities, and the overall (macroscopic) behaviour of the system is governed by inter-particle contacts over time, making it superior to address the large-deformation problem.
\begin{figure}[H]
\centering 
\includegraphics[width =0.99 \linewidth,angle=0,clip=true]{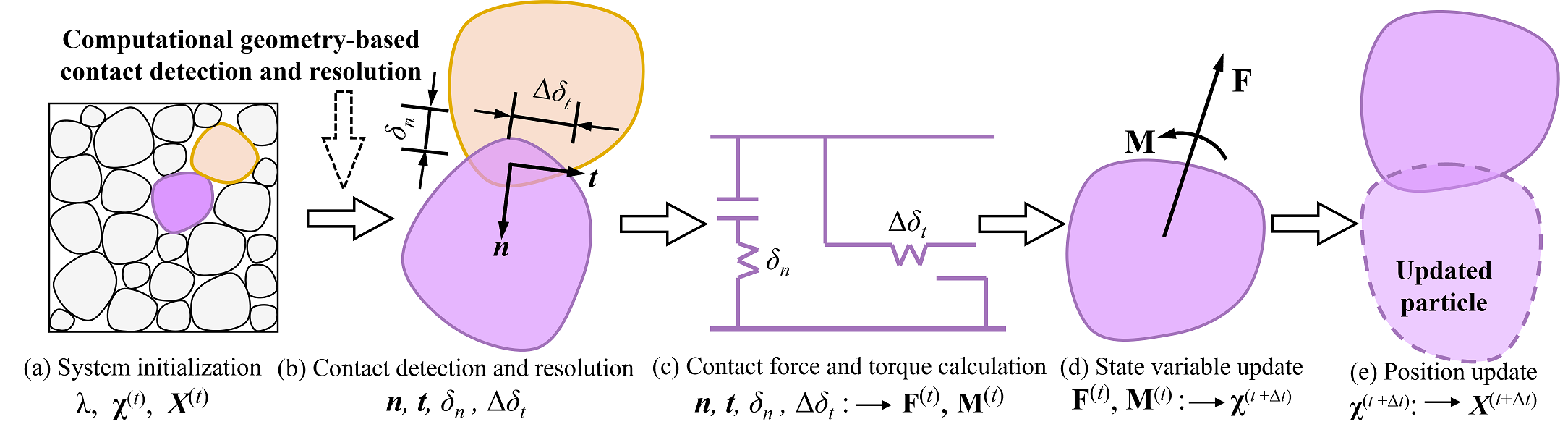}
\caption{{\color{red}The computation process in DEM.}}
\label{DEM}
\end{figure}

Fig.~\ref{DEM} illustrates the fundamental computational steps of the DEM, encompassing the following stages:

a) System initialization: This step involves 1) assigning material properties  $\lambda$  (e.g. mass $m$  and friction coefficient  $\mu$  ) to particles within the current system; and 2) initializing the geometry features  $\boldsymbol{X}^{(t)}$  (e.g. shape, size, and positions)~\citep{cleary2002modelling,peters2009poly,lim2014granular,feng2017generic,lai2020fourier} and state parameters  $\boldsymbol{\chi}^{(t)}$  (e.g. velocity  $\boldsymbol{v}^{(t)}$, angle velocity  \(\boldsymbol{w}^{(t)}\), acceleration  \(\boldsymbol{a}^{(t)}\), and angle acceleration  \(\boldsymbol{\alpha}^{(t)}\)  to each particle.

b) Contact detection and resolution: identifying pairs of particles that are in contact through collision detection algorithms~\citep{feng2023thirty}. Subsequently, computing the contact features (e.g. contact norm  $\boldsymbol{n}$, tangent  $\boldsymbol{t}$, and corresponding inter-particle overlaps  $\delta_{n}$, $\Delta \delta_{t}$  ) between each pair using collision resolution algorithms.

c) Contact force and torque calculation: utilizing the acquired contact features and contact model~\citep{hertz1882ueber,cundall1979discrete,johnson1987contact}, the resultant contact force  $\mathbf{F}^{(t)}$  and contact torque  $\mathbf{M}^{(t)}$, which respectively govern the translational and rotational motion of particles, are calculated.

d) State variable update: the acceleration  $\boldsymbol{a}^{(t+\Delta t)}$  and the angular acceleration  $\boldsymbol{\alpha}^{(t+\Delta t)}$  are respectively updated with Newton's second law with obtained contact force and torque. This update process further refreshes the state variables, including both the  $\boldsymbol{v}^{(t+\Delta t)}$  and  $\boldsymbol{w}^{(t+\Delta t)}$.

e) Position update: the geometry information of the particle (i.e. their location) is updated based on newly obtained state variables. After the particle positions are updated within the current time step, a new iteration will begin.

\subsection{The ML-aided discrete element modeling}
\label{3.2}

In the traditional DEM calculation process, the contact detection and resolution process are the most computationally intensive steps~\citep{williams1999discrete,lai2022machine}. Therefore, leveraging the superior computational efficiency inherent in ML models, the development of ML-based models for contact detection and resolution shows significant potential in accelerating the calculation process of the DEM.

\begin{figure}[H]
\centering 
\includegraphics[width =0.99 \linewidth,angle=0,clip=true]{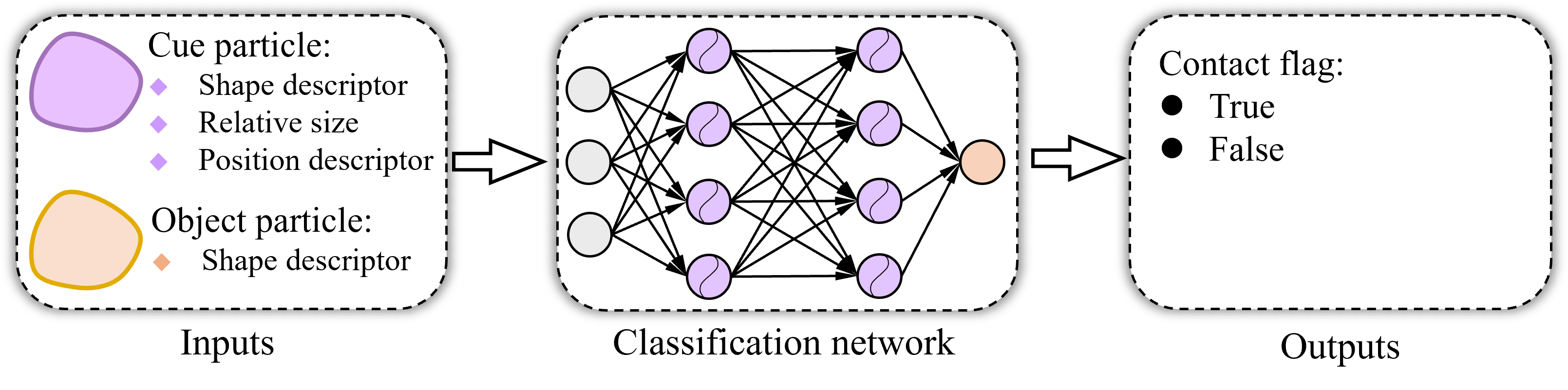}
\caption{{\color{red}The classification neural network for contact detection}~\citep{lai2022machine}}
\label{ML_contact_detection}
\end{figure}

\begin{figure}[H]
\centering 
\includegraphics[width =0.99 \linewidth,angle=0,clip=true]{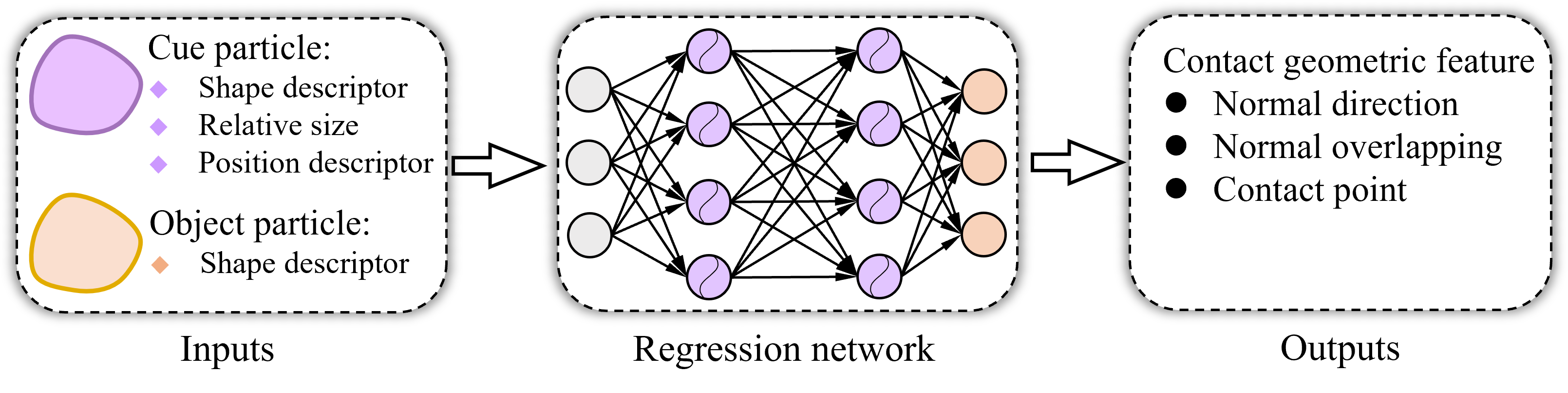}
\caption{{\color{red}The regression neural network for contact resolution}~\citep{lai2022machine}}
\label{ML_contact_resolution}
\end{figure}

Within the ML-enhanced DEM framework, the contact detection and resolution processes are achieved through a classification and regression neural network, respectively. As shown in Fig.~\ref{ML_contact_detection} and Fig.~\ref{ML_contact_resolution}, the key difference between these two networks is their output layers. In the classification network, the output of the final layer is 0 or 1, indicating the contact status of two particles (one considered as an object grain and the other as a cue particle). While the regression network outputs contact features such as contact point, normal, and overlap between two particles given their requisite geometric features.

Lai et al.~\citep{lai2022machine} extracted the contact status and contact features of particles with elliptical and arbitrary shapes by a classification network and a regression network, respectively. Both these two networks take the same parameters as input, including the shape parameters of the object grain, and the size, shape as well as position parameters of the cue particles. The obtained two networks are then embedded into the DEM algorithm to model cases, e.g. random packing, iodometric compression, angle of repose, packing, and compression of arbitrarily irregular-shaped particles. Similarly, Hwang et al.~\citep{hwang2022machine} used one classification and one regression ANN to respectively predict the contact states (i.e. detached or intersected) and contact features of two identical irregular particles, including the mean contact point, averaged norm vector of overlapped vertices, and inter-penetration depth. In their work, to train these two networks, the labelled contact properties of two particles were computed by the deepest point method according to their relative position and orientation which were fed into ML models as the input. 

\subsection{The ML-based grain-level kinematic features simulations}\label{section3.3}

The evolution of the microstructure in granular materials, influenced by the kinematic features of particles, significantly affects the mechanical behaviour of the material. Therefore, many researchers have endeavoured to develop ML models that can directly capture the particle motion laws without calculating the contact forces using empirical or analytical contact models. 

Based on the ML method, both the collision law of sparse objects and the deformation of dense grain media have been widely investigated. For instance, in the work of Katerina et al~\citep{fragkiadaki2015learning}, the trajectory of the billiard is predicted via a CNN which takes the current and previous image information of the system as input and outputs the velocity of the goal object in the future time steps. Meanwhile, a novel object-centric prediction method is used to enhance the translational invariance of the acquired physical laws by the ML model. Wu et al.~\citep{wu2015galileo} integrated the ML model into the physical engine to perceive dynamic features of objects in the future time step according to received physical information (e.g. position, friction coefficient, shape) of objects at previous time steps. Battaglia et al~\citep{battaglia2016interaction} embedded the physical state of the system at the previous step into one graph to construct the interaction network which can reason the collision rules of objects in the complicated system interact in the future step. Based on the ML method, Chang et al.~\citep{chang2016compositional} built a Neural Physics Engine (NPE) which takes the past pair velocity of both the goal and its adjacent objects as input to predict the velocity of objects at the next time step in the system. 

On the other hand, ML models trained with the data generated by DEM simulations have been employed to accelerate the computing of DEM simulations by replacing the contact models with ML models. In the work of Ummenhofer et al~\citep{ummenhofer2019lagrangian} and Lu et al ~\citep{lu2021machine}, a CNN that can predict the collision laws of inter-particles was constructed with continuous filters to substitute the direct computation of particle-particle/boundary collisions in DEM modelling to accelerate the simulation of grains flow. Li et al~\citep{li2023prediction} designed a GNN that takes the acquired static microstructure of grain packing from DEM simulation as input to predict the contact force of the grain system under the uniaxial compression condition. Cheng et al~\citep{cheng2022estimation} used the particle and contact kinematical data obtained from a DEM simulator to train the GNN which can estimate the contact forces of granular assemblies under the uniaxial compression condition. Bapst et al~\citep{bapst2020unveiling} trained GNN with the solely initial position information of the grains system to represent the long-term evolution of a glassy system under the shear condition. Mayr et al~\citep{mayr2023boundary} develop a Boundary-Graph Neural Network (BGNN) to model the interaction of particles with complex boundary conditions. Virtual nodes are inserted into the graph dynamically to represent the boundary surface regions within the cutoff radius of particles and have features encoding triangle orientation representing the boundary. The authors successfully replicate the simulation of 3D granular flows through hoppers, rotating drums, and mixtures. Kumar et al.~\citep{kumar2023accelerating}hypothesize that the GNN messages encode latent representations that preserve the underlying interaction laws between particles in a DEM simulation. The sparse representation of the GNN messages \(\left(\mathrm{e}_{\mathrm{k}} \leftarrow \phi\left(\mathrm{e}_{\mathrm{k}}, \mathrm{v}_{\left\{\mathrm{r}_{\mathrm{k}}\right\}}, \mathrm{v}_{\left\{\mathrm{s}_{\mathrm{k}}\right\}}, \mathrm{u}\right)\right)\) is a learned linear combination of the true forces. To learn a maximally sparse message representation, the authors sort the message vector components by standard deviation and enforce an L1 regularization, forcing the GNN to describe the messages in a minimal vector space. With this approach, they successfully recover the fundamental linear spring interaction law \( F_n = k_n \cdot \text{abs}(\Delta x - r_i - r_j) \) based only on the kinematics of the DEM particles.

We can see most kinematic features-based ML models take the position and velocity of individual grain at previous and current time steps as input and directly output the acceleration of the corresponding particle in the next step to update the current grain state of the whole system, which bypasses the contact force and torque calculation; thereby significantly improve the computational efficiency. However, the existing research indicates that the majority of obtained models are trained using data collected from granular systems composed of circular or spherical particles, and thus ignore the influence of particle shape on the grain trajectory.

\subsection{Summary}

{\color{red}This section summarizes the work related to the grain information-based ML models from two aspects. The first is developing the ML-based contact model based on contact state and contact geometric features in particle-based numerical methods, which has received comparatively less attention from researchers. The second aspect is leveraging appropriate ML models, e.g. (CNN or GNN), to predict the kinematic features for both sparse and dense grain systems at the microscopic scale. Both of these two directions show promise in reducing the intensive computation cost in particle/grain-based numerical techniques. A summary of the advantages and limitations of ML-based microscopic grain information models is provided in Table~\ref{summary3}.}

\begin{table}[H]
\centering
\caption{{\color{red}An overview for the ML-based microscopic grain information models}}
\label{summary3}
\scalebox{0.85}{
\begin{tabular}{c|l|l}
\hline
\begin{tabular}[c]{@{}c@{}}ML-aided\\ microscopic modelling\end{tabular}                         & \multicolumn{1}{c|}{Advantages}                                                                                                                                                                                 & \multicolumn{1}{c}{Disadvantages}                                \\ \hline
\begin{tabular}[c]{@{}c@{}}ML-based\\ contact models\end{tabular}                                & \begin{tabular}[c]{@{}l@{}}1). High computational efficiency\\ compared to traditional contact models.\\ 2). Strong ability to account for the\\ influence of particle shape.\\ 3). User friendly.\end{tabular} & \begin{tabular}[c]{@{}l@{}}1). The training data normally\\ incorporates contact assumptions\\ 2). Limited by the completeness \\ of the training data\end{tabular}                                      \\ \hline
\begin{tabular}[c]{@{}c@{}}ML-aided \\ grain-level\\ kinematic feature\\ simulation\end{tabular} & \begin{tabular}[c]{@{}l@{}}1). Lower computational complexity\\ 2). High computational efficiency\end{tabular}                                                                                                  & \begin{tabular}[c]{@{}l@{}}1). Error accumulation problem\\ 2). Ignoring the influence\\ of particle rotation\\ 3). Weaker suitability to the system\\ where particles are added or removed\end{tabular} \\ \hline
\end{tabular}}
\end{table}

{\color{red}Compared to the traditional contact models, ML-based contact models offer one significant advantage by directly outputting contact features and relationships between two particles given their positions, bypassing the traditional contact detection and resolution process, which accelerates the overall computational process of the DEM. Furthermore, the ML-based contact models have a strong capability to account for the influence of particle shape, as the trained ML model can instantly provide contact features when given the necessary geometric information of particles, regardless of their shape. This represents an improvement over traditional contact detection methods, which often simplify grains into spheres (3D) or circles (2D) to improve computational efficiency but overlook the influence of the rotation resistance of particles. Additionally, although more advanced contact detection algorithms and contact models with rigorous theoretical foundations have been developed, their implementation remains challenging for engineers and researchers unfamiliar with computational geometry ~\citep{feng2021energy, feng2021generic}.

In ML-aided grain-level kinematic feature simulations, the fine-tuned geometry information networks can simultaneously predict the acceleration of all grains based on their positions from the previous time step. This approach eliminates the need for the iterative calculation of particle accelerations typically required in traditional discrete element modeling, thereby significantly enhancing simulation efficiency.

However, it is also important to acknowledge the limitations associated with each aspect. The development of ML-based contact models is primarily hindered by the completeness and quality of the training data, including contact geometry information. The contact state and contact features are directly related to the relative position and geometry shape of two particles. To create an ML model that can accurately predict contact information for any contact position between two particles, the training data must encompass a wide range of contact conditions, which typically requires significant time and effort to generate, especially in grains with complex shapes. In addition, the training data generation relies on the theoretical contact models used. For example, the definition of contact geometry features, e.g. the overlap distance and the direction of contact norm, are empirical and can vary in different contact models. This disparity further limits the development of ML-based contact models. 

On the other hand, when employing GNNs or CNNs to predict the rollouts of particle systems, error accumulation becomes unavoidable. This occurs because the dynamic update at the current time step depends on the particle positions from the previous step. Furthermore, apart from the position and translation velocity, particle rotation, and angular velocity are freedoms of individual particle/grain of granular materials. It seems that current kinematic features-based ML models are unable to reflect these fundamental physics. Additionally, it should be noted that geometry information-based networks, such as GNNs, are not well-suited for systems in which the number of particles changes during the simulation. This limitation arises from the inherent structure of graph-based networks, which require a fixed number of nodes to maintain consistency throughout the simulation.}

{\color{red}
In addition to the aforementioned discussion, several promising avenues for future research are also proposed in this section, aiming to address current limitations and enhancing existing methodologies:

1). Enhancing training data for ML-based contact models: The critical challenge in developing one universal ML-based contact model is the quality and completeness of the training data. Consequently, future research could focus on developing new methodologies to define contact features beyond traditional empirical models, thereby minimizing inherent assumptions in the training datasets and improving the overall data quality. Additionally, efforts could be directed toward generating more comprehensive and diverse training datasets to enhance the generalization capability of ML-based contact models across a wider range of contact scenarios.

2). Mitigating error accumulation in particle system rollouts: Error accumulation is a key limitation when utilising geometry information-based networks, particularly in long-term simulations of particle systems. Therefore, future works could explore error correction mechanisms, such as active learning, to dynamically adjust the model prediction during simulations to eliminate error propagation.

3). Incorporating rotational dynamics into ML-based kinematic feature models: Current ML-based models for grain-level kinematic feature simulations focus primarily on positional and translational velocities, ignoring rotational dynamics. Future work could focus on integrating rotational degrees of freedom, such as particle rotation and angular velocity, into ML models, which may involve extending current CNN or GNN architectures to account for these additional physical factors.}

\section{The ML-based constitutive models of granular materials}\label{section4}

The development of the ML-based constitutive model of granular materials can be traced back to the 1990s~\citep{sidarta1998constitutive,ghaboussi1998new,penumadu1999triaxial}, and the development of the ML-based constitutive model of granular materials is undeniably one of the most prominent subjects in ML-assist numerical methods. The construction of ML surrogate stress-strain models for grain media depends on numerous factors, such as the hyper-parameters, optimization algorithm, loss function, etc., used in neural networks~\citep{zhang2021state}. However, the two fundamental factors that govern the performance of the ML-based constitutive model are the data resource and the feature selection of input-output corresponding to the used networks.

\subsection{The data sources}\label{section4.1}

Table \ref{table2} offers a summary of partial works on ML-based constitutive models of granular materials over the past decades, there are mainly two types of data resources used when developing the ML constitutive models of granular materials. One is the experiment data, and the other is synthetic data.

\subsubsection{The experiment data}
\label{section4.1.1}

The experiment data, which directly reflects the stress-strain response of granular materials, implicitly encapsulates the most authentic constitutive laws without any assumption, and thus the ML models developed from the experiment data can reveal the most essential mechanical features of granular materials. As listed in Table \ref{table2}, the mechanical response of different granular materials, such as soil, sand, clay, ballast, and rockfill, has been investigated by different neural networks, where most ML models focus on the mechanical behaviour of granular materials under the drained and undrained triaxial test, and the remaining research dedicates to develop the ML surrogate models which can represent the mechanical features of granular media under direct shearing~\citep{sezer2011prediction}, simple shearing~\citep{wang2018multiscale},  and tension-shear~\citep{wang2019meta}.

While the experiment data can provide reliable inputs for neural networks to extract underlying principles governing the behaviour of materials, the limitations of the experiment data should also be taken into consideration. The training of neural networks normally requires a sufficient amount of data samples, making a purely experimental data-driven approach expensive. Additionally, restricted by the experimental facility, most experiment data used for training ML models are generated under specific shearing or triaxial compression conditions, covering only a partial range of stress-strain space and material types, and thus the robustness of trained ML models is limited.

\subsubsection{The synthetic data}
\label{section4.1.2}

Compared to experiment data, synthetic data can be a cost-effective alternative to experimental data to develop ML-based constitutive models for granular material, as experimental constraints do not bind it and can span a wider range of stress-strain space. As demonstrated in Table \ref{table2}, the synthetic data generally can be acquired in two approaches. The first is the phenomenological constitutive models, such as the critical state-based models~\citep{yao2009uh,yin2011modeling} and the deviatoric hardening model~\citep{poorooshasb1985yielding, pande2020role}, and the other is the particle-based numerical techniques, such as the discrete element method (DEM).

Given these advantages, it is possible to generate extensive amounts of synthetic data encompassing various materials and more extensive stress-strain space to establish more robust machine learning models. Ma et al.~\citep{ma2022predictive} provides an example of this, where one ML model which is capable of simulating the stress-strain response of granular materials with different particle size distributions (PSDs) and initial void ratio (\(e_0\)) under random loading paths was obtained through DEM-generated data. In addition, the development of the synthetic data-based ML model can also provide prior knowledge for constructing experiment data-based machine learning models. In reference~\citep{basheer2002stress}, several mapping methods based on synthetic data were compared before selecting the true sequential dynamic mapping method for simulating the cyclic behaviour of soils with experiment data.

While ML models derived from synthetic data can capture the fundamental mechanical characteristics of granular materials under specific loading paths, they are limited in uncovering deeper constitutive laws, since synthetic data are generated under some assumptions (e.g. the homogenization in theoretical models) and simplification (e.g. the simplified shape of grain in DEM), which results in the loss of some intrinsic physical information of granular materials. However, there is no doubt that synthetic data could be a cost-effective supplement to experimental data in the development of ML-based constitutive models.

\begin{table}[]
\centering
\caption{The development of ML-based constitutive models for granular materials}
\label{table2}
\scalebox{0.85}{
\begin{tabular}{ccccccc}
\hline
Material                                                   & Reference                                         & Experiment    & Loading & Drained & \begin{tabular}[c]{@{}c@{}}Data\\ source\end{tabular}                    & \begin{tabular}[c]{@{}c@{}}Neural\\ network\end{tabular} \\ \hline
Sand                                                       & \citep{sidarta1998constitutive}  & Triaxial      & M       & D       & Experiment                                                               & MLP                                                     \\
Sand                                                       & \citep{ghaboussi1998new}         & Triaxial      & M       & D+U     & Experiment                                                               & MLP                                                     \\
Soil                                                       & \citep{zhu1998modelling}         & Triaxial      & M       & D       & Experiment                                                               & MLP                                                     \\
Sand                                                       & \citep{penumadu1999triaxial}     & Triaxial      & M       & D       & Experiment                                                               & MLP                                                     \\
/                                                          & \citep{basheer2000selection}     & /             & C       & /       & \begin{tabular}[c]{@{}c@{}}Synthetic data,\\ and experiment\end{tabular} & MLP                                                     \\
\begin{tabular}[c]{@{}c@{}}Coarse\\ sand\end{tabular}      & \citep{romo2001recurrent}        & Triaxial      & M       &         & Experiment                                                               & RNN                                                      \\
/                                                          & \citep{basheer2002stress}        & /             & C       & /       & Synthetic data                                                           & MLP                                                     \\
\begin{tabular}[c]{@{}c@{}}Lateritic\\ gravel\end{tabular} & \citep{habibagahi2003neural}     & Triaxial      & M       & D       & Experiment                                                               & MLP                                                     \\
Sand                                                       & \citep{banimahd2005artificial}   & Triaxial      & M       & U       & Experiment                                                               & MLP                                                     \\
Ballast                                                    & \citep{shahin2006modeling}       & Triaxial      & M       & D       & Experiment                                                               & MLP                                                     \\
/                                                          & \citep{fu2007integration}        & Triaxial      & M       & U       & \begin{tabular}[c]{@{}c@{}}Synthetic data\\ (MCC)\end{tabular}           & MLP                                                     \\
Sand                                                       & \citep{hashash2008integration}   & Triaxial      & M       & D       & Experiment                                                               & MLP                                                     \\
Soil                                                       & \citep{he2009modeling}           & Triaxial      & M       & U       & Experiment                                                               & MLP                                                     \\
\begin{tabular}[c]{@{}c@{}}Lateritic\\ gravel\end{tabular} & \citep{johari2011modelling}      & Triaxial      & M       & D       & Experiment                                                               & MLP                                                     \\
Soil                                                       & \citep{lv2011study}              & Triaxial      & M       & U       & Experiment                                                               & MLP                                                     \\
Sand                                                       & \citep{sezer2011prediction}      & Direct shear  & M       & /       & Experiment                                                               & MLP                                                     \\
Rockfill                                                   & \citep{araei2014artificial}      & Triaxial      & M       & D       & Experiment                                                               & MLP                                                     \\
Sand                                                       & \citep{rashidian2014application} & Triaxial      & M       & D       & Experiment                                                               & MLP                                                     \\
/                                                          & \citep{stefanos2015neural}       & Triaxial      & M       & D       & \begin{tabular}[c]{@{}c@{}}Synthetic data\\ (HS)\end{tabular}            & MLP                                                     \\
Sand                                                       & \citep{kohestani2016modeling}    & Triaxial      & M       & D       & Experiment                                                               & MLP                                                     \\
/                                                          & \citep{li2017applying}           & /             & M       & /       & \begin{tabular}[c]{@{}c@{}}Synthetic data\\ (DEM)\end{tabular}           & MLP                                                     \\
/                                                          & \citep{wang2018multiscale}       & Simple shear  & C       & /       & \begin{tabular}[c]{@{}c@{}}Synthetic data\\ (DEM)\end{tabular}           & LSTM                                                     \\
/                                                          & \citep{wang2019meta}             & Tension-shear & C       & /       & \begin{tabular}[c]{@{}c@{}}Synthetic data\\ (DEM)\end{tabular}           & GRU                                                      \\
/                                                          & \citep{zhang2019investigation}   & /             & M       & /       & \begin{tabular}[c]{@{}c@{}}Synthetic data\\ (MCC)\end{tabular}           & LSTM                                                     \\
/                                                          & \citep{zhang2020ai}              & Triaxial      & C       & D + U   & \begin{tabular}[c]{@{}c@{}}Synthetic data\\ (EM)\end{tabular}            & LSTM                                                     \\
/                                                          & \citep{qu2021towards}            & Triaxial      & C       & /       & \begin{tabular}[c]{@{}c@{}}Synthetic data\\ (DEM)\end{tabular}           & GRU                                                      \\
/                                                          & \citep{wang2022data}             & Triaxial      & C       & /       & \begin{tabular}[c]{@{}c@{}}Synthetic data\\ (DEM)\end{tabular}           & TCNN                                                     \\
/                                                          & \citep{ma2022predictive}         & Triaxial      & M       & /       & \begin{tabular}[c]{@{}c@{}}Synthetic data\\ (DEM)\end{tabular}           & LSTM                                                     \\ \hline
\end{tabular}}
\end{table}

\subsection{The training strategy for different ML-based constitutive models}\label{section4.2}

\subsubsection{The history-dependent features of ML-based constitutive models}\label{section4.2.1}

\begin{figure}[H]
  \centering
  \begin{subfigure}[b]{0.47\linewidth}
    \includegraphics[width=\linewidth]{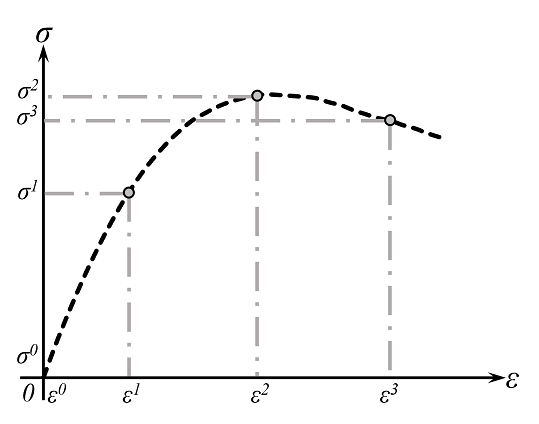}
    \caption{ The monotonous loading case.}
    \label{mono_loading}
  \end{subfigure}
  \hfill
  \begin{subfigure}[b]{0.47\linewidth}
    \includegraphics[width=\linewidth]{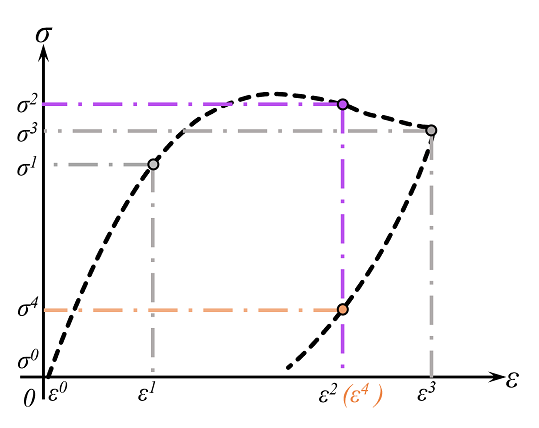}
    \caption{ The reversal loading case }
    \label{reversal_loading}
  \end{subfigure}
  \caption{Different loading paths in 1D plastic deformation}
  \label{mono_reversal}
\end{figure}

Under the quasi-static condition, as shown in Fig.~\ref{mono_loading}, the relationship between the strain tensor \( \boldsymbol{\varepsilon}^{(t)} \) and stress tensor \( \boldsymbol{\sigma}^{(t)} \) of each time step in the monotonous loading cases keeps the one-to-one mapping which can be pursued as:
\begin{equation}
\boldsymbol{\sigma}^{(t)}=\boldsymbol{f}\left(\boldsymbol{\varepsilon}^{(t)}, \mathbf{W}, \mathbf{b}\right)
\end{equation}

However, when subjected to loading reversal, as demonstrated in Fig.~\ref{reversal_loading}, a reversal loading may result in two identical strains (e.g. \( \varepsilon^{2} = \varepsilon^{4} \)) corresponding to different stress states, and therefore, it is necessary to introduce the history variables to differentiate the loading state (i.e. the unloading and reloading).

Different from the traditional theoretical constitutive models, in which loading states can be explicitly formulated, ML models depend on other methods to differentiate the loading history under cycling loading conditions. Time-sequence neural networks draw history information from their input data consisting of discrete strain-stress data sequences: 
\begin{equation}
    \hat{\boldsymbol{\sigma}}^{(t)}=\boldsymbol{f}^{M L}\left(\left\{\boldsymbol{\varepsilon}^{(t-n)}, \ldots, \boldsymbol{\varepsilon}^{(t-2)}, \boldsymbol{\varepsilon}^{(t-1)}, \boldsymbol{\varepsilon}^{(t)}\right\}, \mathbf{W}, \mathbf{b}\right)
\end{equation}
For instance, RNNs learn loading history by recording the hidden state encoded in the input data with RNN neurons, while TCNN identifies different loading states by the causal convolution calculation with trainable filters. 

However, single-step-based neural networks like the MLP cannot distinguish the loading history itself. Therefore, they need to introduce artificially added history/internal variables as extra input to transform the one-to-many mapping into a surjective between the input and output variables, which can be expressed as:

\begin{equation}
\hat{\boldsymbol{\sigma}}^{(t)}=f^{\text{MLP}}\left(\boldsymbol{\varepsilon}^{(t)}, \varphi^{(t)}, \mathbf{W}, \mathbf{b}\right)
\end{equation}
where \( \boldsymbol{\sigma}^{(t)} \) and \( \boldsymbol{\varepsilon}^{(t)} \) represent the current stress and strain states of granular materials, \( \boldsymbol{\varphi}^{(t)} \) means the history variables and consists of \( \{ \varphi_1^{(t)}, \ldots, \varphi_m^{(t)} \} \); \( \textbf{W} \) and \( \textbf{b} \) mean trainable parameters of the ML model.

\subsubsection{The selection of input features in ML models}\label{section4.2.2}

In addition to the data resource, selecting the necessary information from the obtained data as the input and output features to train the neural network is also a key aspect in the development of ML models. Generally, the input variable of the neural network is normally a combination of the following three types of parameters: 1) the material parameters, such as the PSDs, and the relative density \( D_r \) of the granular assembly, representing the basic intrinsic properties of materials; 2) the state parameters, e.g. the void ratio \( e \), the mean effective stress \( p \), the deviatoric stress \( q \), the stress \( \boldsymbol{\sigma} \), and strain \( \boldsymbol{\varepsilon} \) at the current time step \( t \), which govern the evolution of the stress-strain relationship; 3) the history parameters which is used to differentiate loading states.

The variables used as the input and output of the data-driven model vary according to specific problems. As listed in Table~\ref{table2}, ML models are constructed with single-step-based or time-sequence neural networks to simulate the behaviour of granular materials with cycling or monotonous loading data. Thanks to their specific architectures, time-sequence neural networks like the LSTM, GRU, and TCNN inherently acquire historical information from their input data, which comprises a sequence of discrete data ~\citep{qu2021towards,ma2022predictive,wang2022data}. Thus their input variables always consist of material and state parameters whether in monotonous or cycling loading 

\begin{figure}[H]
  \centering
  \begin{subfigure}[b]{0.42\linewidth}
    \includegraphics[width=\linewidth]{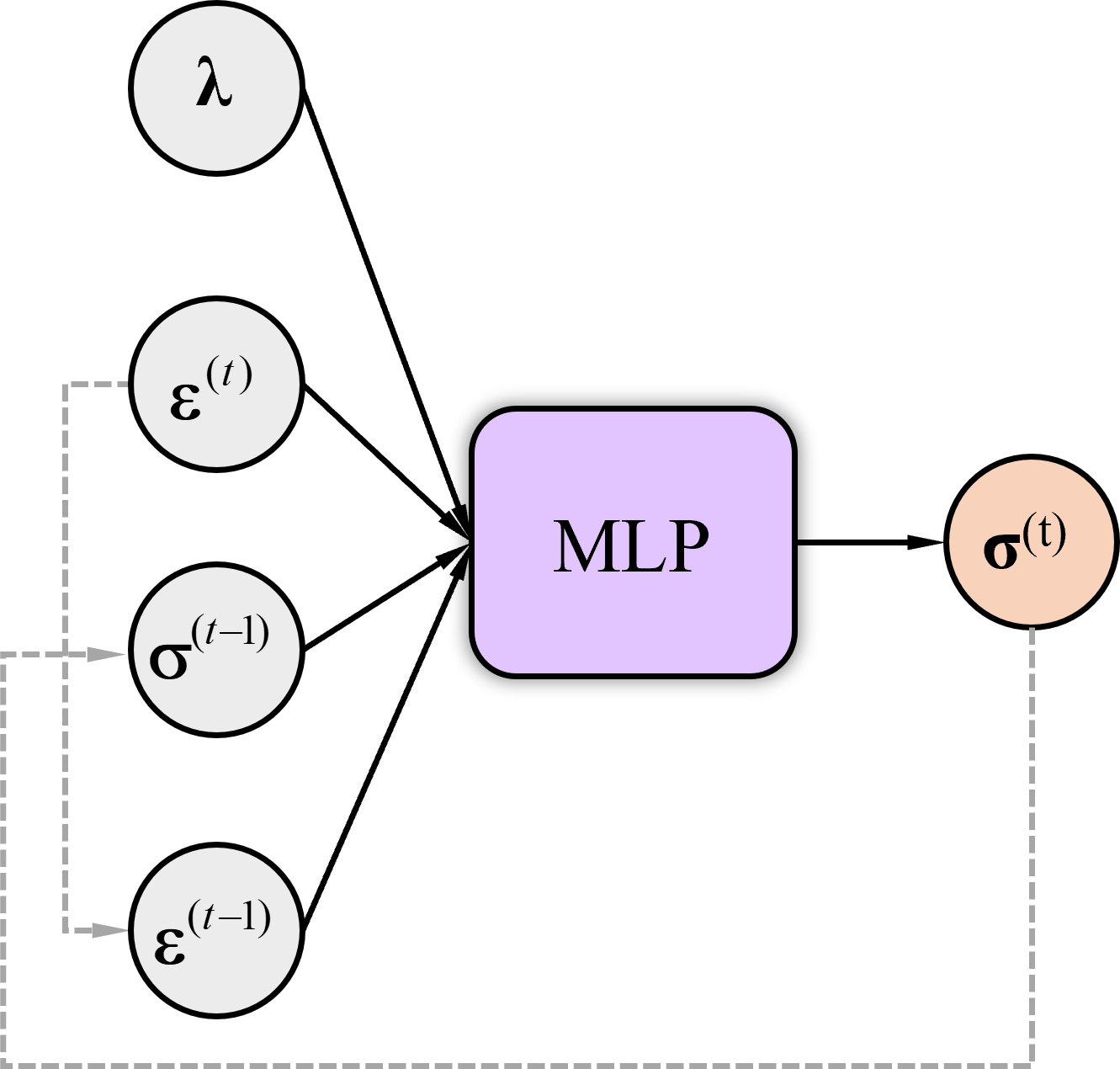}
    \caption{ Regarding state variables as history variables~\citep{hashash2004numerical}.}
    \label{total_BPNN}
  \end{subfigure}
  \hfill
  \begin{subfigure}[b]{0.42\linewidth}
    \includegraphics[width=\linewidth]{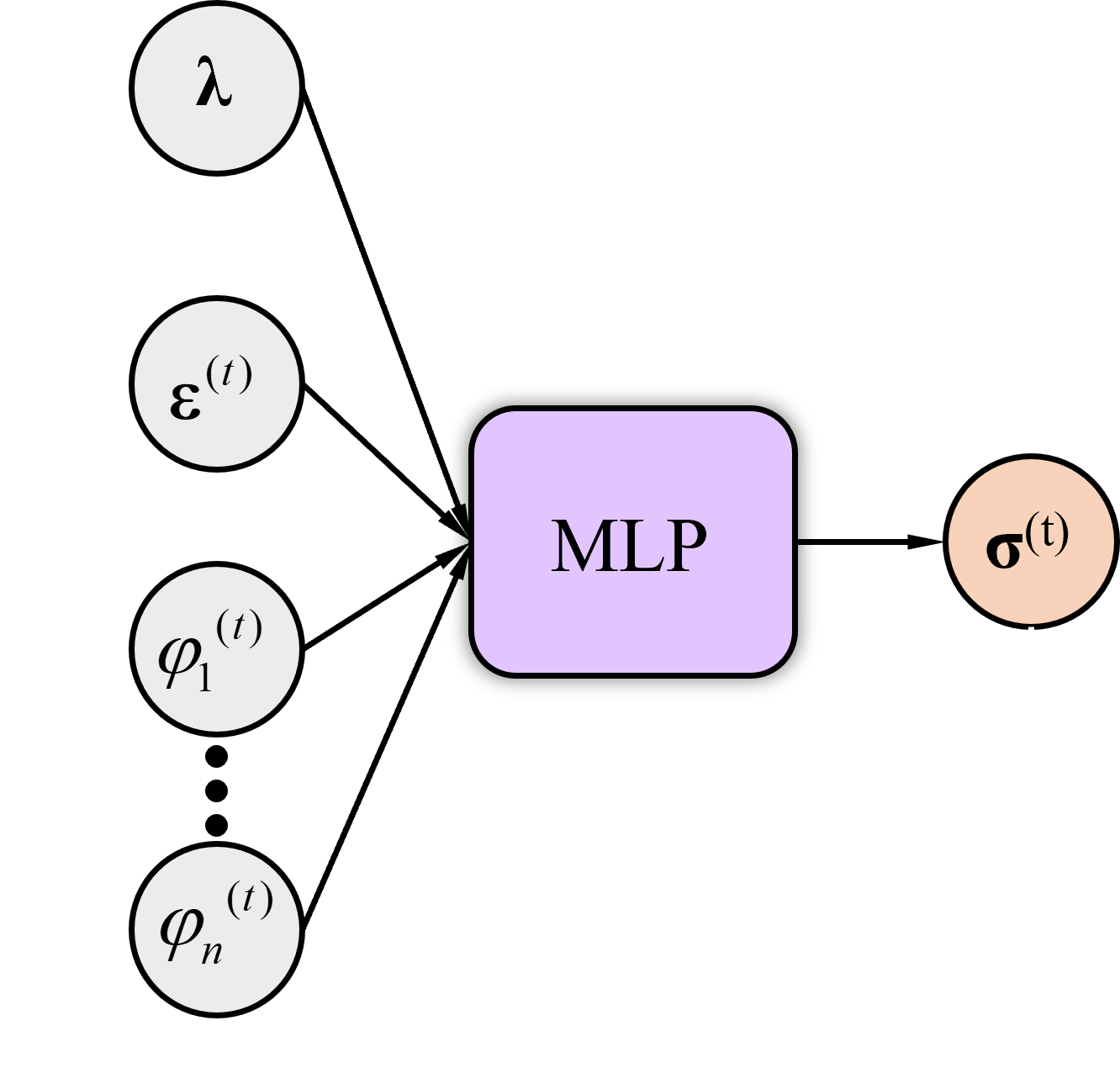}
    \caption{Regarding parameterized variables as history variables~\citep{huang2020machine}}
    \label{featured_hist}
  \end{subfigure}
  \caption{Different two methods for the selection of history variables in the MLP}
  \label{fig11}
\end{figure}

In single-step-based networks such as MLP, to consider the influence of the loading history on the constitutive relationship, there are generally two approaches to selecting history variables. The first one is regarding the predicted state parameters of the last time step as the history variables to predict the current state, such as works~\citep{sidarta1998constitutive,basheer2002stress,hashash2008integration,stefanos2015neural}. As illustrated in Fig.~\ref{total_BPNN}, the current prediction stress $\hat{\boldsymbol{\sigma}}^{(t)}$ not only relies on the current state variable strain $\boldsymbol{\varepsilon}^{(t)}$ but also the state variable of the strain $\boldsymbol{\varepsilon}^{(t-1)}$ and the predicted stress $\hat{\boldsymbol{\sigma}}^{(t)}$ at the $(t-1)^{th}$ time step. The process can be formulated as:
\begin{equation}
\hat{\boldsymbol{\sigma}}^{(t)}=f^{\mathrm{ML}}\left(\boldsymbol{\lambda}, g(\boldsymbol{\varepsilon}^{(t)},\boldsymbol{\sigma}^{(t)}), \varphi\left\{\boldsymbol{\varepsilon}^{(t-1)}, \hat{\boldsymbol{\sigma}}^{(t-1)}\right\}, \mathbf{W}, \mathbf{b}\right)
\end{equation}
where $\boldsymbol{\lambda}$ represents the vector consisting of material parameters. However, one inevitable issue is that any predicted error of the previous stress state by neural network can lead to error accumulation in the ML system.

As demonstrated in Fig.~\ref{featured_hist}, an alternative scheme is directly extracting history variables $\boldsymbol{\varphi}^{(\mathrm{t})}$ from the state variable of the current step, such as using the absolute accumulated strain increment $\Delta \boldsymbol{\varepsilon}^{(\mathrm{t})}$ of the current time step as the history variable~\citep{huang2020machine,guan2023machine}, which can be computed as:
\begin{equation}
    \boldsymbol{\varphi}^{(t)}= 
    \begin{cases}\Delta \boldsymbol{\varepsilon}^{(t)}=0 & t=0 \\ \Delta \boldsymbol{\varepsilon}^{(t)}=\sum_{k=1}^t\left|\boldsymbol{\varepsilon}^{(k)}-\boldsymbol{\varepsilon}^{(k-1)}\right|, & t \geq 1\end{cases}
\end{equation}
where $\boldsymbol{\varepsilon}_{\mathrm{ij}}{ }^{(\mathrm{t})}$ represents the strain component at the current time step of $t$.

\subsection{Examples of the ML-based stress-strain models derived from DEM data}\label{section4.3}

This section provides some intuitive demonstration of the performance of different neural networks in predicting the stress-strain response of granular materials. Four different loading paths: 1) the conventional triaxial compression (CTC), 2) the isobaric axisymmetric triaxial loading, i.e., constant-\( p \) (CP), 3) the intermediate principal stress coefficient of tri-axial loading, i.e., constant-\( b \) (CB), and 4) random strain loading, are leveraged to generate the training data with the same RVE, where \( p = \frac{\sigma_{1}+\sigma_{2}+\sigma_{3}}{3} \) and \( b = \frac{\sigma_{2}-\sigma_{3}}{\sigma_{1}-\sigma_{3}} \). The relevant work ~\citep{wang2022data} shows that time-sequential neural networks, including GRU, LSTM, and TCNN have similar performance in predicting the constitutive laws of granular materials. Therefore, only one of the time-sequential neural networks (GRU) and the single-step-based MLP are trained and tested with the generated data.

\subsubsection{The stress-strain space of the training data}\label{section4.3.1}

\begin{figure}[H]
\centering 
\includegraphics[width =0.62 \linewidth,angle=0,clip=true]{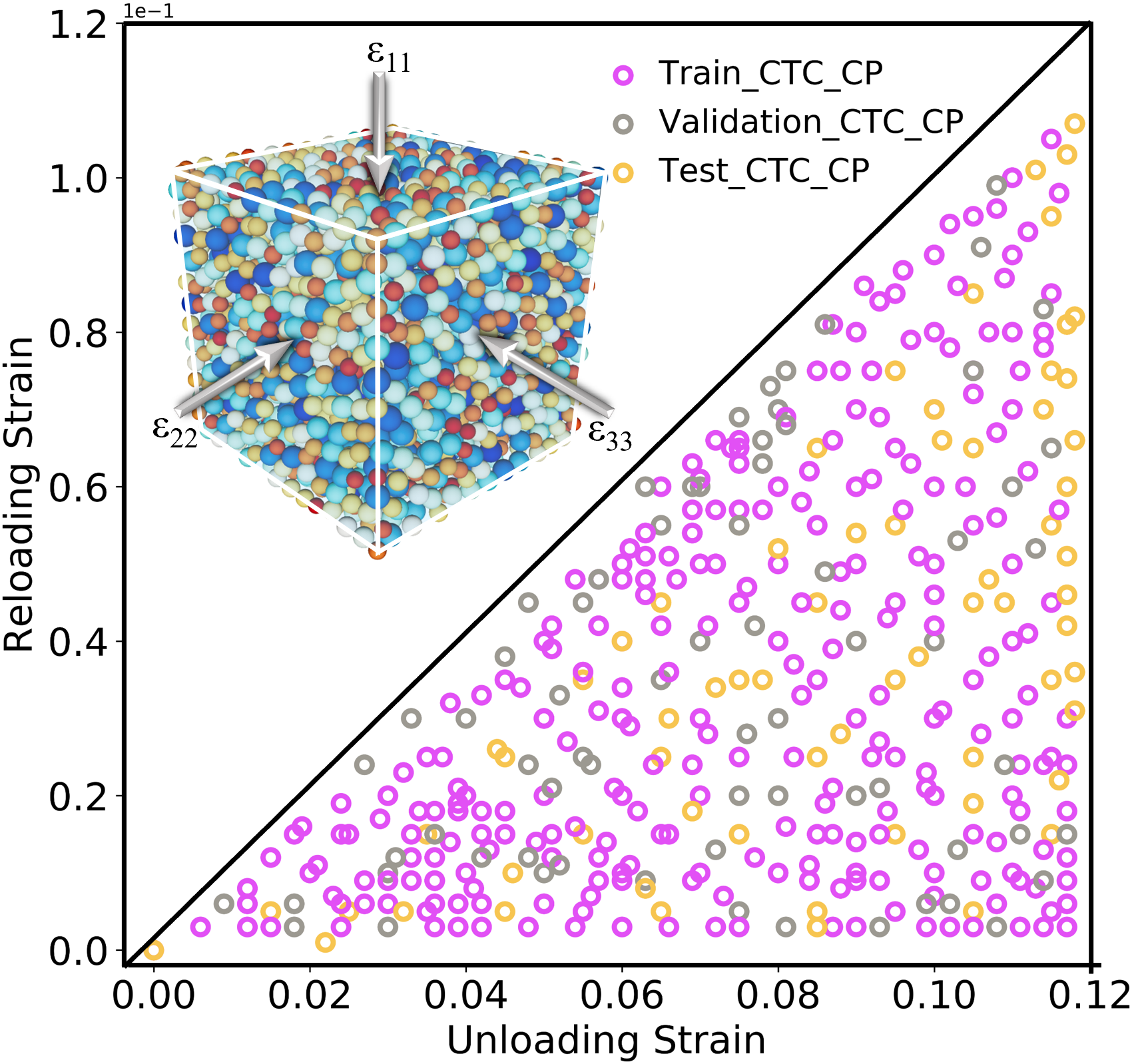}
\caption{{\color{red}The stress-strain space of cases with one-cycle CTC and CP loading path.}}
\label{pic12}
\end{figure}

Both the CTC and CP cases include one-cycle loading. As shown in Fig.~\ref{pic12}, a set of unloading-reloading points are adopted based on the principle that the unloading strain is bigger than the reloading strain to control the position of the hysteresis loop. In CB loading cases, only the monotonous loading condition is involved and the $b$-value is set to be a different value, increasing from 0 to 1 with an interval of 0.05, to control the loading paths.

Eventually, there are 440 loading cases generated under the CTC, CP, and CB loading conditions. Wherein 360 loading curves, involving these three loading conditions are randomly selected as the training and validation sets; the left 80 cases comprise the test set.

\subsubsection{The prediction results}\label{section4.3.2}

As mentioned in Section~\ref{section4.2.2}, there are normally two types of history variables available for the MLP to distinguish from different loading states: 1) one is the predicted state variables from the last time step, and 2) the other is the parameterized variables from the current state variables. To demonstrate the performance of different history variables in MLP, two MLPs are trained and tested with the same data but different history variables in this section. Wherein the first MLP employs the strain $\boldsymbol{\varepsilon}^{(t-1)}$ and predicted stress tensor $\hat{\boldsymbol{\sigma}}^{(t-1)}$ at the last time step as the history variables to predict the current stress~\citep{hashash2004numerical}, and the other utilizes the absolute accumulated strain increment $\Delta \boldsymbol{\varepsilon}_{\mathrm{ij}}{ }^{(\mathrm{t})}$ as the history variables~\citep{huang2020machine}, which can be respectively formulated as:
\begin{equation}
    \hat{\boldsymbol{\sigma}}^{(t)}=f^{\mathrm{MLP} 1}\left(\boldsymbol{\varepsilon}^{(t)},\left\{\boldsymbol{\varepsilon}^{(t-1)}, \hat{\boldsymbol{\sigma}}^{(t-1)}\right\}, \mathbf{W}, \mathbf{b}\right)
\end{equation}
\begin{equation}
    \hat{\boldsymbol{\sigma}}^{(t)}=f^{\mathrm{MLP} 2}\left(\boldsymbol{\varepsilon}^{(t)}, \Delta \boldsymbol{\varepsilon}^{(t)}, \mathbf{W}, \mathbf{b}\right)
\end{equation}

Furthermore, one of the time-sequence neural networks (i.e. GRU) is also trained and tested with the same data. As a time-sequential network, the history-loading information is encapsulated in the input strain sequence $\left\{\varepsilon^{(\mathrm{t}-\mathrm{n}+1)}, \boldsymbol{\varepsilon}^{(\mathrm{t}-\mathrm{n}+2)}, \ldots, \boldsymbol{\varepsilon}^{(\mathrm{t})}\right\}$ of GRU and can be identified by its RNN cells. Therefore, there is no need for the assistance of the extra history variable to learn the constitutive laws of granular materials, which can be expressed as:

\begin{equation}
\hat{\boldsymbol{\sigma}}^{(t)}=\boldsymbol{f}^{\mathrm{GRU}}\left(\left\{\boldsymbol{\varepsilon}^{(t-n+1)}, \boldsymbol{\varepsilon}^{(t-n+2)}, \ldots, \boldsymbol{\varepsilon}^{(t)}\right\}, \mathbf{W}, \mathbf{b}\right)
\end{equation}
where $n$ represents the time step of the input strain sequence. The hyperparameters used in GRU and MLP are listed in Table \ref{table3}.

\begin{table}[H]
\centering
\caption{Hyperparameters used in GRU and MLP} \label{table3}
\scalebox{0.75}{
\begin{tabular}{cccccccccc}
\hline
\begin{tabular}[c]{@{}c@{}}ML \\ models\end{tabular} & \begin{tabular}[c]{@{}c@{}}Model \\ architecture\end{tabular}                                                                         & \begin{tabular}[c]{@{}c@{}}The number \\ of $\mathbf{W}$, $\mathbf{b}$\end{tabular} & \begin{tabular}[c]{@{}c@{}}Learning\\ rate\end{tabular} & \begin{tabular}[c]{@{}c@{}}Batch \\ size\end{tabular} & \begin{tabular}[c]{@{}c@{}}Time steps \\ of input\end{tabular} & \begin{tabular}[c]{@{}c@{}}Active\\ function\end{tabular} & Optimizer & epoch & \begin{tabular}[c]{@{}c@{}}Time\\ (s/epoch)\end{tabular} \\ \hline
GRU                                                  & \begin{tabular}[c]{@{}c@{}}3 GRU Layers \\ /50 Units\\  + \\ 1 Dense layer \\ /50 Units\\  + \\ 1 Dense layer\\ /3 Units\end{tabular} & 39,003                                                                              & 0.001                                                   & 128                                                   & 40                                                             & Relu                                                      & Adam      & 1000  & 24                                                       \\ \hline
MLP                                                 & \begin{tabular}[c]{@{}c@{}}9 Dense Layers \\ /20 Units\\  +  \\ 1 Dense layer \\ /3 Units\end{tabular}                                & 22703                                                                               & 0.001                                                   & 128                                                   & /                                                              & Relu                                                      & Adam      & 1000  & 4                                                        \\ \hline
\end{tabular}
}
\end{table}

\begin{figure}[ht]
\centering

% 第一排图片
\begin{subfigure}{.33\textwidth}
  \centering
  \includegraphics[width=\linewidth]{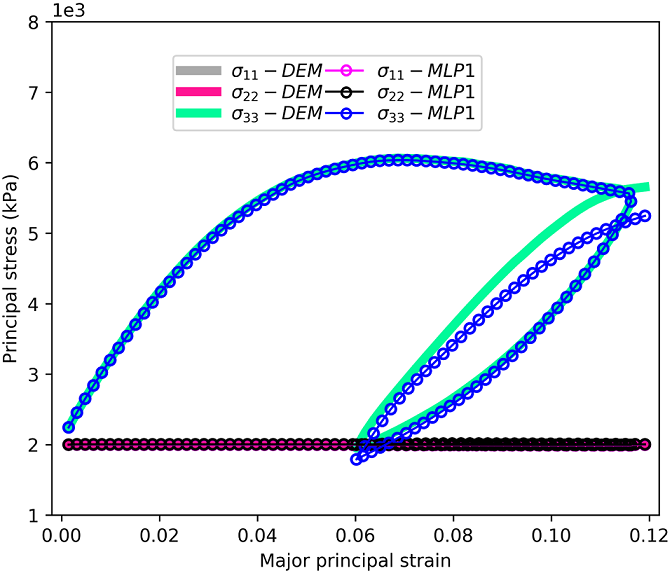}
  \caption{CTC case 1}
  \label{pic13a}
\end{subfigure}%
\hfill
\begin{subfigure}{.34\textwidth}
  \centering
  \includegraphics[width=\linewidth]{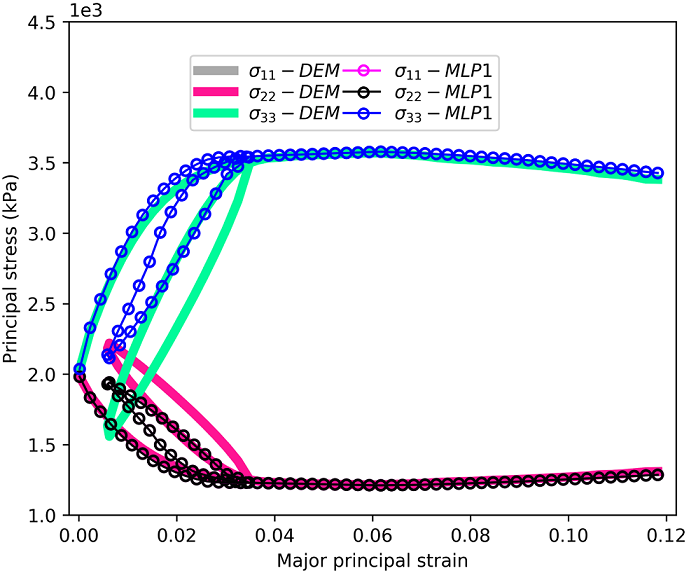}
  \caption{CP case 1}
  \label{pic13b}
\end{subfigure}%
\hfill
\begin{subfigure}{.33\textwidth}
  \centering
  \includegraphics[width=\linewidth]{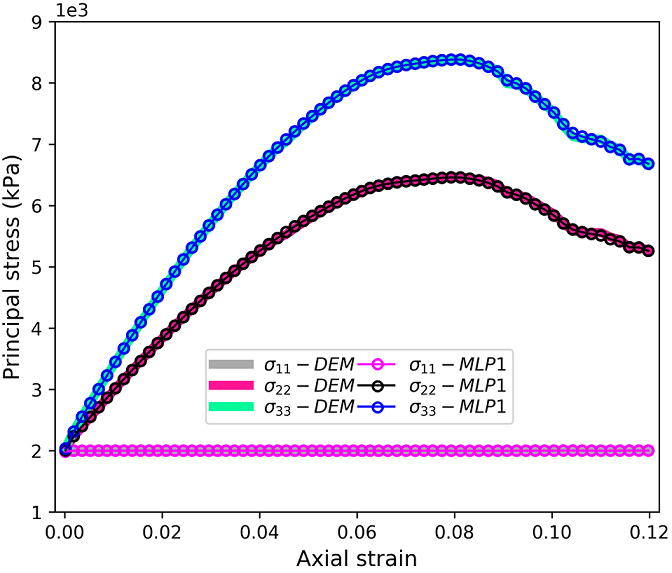}
  \caption{CB case 1}
  \label{pic13c}
\end{subfigure}

\vspace{1em} % 添加一些垂直空间

% 第二排图片
\begin{subfigure}{.33\textwidth}
  \centering
  \includegraphics[width=\linewidth]{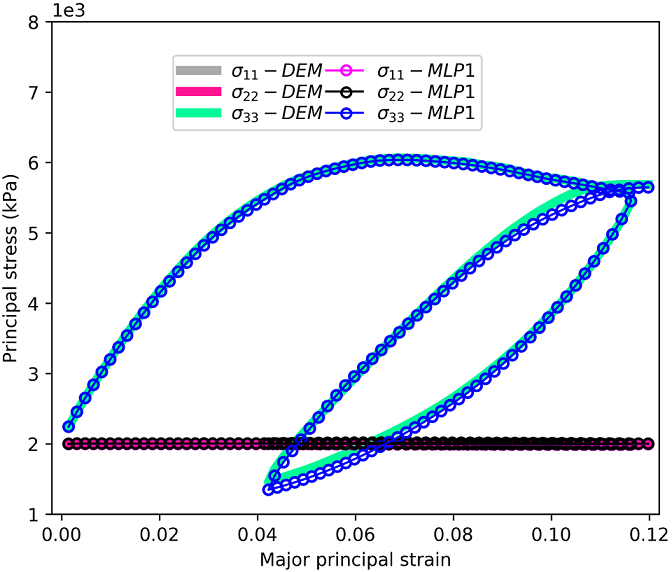}
  \caption{CTC case 2}
  \label{pic13d}
\end{subfigure}%
\hfill
\begin{subfigure}{.34\textwidth}
  \centering
  \includegraphics[width=\linewidth]{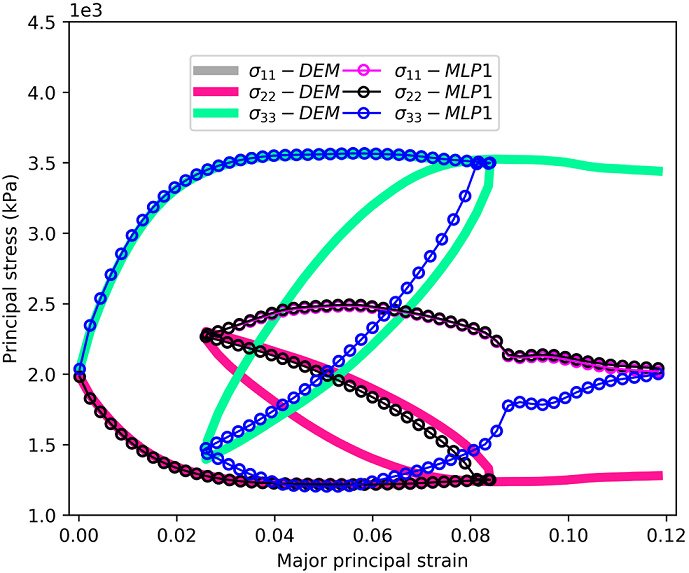}
  \caption{CP case 2}
  \label{pic13e}
\end{subfigure}%
\hfill
\begin{subfigure}{.33\textwidth}
  \centering
  \includegraphics[width=\linewidth]{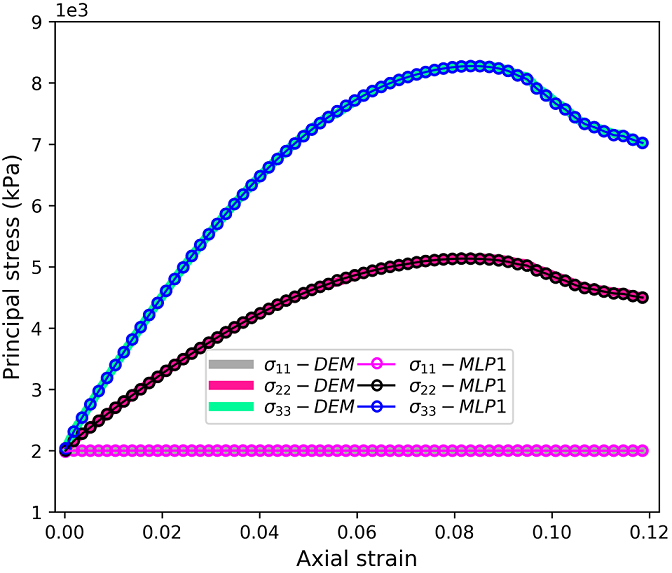}
  \caption{CB case 2}
  \label{pic13f}
\end{subfigure}

\caption{The prediction result of MLP1}
\label{pic13}
\end{figure}

Fig.\ref{pic13} gives an intuitive demonstration of the performance of the trained MLP1 by six representative prediction cases. In CTC and CP loading cases, in which one reversal loading is included, although the MLP1 can capture the constitutive laws of granular materials in partial cases (CTC case 2), an error accumulation phenomenon as mentioned in Section~\ref{section4.2.2} occurs, especially in reloading stage, see CTC case 1, CP case 1, and CP case 2. Furthermore, it is also found that uncontrollable accumulated errors can eventually result in the breakdown of the entire prediction system (see CP case 2), while controllable accumulated errors can be gradually diminished (see CP case 1) due to the relationship between the strain and stress changes from the one-to-many to a surjective after the reversal loading. Additionally, the trained MLP1 performs well in all CB cases, as the relationship between stress and strain remains strictly surjective in these cases, and thus the selection of the history variables has less influence on the performance of the MLP1.

\begin{figure}[ht]
\centering

% 第一排图片
\begin{subfigure}{.33\textwidth}
  \centering
  \includegraphics[width=\linewidth]{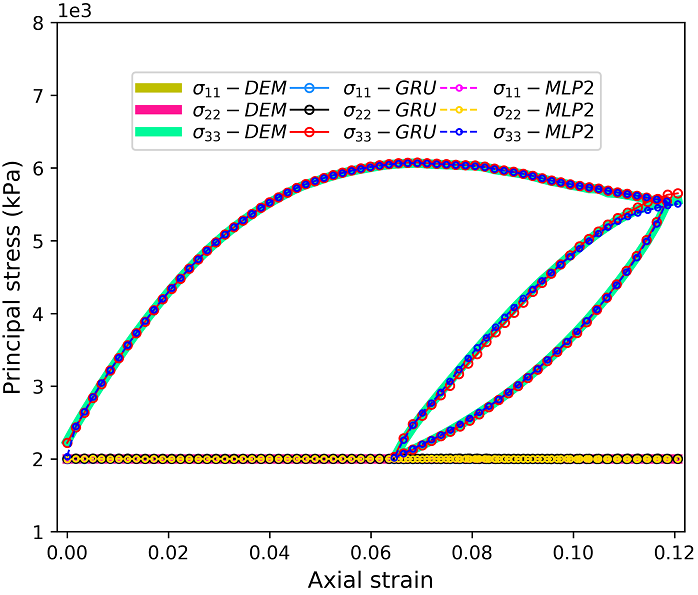}
  \caption{CTC case 1}
  \label{pic14a}
\end{subfigure}%
\hfill
\begin{subfigure}{.34\textwidth}
  \centering
  \includegraphics[width=\linewidth]{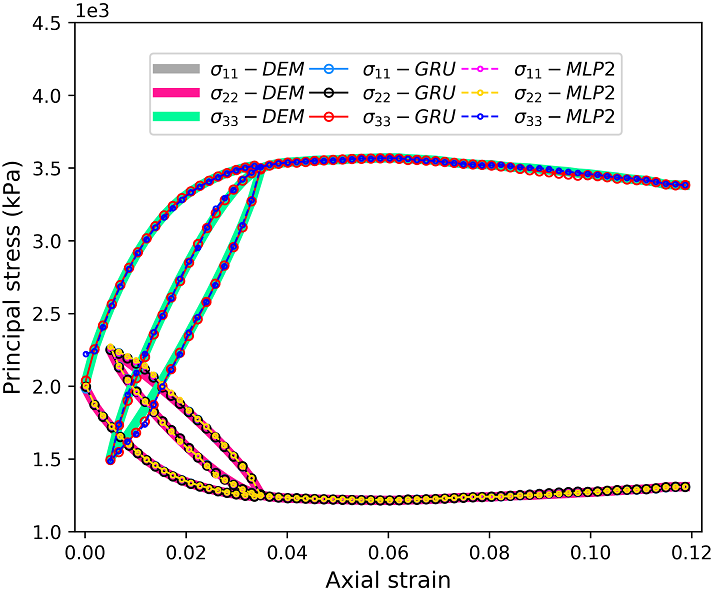}
  \caption{CP case 1}
  \label{pic14b}
\end{subfigure}%
\hfill
\begin{subfigure}{.33\textwidth}
  \centering
  \includegraphics[width=\linewidth]{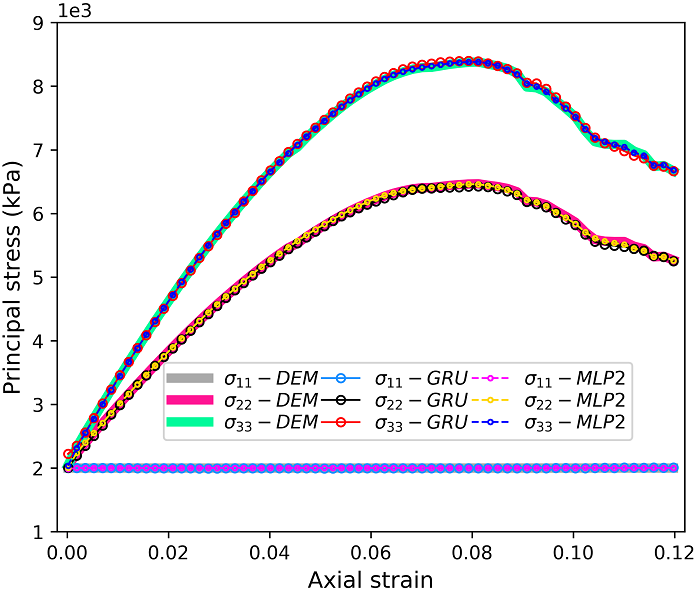}
  \caption{CB case 1}
  \label{pic14c}
\end{subfigure}

\vspace{1em} % 添加一些垂直空间

% 第二排图片
\begin{subfigure}{.33\textwidth}
  \centering
  \includegraphics[width=\linewidth]{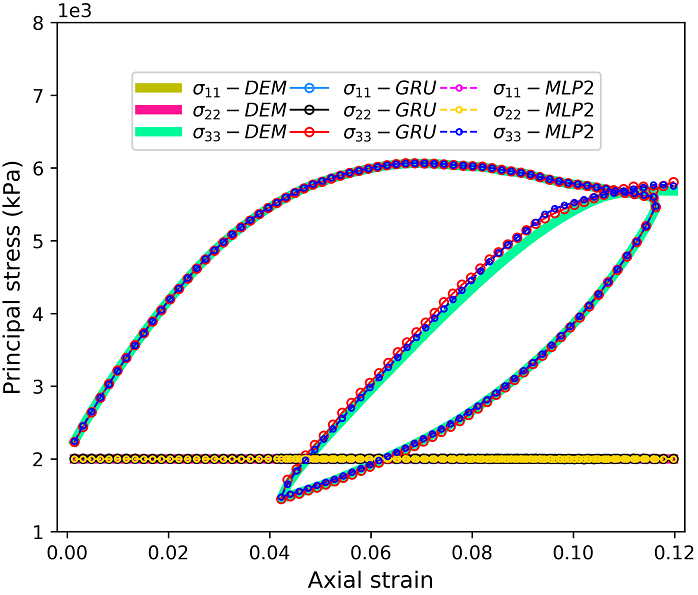}
  \caption{CTC case 2}
  \label{pic14d}
\end{subfigure}%
\hfill
\begin{subfigure}{.34\textwidth}
  \centering
  \includegraphics[width=\linewidth]{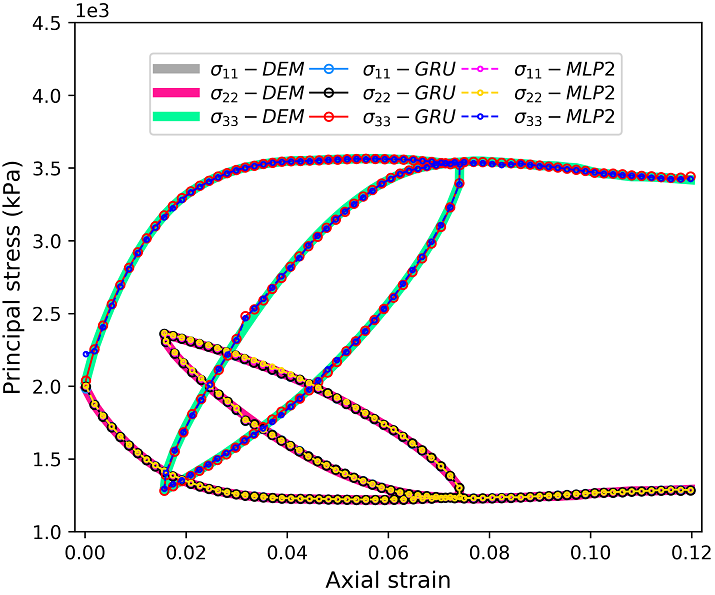}
  \caption{CP case 2}
  \label{pic14e}
\end{subfigure}%
\hfill
\begin{subfigure}{.33\textwidth}
  \centering
  \includegraphics[width=\linewidth]{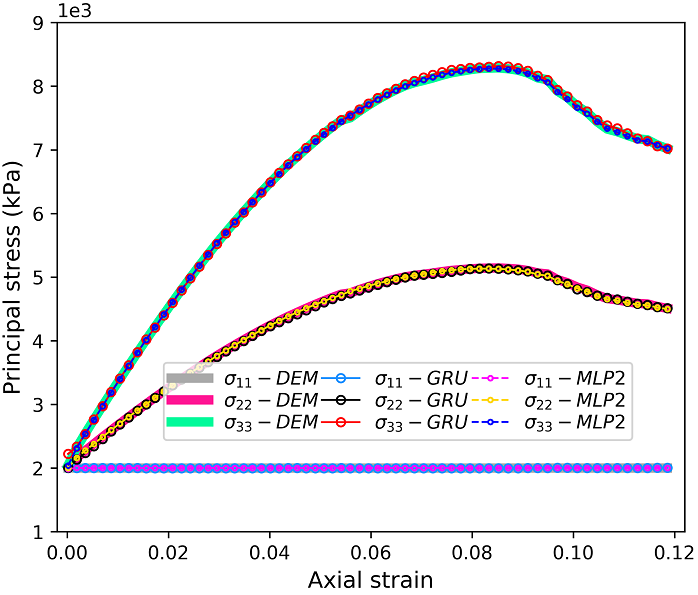}
  \caption{CB case 2}
  \label{pic14f}
\end{subfigure}

\caption{The prediction result of GRU and MLP2}
\label{pic14}
\end{figure}

The identical cases in Fig.~\ref{pic13} are also employed to test the trained MLP2 and GRU. The obtained results are showcased in Fig.~\ref{pic14}, and the predicted average mean absolute error (MAE) of these two models for the 80 test cases is listed in Table~\ref{table4}. The outcomes, as illustrated in Fig.~\ref{pic14} and detailed in Table~\ref{table4}, reveal the MLP can achieve comparable performance to the GRU using the parameterized history variables. Furthermore, a comparison between the prediction results in Fig.~\ref{pic14} and their counterparts in Fig.~\ref{pic13} indicates that the use of parameterized history variables effectively mitigates the problem of error accumulation.

\begin{table}[H]\centering
\caption{The prediction results of all test cases with the GRU and MLP} \label{table4}
\begin{tabular}{ccc}
\hline
Loading paths        & ML models & Average MAE \\ \hline
\multirow{2}{*}{CB}  & GRU       & 0.0140      \\ \cline{2-3} 
                     & MLP2     & 0.0159      \\ \hline
\multirow{2}{*}{CP}  & GRU       & 0.0174      \\ \cline{2-3} 
                     & MLP2     & 0.0183      \\ \hline
\multirow{2}{*}{CTC} & GRU       & 0.0168      \\ \cline{2-3} 
                     & MLP2     & 0.0185      \\ \hline
\end{tabular}
\end{table}

\begin{figure}[ht]
\centering

% 第一排图片
\begin{subfigure}{.33\textwidth}
  \centering
  \includegraphics[width=\linewidth]{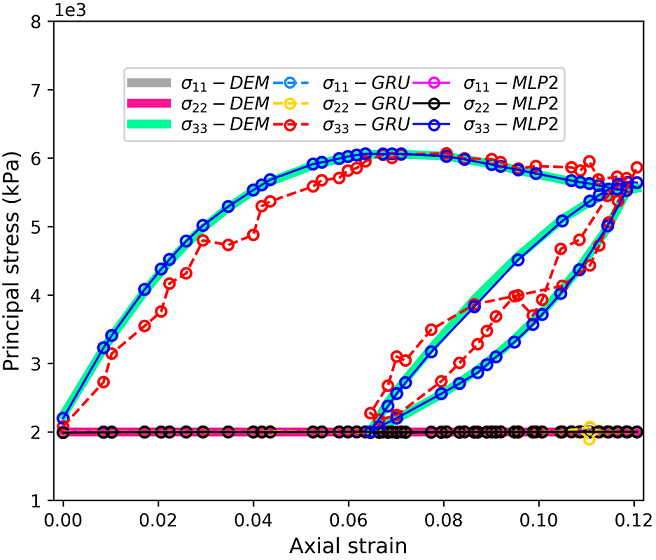}
  \caption{CTC case 1}
  \label{pic15a}
\end{subfigure}%
\hfill
\begin{subfigure}{.34\textwidth}
  \centering
  \includegraphics[width=\linewidth]{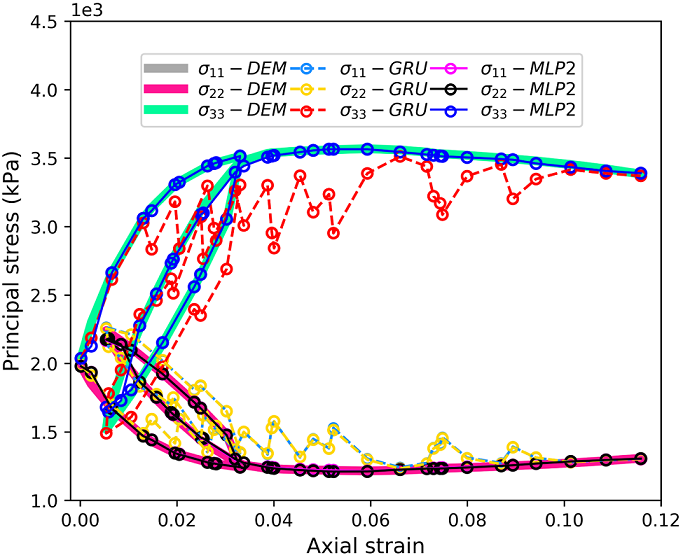}
  \caption{CP case 1}
  \label{pic15b}
\end{subfigure}%
\hfill
\begin{subfigure}{.33\textwidth}
  \centering
  \includegraphics[width=\linewidth]{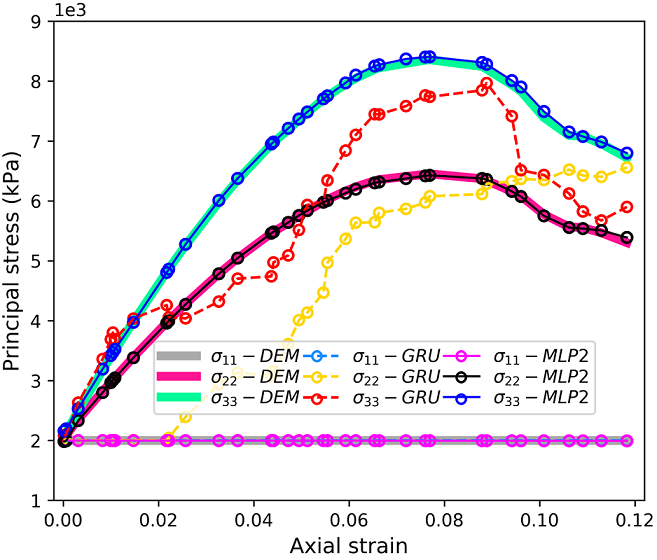}
  \caption{CB case 1}
  \label{pic15c}
\end{subfigure}

\vspace{1em} % 添加一些垂直空间

% 第二排图片
\begin{subfigure}{.33\textwidth}
  \centering
  \includegraphics[width=\linewidth]{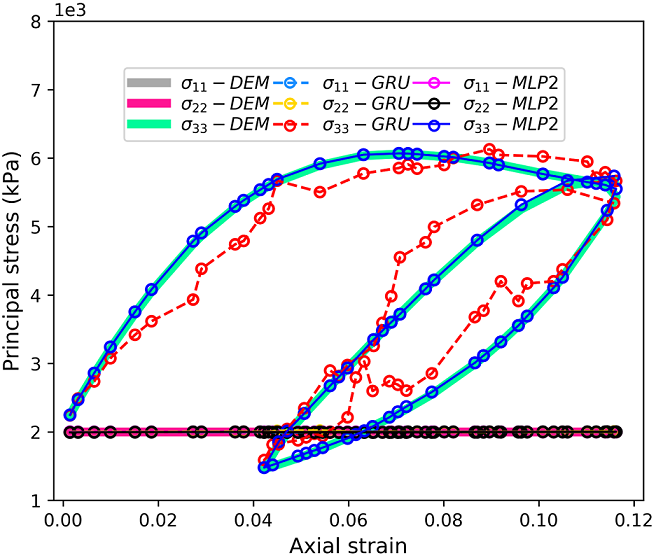}
  \caption{CTC case 2}
  \label{pic15d}
\end{subfigure}%
\hfill
\begin{subfigure}{.34\textwidth}
  \centering
  \includegraphics[width=\linewidth]{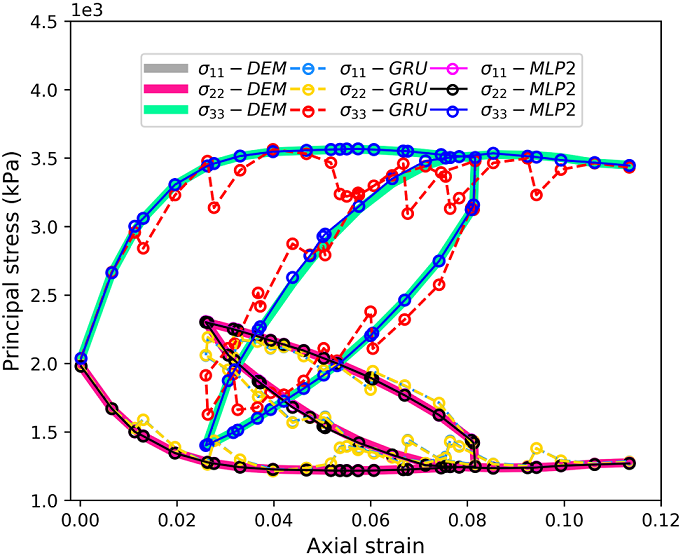}
  \caption{CP case 2}
  \label{pic15e}
\end{subfigure}%
\hfill
\begin{subfigure}{.33\textwidth}
  \centering
  \includegraphics[width=\linewidth]{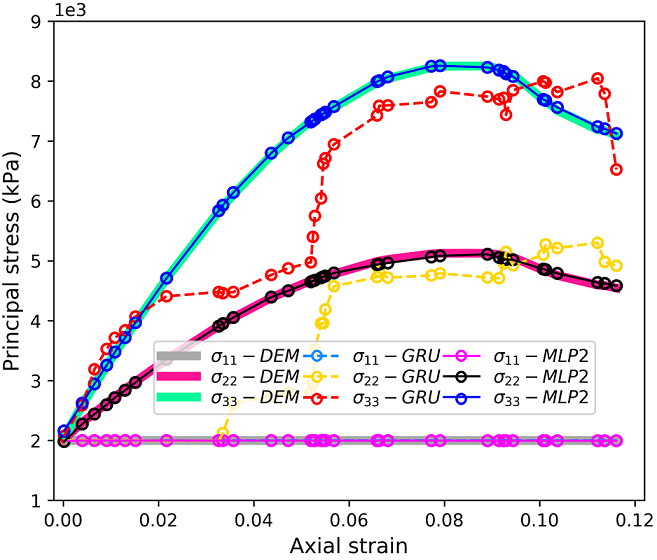}
  \caption{CB case 2}
  \label{pic15f}
\end{subfigure}

\caption{ The prediction results of GRU and MLP2 in cases with different loading intervals}
\label{pic15}
\end{figure}

From the perspective of engineering application, the strain intervals are normally different and random. Therefore, the sensitivity analysis of different neural networks to the loading strain interval is also conducted in this work. The strain intervals of 80 test cases are randomized, and the trained MLP2 and GRU are tested again with the newly generated test data. The prediction results of the same cases in Fig.~\ref{pic13} and Fig.~\ref{pic14} are plotted in Fig.~\ref{pic15}. The results illustrate that the single-step-based MLP remains effective in capturing the stress-strain response despite variations in the loading interval. In contrast, the time-sequential neural network GRU exhibits notable sensitivity to changes in the loading interval, resulting in poorer performance under such conditions.

\subsection{Summary}\label{section4.4}

{\color{red}This section provides an overview of ML-based constitutive models for granular materials, focusing on two key aspects: the sources of training data and the training strategies employed by different neural networks, and a concise summary of the advantages and limitations of each aspect is provided in Table~\ref{summary4}. In addition, Section~\ref{section4.3} presents a detailed case study that illustrates the predictive capabilities of various neural networks under different loading conditions. The following conclusions can be drawn from the analysis:}

\begin{table}[H]
\centering
\caption{{\color{red}A concise summary of the ML-based constitutive models for granular materials}}
\label{summary4}
\scalebox{0.7}{
\begin{tabular}{c|l|l|l}
\hline
\begin{tabular}[c]{@{}c@{}}ML-based\\ constitutive models\end{tabular}
& \multicolumn{1}{c|}{Types} 
& \multicolumn{1}{c|}{Advantages}
& \multicolumn{1}{c}{Disadvantages} \\ 

\hline

\multirow{2}{*}{\begin{tabular}[c]{@{}c@{}}Data sources\\ used\end{tabular}}
& Experiment data 
& \begin{tabular}[c]{@{}l@{}}
1). Reflecting the most authentic and realistic\\ constitutive laws of granular materials\\ 
2). Representative of real-world loading conditions\end{tabular}
& \begin{tabular}[c]{@{}l@{}}
1). High cost and time-consuming\\
2). Incorporating noise\\
3). Incomplete datasets\end{tabular} \\ \cline{2-4} & Synthetic data 
& \begin{tabular}[c]{@{}l@{}}
1). Easily generated in large quantities\\ 
2). Wide coverage of different loading\\ scenarios\end{tabular} 
& \begin{tabular}[c]{@{}l@{}}
1). Smearing out some intrinsic\\ features of granular materials\\ 
2). Lack of authenticity of\\ experimental data\end{tabular}\\

\hline

\multirow{2}{*}{\begin{tabular}[c]{@{}c@{}}Neural network\\type \end{tabular}}
& Single-step-based
& \begin{tabular}[c]{@{}l@{}}
1). Simple architecture\\ 
2). High training and prediction efficiency\\ 
3). Robust to the change of loading step\end{tabular} 

& \begin{tabular}[c]{@{}l@{}}
1). Requirement for artificially\\ added internal/history variables\\ 
2). Error accumulation problem\end{tabular} \\ \cline{2-4} 
& Multi-step-based 
& \begin{tabular}[c]{@{}l@{}}
1). No extra history variables are\\ required in identifying the loading history\\ 
2). Stronger non-linear mapping capability\\ than single-step-based networks\end{tabular}
& \begin{tabular}[c]{@{}l@{}}
1). Complex structure\\ 
2). Lower training efficiency\\ 
3). Sensitive to the change of\\ the loading interval\end{tabular}\\ 
\hline
\end{tabular}}
\end{table}

{\color{red}The datasets used to train ML-based constitutive models can be categorized into two types: experimental data and synthetic data. Experimental data are normally generated under conditions that closely resemble real-world loading scenarios. However, these datasets are typically costly and time-consuming to collect, resulting in limited coverage of the sampling space. Additionally, experimental data often contain noise introduced by measurement devices, which can adversely affect the training of neural networks.

In contrast, the data obtained from existing phenomenon-based constitutive models are more cost-effective, enabling the generation of diverse and extensive datasets that cover a wide range of loading scenarios. This diversity can help produce more general neural network models. However, their fidelity fully depends on the selected constitutive models, which may have restrictions in reproducing certain characteristics of granular materials. To the best knowledge of the authors, no constitutive model is perfect. For the data obtained from DEM simulations, reliability is also a problem that we need to pay attention to. Apart from the accuracy of the method itself, the ad-hoc/arbitrary selection of microscopic parameters for the grain/particle is primarily calibrated based on certain cases, which may be not applicable to a generic problem.

On the other hand, the constitutive law of granular materials is inherently state and history-dependent, and both single-step and time-sequence (multi-step) neural networks can be employed to predict such constitutive behavior. Single-step-based neural networks, such as MLPs, offer the advantage of a simple architecture, which typically leads to high training and prediction efficiency. However, these models lack the inherent ability to capture loading history, necessitating the inclusion of artificially added internal or history variables. There are generally two approaches for incorporating history variables in MLPs: (a) state parameters predicted at the previous time step, which can lead to an accumulation of errors over time, and (b) variables extracted directly from the current state parameters, which aim to provide more immediate historical context without the propagation of previous errors.

Due to the unique architecture, time-sequence neural networks are inherently capable of distinguishing different loading histories without requiring additional assistance. This architecture enhances their ability to capture long-term historical dependencies and perform complex non-linear mappings. However, the increased complexity of these networks, compared to single-step-based models, results in more training parameters. Consequently, time-sequence neural networks are more challenging to tune and exhibit lower training and prediction efficiency. Additionally, according to the prediction result showcased in Section~\ref{section4.3}, it is also found that time-sequential networks are sensitive to the change of strain intervals, while the single-step-based network still performs well under the same loading cases, indicating that time-sequential networks are more suitable to be applied in the data with a fixed loading step.}

{\color{red}
Based on the discussion of ML-based constitutive models of granular materials, some promising avenues for future research, particularly from the perspectives of transfer learning and optimization for training strategy, could be potential ways to solve the challenges mentioned above:

1). Application of transfer learning in ML-based constitutive Modeling: Given the high cost and limited availability of experimental data, transfer learning offers a promising approach to enhance the generalization capabilities of ML-based constitutive models. By leveraging knowledge from pre-trained models on synthetic datasets, transfer learning can fine-tune these models using smaller amounts of high-fidelity experimental data, which has the potential to improve the accuracy and authenticity of ML-based constitutive models across a broader range of loading conditions.

2). Optimization of training strategies: Single-step-based neural networks are inherently limited in their ability to capture loading history, necessitating the development of parameterization schemes for internal variables which aim to represent loading history using the fewest physically meaningful variables possible. Additionally, future research could focus on improving the performance of time-sequence neural networks under varying loading intervals. Approaches such as interpolation or data augmentation may serve as effective strategies to enhance the resilience of these networks to non-uniform loading intervals, thereby broadening their applicability in real-world scenarios.}

\section{The ML-aided macroscopic simulation of granular materials}\label{section5}

Apart from the use in depicting the microscopic interparticle contact characteristics and corresponding contact forces, ML models have also been integrated with continuum-based particle methods or mesh-based simulation approaches to predict the macroscale deformation of granular systems. According to the coupled numerical methods, the ML-aided macroscopic simulation can be categorized into two types: 1) ML simulation based on the interpolation of macroscopic particle kinematic features, and 2) the macroscopic simulation using ML models as constitutive models (e.g. the MPM/FEM-ML framework) in this review.

\subsection{The macroscopic particle kinematic features-based ML simulation} \label{section5.1}

Over the past decades, various continuum-based particle methods, such as the particle finite element method (PFEM)~\citep{onate2004particle}, element-free Galerkin method (EFGM)~\citep{belytschko1994element}, smoothed particle hydrodynamics (SPH), and material point method (MPM), have been developed to model the deformation features of either fluid or solid. Among these methods, the SPH and MPM stand out as particularly representative and have been extensively utilized across a broad range of engineering applications. Therefore, our primary focus in this section centres on reviewing the integration of data-driven methods with SPH and MPM to depict the macroscopic deformation of granular materials. Before that, a concise introduction to SPH and MPM will be given.

\subsubsection{The continuum-based particle numerical methods}\label{section5.1.1}

In both MPM and SPH, the kinematic features of macroscopic material points or SPH particles govern the macroscale deformation of the material under consideration. The standard MPM was originally proposed in the literature~\citep{sulsky1995application} and is a particle-based Eulerian-Lagrangian method that has gained increased attention in geomechanics~\citep{bardenhagen2000material,solowski2015evaluation,fern2019material}, especially in addressing large-strain problems of granular materials~\citep{wikeckowski2004material, beuth2008large, beuth2011solution}.

\begin{figure}[H]
\centering 
\includegraphics[width =0.6 \linewidth,angle=0,clip=true]{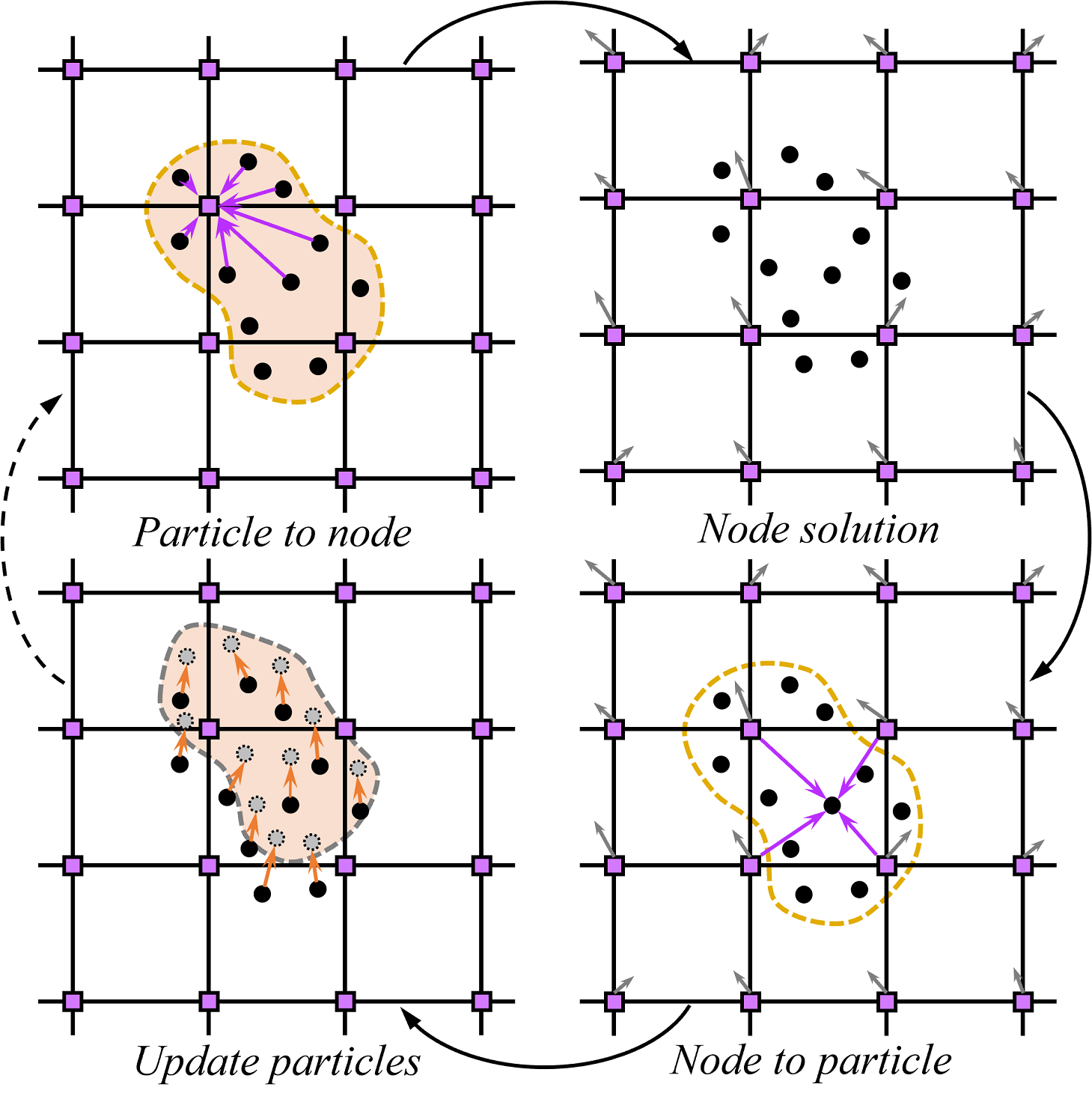}
\caption{{\color{red}The calculation process of the MPM.}}
\label{pic16}
\end{figure}

As shown in Fig.~\ref{pic16}, the computational process of the MPM consists of four steps: 1) In the
\textit{\textbf{particle-to-node}} stage, the continuum body $\Omega$ is discretized into a total of $n_p$ material points located on the background mesh with a total of $n_n$ nodes. The physical variables encapsulated in material points at the time step $t$, e.g. mass $m_p$, velocity $\mathbf{v}_p^t$, and acceleration $\mathbf{a}_p^t$, are projected to the corresponding nodes via the shape function $N_i(\mathbf{x})(i=1,2,3,...,n_n)$
to compute nodal mass $m_i^t=\sum_{p=1}^{n^p} N_i\left(\mathbf{x}_p^t\right) m_p$, nodal momentum $\mathbf{q}_i^t=\sum_{p=1}^{n^p} m_p \mathbf{v}_p^t N_i\left(\mathbf{x}_p^t\right)$, nodal force $\mathbf{f}_i^t=\sum_{p=1}^{n^p} m_p \mathbf{a}_p^t N_i\left(\mathbf{x}_p^t\right)$ as well as nodal velocity $\mathbf{v}_i^t=\mathbf{q}_i^t / m_i^t$ and acceleration $\mathbf{a}_i^t=\mathbf{f}_i^t / m_i^t$ at the time step of $t$. Where $\mathbf{x}_p^t$ represents the location of the $p^{t h}$ material point at the $t$ time step. Then, 2) the following time step's nodal displacement $\mathbf{u}_{\mathrm{i}}^{\mathrm{t}+1}$, velocity $\mathbf{v}_{\mathrm{i}}^{\mathrm{t}+1}$, and $\mathbf{a}_{\mathrm{i}}^{\mathrm{t}+1}$ are solved with the governing equation in the \textit{\textbf{node solution}} step, which is similar to the updated Lagrangian FEM process~\citep{wriggers2008nonlinear,bandara2015coupling}. 3) the obtained solution is interpolated back to the material points in the \textit{\textbf{node-to-particle}} stage and 4) the dynamic information of material points including their position $\mathbf{x}_{p}{ }^{t+1}$, velocity $\mathbf{v}_p{ }^{t+1}$, and acceleration $\mathbf{a}^{t+1}$, is calculated and the Lagrangian mesh is reset for the next iteration in the \textit{\textbf{update particles}} step.

\begin{figure}[ht]
\centering

% 第一排图片
\begin{subfigure}[b]{0.46\textwidth}
  \centering
  \includegraphics[width=\linewidth]{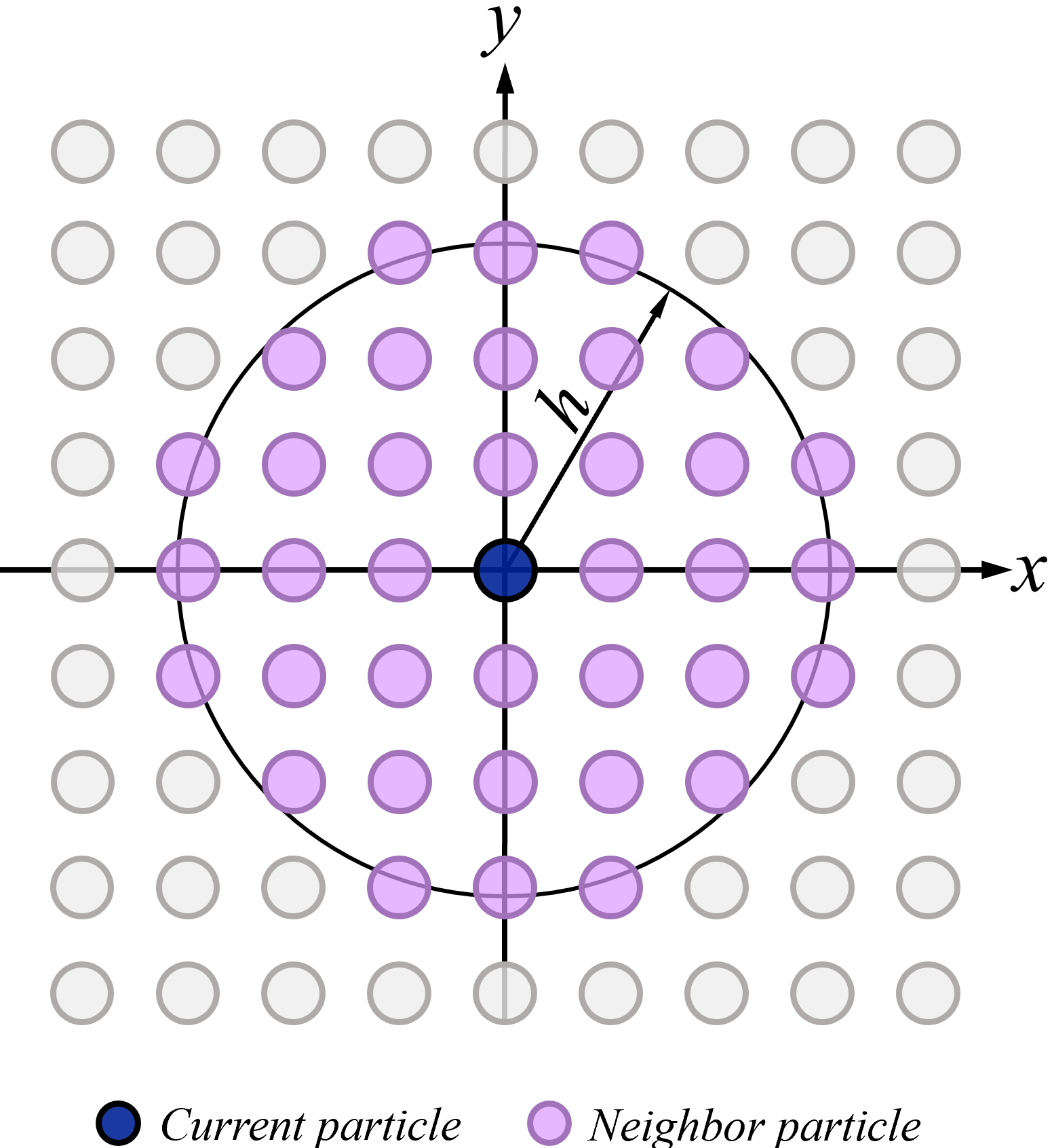}
  \caption{ The discretized system with an interaction radius $h$.}
  \label{pic17a}
\end{subfigure}
\hfill
\begin{subfigure}[b]{0.51\textwidth}
  \centering
  \includegraphics[width=\linewidth]{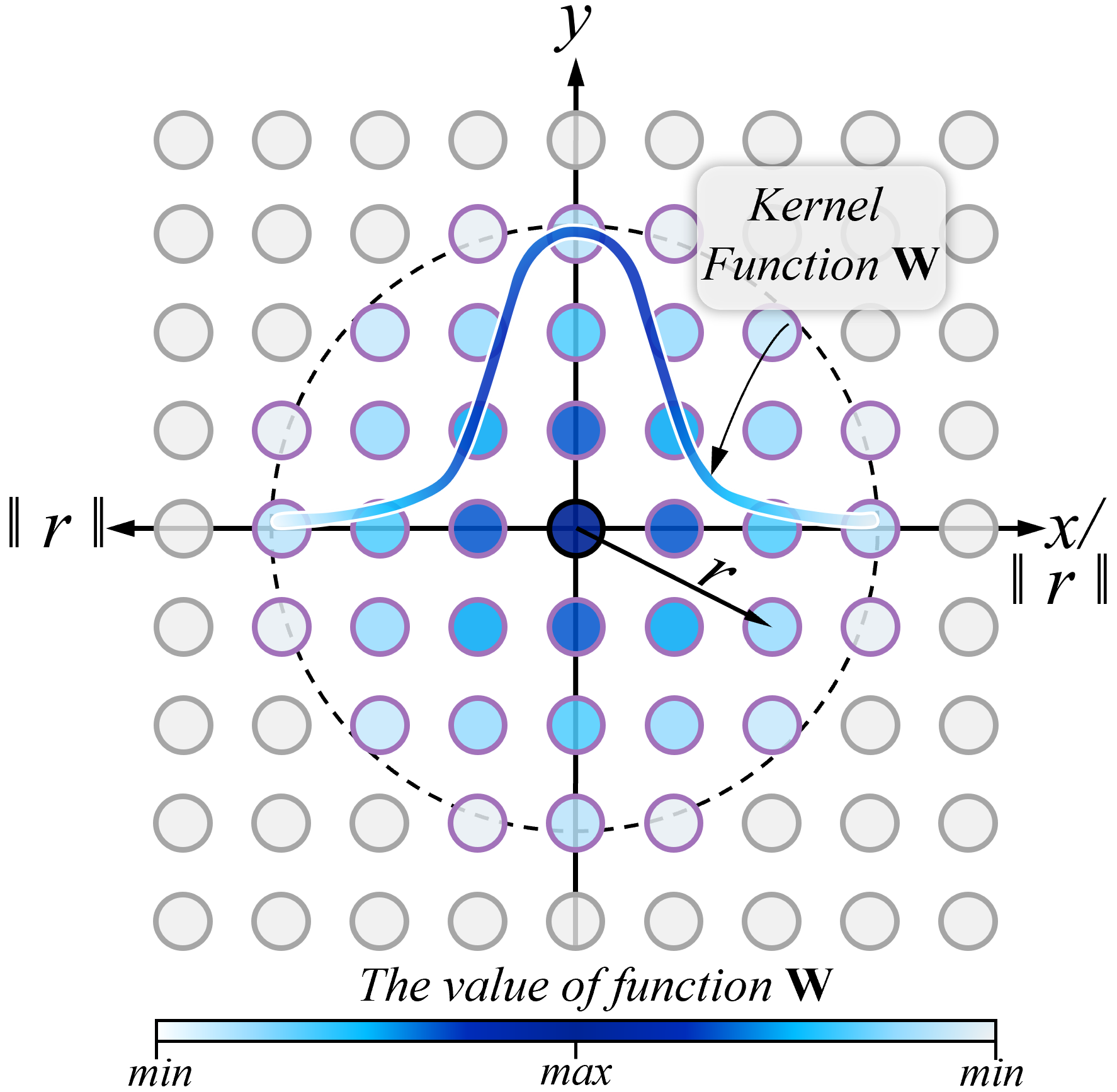}
  \caption{The weight contribution following the kernel function $\mathbf{W}$}
  \label{pic17b}
\end{subfigure}%

\caption{{\color{red}2D SPH particles.}}
\label{pic17}
\end{figure}

SPH, initially developed for astrophysical problems~\citep{gingold1977smoothed,lucy1977numerical}, is another commonly used mesh-free Lagrangian method and has also been used to model the deformation of granular materials~\citep{bui2007numerical,bui2008lagrangian,bui2008sph,wang2014frictional}. Similar to the MPM, as demonstrated in Fig.~\ref{pic17a}, the SPH method discretizes the research domain into a set of macroscopic interpolation particles to represent physical properties $\mathrm{A}\left(\mathbf{x}_{\mathrm{i}}{ }^t\right)$ of specific point $\mathbf{x}_{\mathrm{i}}{ }^{\mathrm{t}}$ at the current time step $t$, and each physical variable, e.g. density $\rho\left(\mathbf{x}_i^t\right)$, velocity $\mathbf{v}\left(\mathbf{x}_i^t\right)$, and pressure $\mathbf{p}\left(\mathbf{x}_i^t\right)$, is impacted by its neighbour particles $\mathbf{x}_j^t$ within an interaction radius (h). By using the smooth function~\citep{liu2010smoothed} (also referred to as kernel function) as depicted in Fig.~\ref{pic17b}, field variables at any given point can be estimated by interpolation calculations:
\begin{equation}
    \mathbf{A}\left(\mathbf{x}_{\mathrm{i}}^{\mathrm{t}}\right)=\sum_j \mathbf{A}\left(\mathbf{x}_{\mathrm{j}}^{\mathrm{t}}\right) \frac{m_j}{\rho_j} \mathbf{W}\left(\mathbf{x}_{\mathrm{i}}^{\mathrm{t}}-\mathbf{x}_{\mathrm{j}}^{\mathrm{t}}, h\right)
\end{equation}
where $m_j, \rho_j$ represent the mass and density of the $j^{t h}$ neighbour particle, and $\|r\|=\left\|\mathbf{x}_{\mathbf{i}}{ }^t-\mathbf{x}_j{ }^t\right\|$. Then, the dynamic information of each SPH particle can be iteratively computed by forces (e.g. body forces, viscosity forces, etc) acting on them.

Relying on macroscopic particle kinematic features to govern the deformation of the target domain makes MPM and SPH effectively avoid mesh entanglement and distortion, showcasing notable superiority in handling large-strain issues. However, updating the dynamic features for each particle is computationally significant, particularly in simulations involving a substantial number of particles.

\subsubsection{The framework of macroscopic kinematic features-based ML models}\label{section5.1.2}

The advancement of data-driven simulators in the past two decades~\citep{spengler1999fast} offers one promising way to solve the abovementioned problem in MPM and SPH methods. Given one specific position, ML-based simulators can output necessary dynamic features simultaneously, making them far superior in computational efficiency~\citep{he2019learning} bypassing the complex physical solution process.

In MPM and SPH methods, the system consists of an unordered set of macroscopic interpolation particles including rich interaction relations with one another. These particle features and particle-particle interaction can be naturally expressed with the graph that can be modelled by the graph network simulators (GNS) but not convenient to be transferred into the regular sequences or vectors which single-step, time-sequential networks~\citep{wiewel2019latent}, and specific tree networks~\citep{ladicky2015data} can extract knowledge from. 

\begin{figure}[H]
\centering 
\includegraphics[width =0.99 \linewidth,angle=0,clip=true]{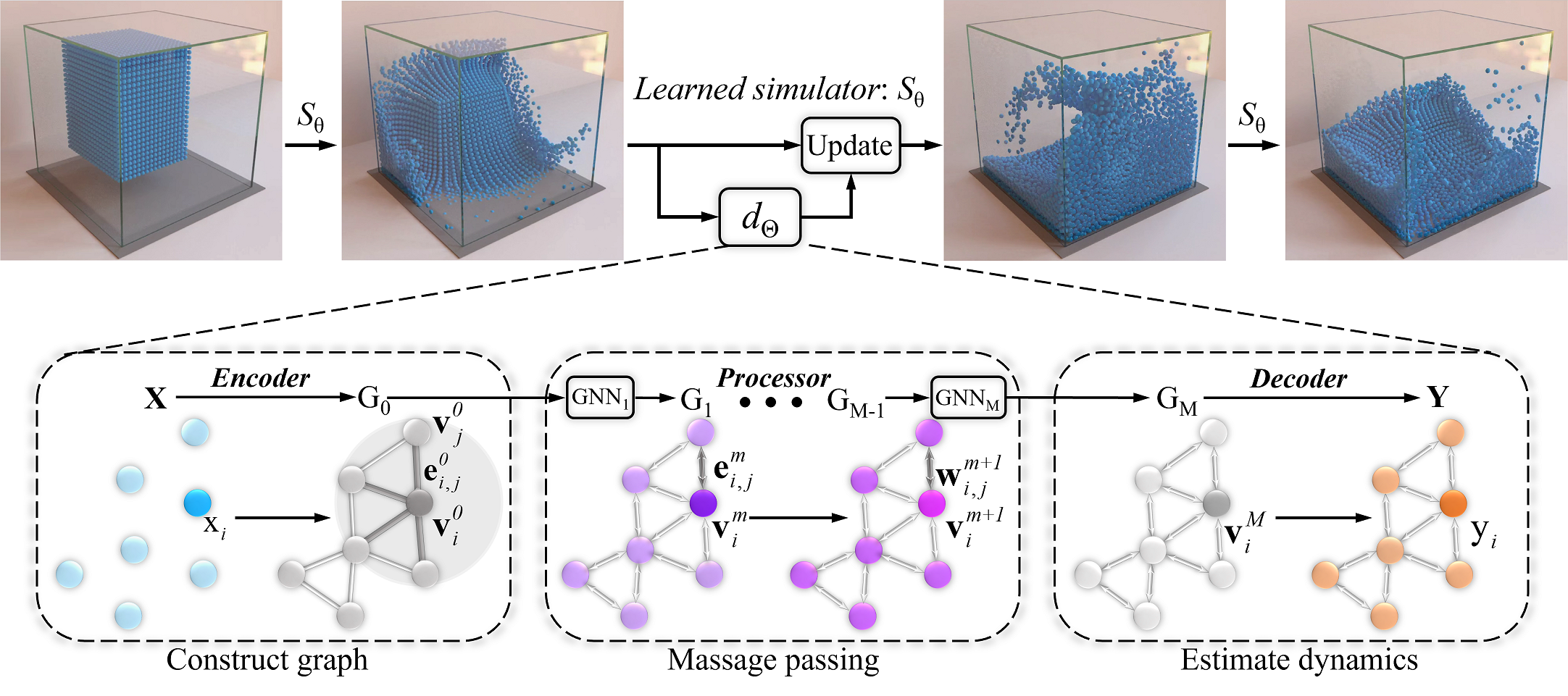}
\caption{{\color{red}The graph network simulators}~\citep{sanchez2020learning}}
\label{pic18}
\end{figure}

GNS operates on graphs to learn the physics of the dynamic system and predict rollouts. The graph network spans the system domain with nodes representing a collection of particles and the links connecting the nodes representing the local interaction between the material points. As shown in Fig.~\ref{pic18}, GNS ($\left.S_\theta\right)$ consists of two parts: one is the learnable approximator $d_{\Theta}$, and the other is the updater. The current physical state of each particle in MPM or SPH simulations is represented as a set of $\mathbf{x}_{\mathrm{i}}^{\mathrm{t}} \in \mathbf{X}^{\mathrm{t}}$ in $S_\theta$. Where $\mathbf{x}_{\mathbf{i}}^{\mathrm{t}}$ is a vector, normally consisting of position $\mathbf{p}_{\mathrm{i}}^{\mathrm{t}}$, velocity $\dot{\mathbf{p}}_{\mathbf{i}}^{\mathrm{t}}$, boundary information $\mathbf{b}_{\mathrm{i}}{ }^{\mathrm{t}}$, type $\mathbf{f}_{\mathrm{i}}$ (i.e. deformable or rigid) of each material or SPH point, which can be expressed as:
\begin{equation}
    \mathbf{x}_{\mathrm{i}}^{\mathrm{t}}=\left[\mathbf{p}_{\mathrm{i}}{ }^{ },\left\{\dot{\mathbf{p}}_{\mathrm{i}}{ }^{\mathrm{t}-\mathrm{n}+1}, \dot{\mathbf{p}}_{\mathrm{i}}{ }^{\mathrm{t}-\mathrm{n}+2}, \ldots, \dot{\mathbf{p}}_{\mathrm{i}}{ }^{\mathrm{t}}\right\}, \mathbf{b}_{\mathrm{i}}{ }^{\mathrm{t}}, \mathbf{f}_{\mathrm{i}}\right]
\end{equation}
where $n$ represents the previous time step. Instead of the physical solving process, the approximator $d_{\Theta}$ takes the current state of the particle system $\mathbf{X}^t$ as input to predict its dynamic feature $\mathbf{Y}^t$. The updater then computes the physical state $\mathbf{X}^{t+1}$ at the next time step with the $\mathbf{X}^t$ and predicted $\mathbf{Y}^t$.

The approximator $d_{\Theta}$ comprises three parts, including \textit{Encoder}, \textit{Processor}, and \textit{Decoder} as demonstrated in Fig.~\ref{pic18}. In the prediction process of $d_{\Theta}$, the \textit{Encoder} is responsible for embedding $\mathbf{X}^t$ into the initial latent graph $G_0\left(\mathbf{V}_0, \mathbf{E}_0\right)$ by assigning the node to each particle (i.e. node encoder $\delta_\theta{ }^v$ ) and adding edges $\mathbf{r}_{\mathrm{i}, \mathrm{j}}{ }^t$ between particles within one interaction radius $\mathrm{R}$ (i.e. edge encoder $\delta_\theta{ }^e$ ), which is formulated as:
\begin{equation}
    \mathbf{v}_{\mathrm{i}}{ }^{\mathrm{t}}=\delta_{\Theta}{ }^v\left(\mathbf{x}_{\mathrm{i}}{ }^{\mathrm{t}}\right) 
\end{equation}
\begin{equation}
    \mathbf{e}_{\mathrm{i}, \mathrm{j}}{ }^{\mathrm{t}}=\delta_{\Theta}{ }^v\left(\mathbf{r}_{\mathrm{i}, \mathrm{j}}{ }^{\mathrm{t}}\right)
\end{equation}
where $\mathbf{v}_i{ }^t \in \mathbf{V}_0$ represents the node state in the graph and $\mathbf{e}_{i, j}{ }^t \in \mathbf{E}_0$ means the pair-wise relationship between nodes in the graph. Then, the \textit{Processor}, consisting of $M$ GNN layers, performs an $M$ times message-passing process on $G_0$. Finally, the Decoder outputs the dynamic features of the system $\mathbf{y}_{\mathrm{i}}{ }^{\mathrm{t}} \in \mathbf{Y}^{\mathrm{t}}$ with the final graph $\mathrm{G}_{\mathrm{M}}$ from the \textit{Processor} to update the velocity $\dot{\mathbf{p}}_{\mathrm{i}}{ }^{\mathrm{t}+1}$ and position $\mathbf{p}_{\mathrm{i}}{ }^{\mathrm{t}+1}$ of each node by :
\begin{equation}
    \dot{\mathbf{p}}_{\mathrm{i}}{ }^{\mathrm{t}+1}  =\dot{\mathbf{p}}_{\mathrm{i}}{ }^{\mathrm{t}}+\mathbf{y}_{\mathrm{i}}{ }^{\mathrm{t}} \Delta t
\end{equation}
\begin{equation}
    \mathbf{p}_{\mathrm{i}}^{\mathrm{t}+1} =\mathbf{p}_{\mathrm{i}}{ }^{\mathrm{t}}+\dot{\mathbf{p}}_{\mathrm{i}}^{\mathrm{t}+1} \Delta t
\end{equation}

\begin{figure}[H]
\centering 
\includegraphics[width=\linewidth,angle=0,clip=true]{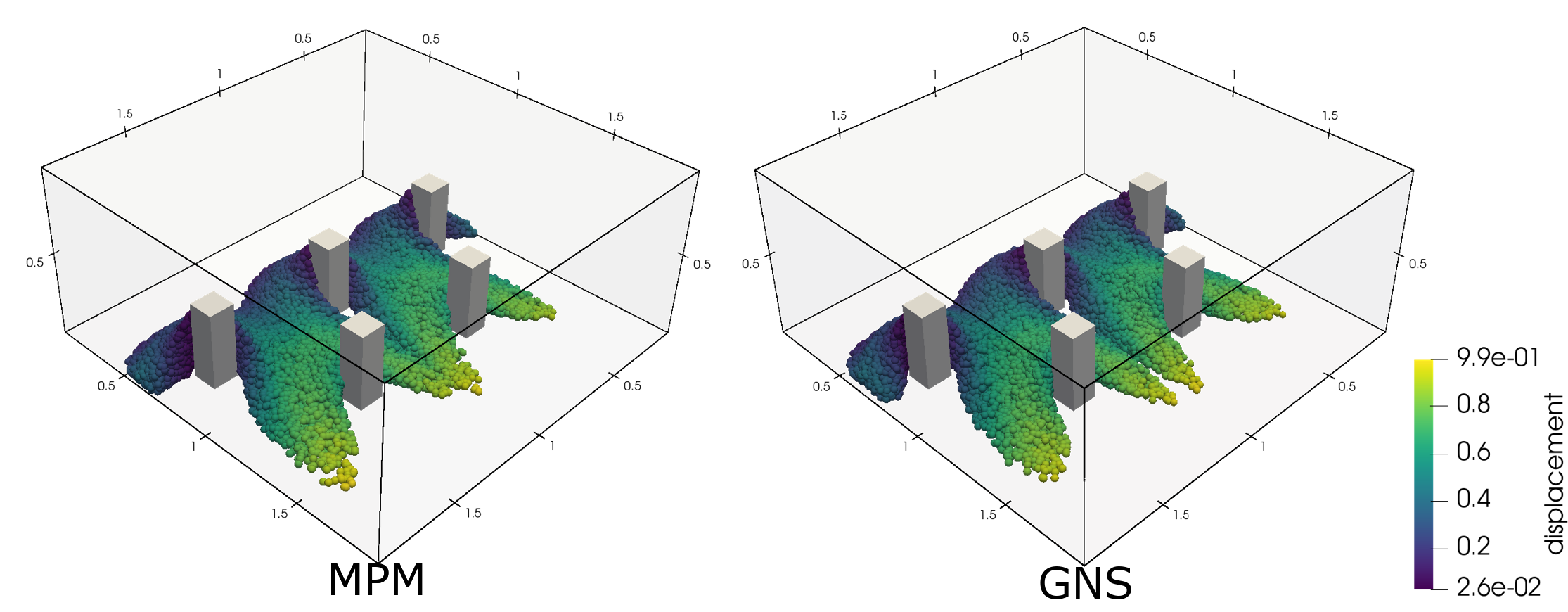}
\caption{{\color{red}Comparison of granular flow interaction with a barrier between MPM and GNS. The colour represents the magnitude of displacement. The GNS prediction is outside the training regime}~\citep{choi2024inverse}.}
\label{pic19}
\end{figure}

So far, the application of GNS in particle-based numerical methods (e.g. SPH and MPM) is still in the nascent stage. For example, ML models trained with the data generated from MPM simulations have been used to effectively predict the macroscale deformation of granular systems. Gonzalez et al.~\citep{sanchez2020learning} developed a graph network-based simulator (GNS) via the ‘encode-process-decode’ scheme to model the dynamic features of physical systems of different materials, including sand and goop, by the data generated from both the MPM and SPH simulations. Aoyama et al.~\citep{aoyama2023optimal} constructed one GNN with the data generated from the MPM simulator to predict the stacked shape of the grain collection within different containers. Choi et al.~\citep{choi2024graph} introduced physics-inspired inductive biases, such as an inertial frame that allows learning algorithms to prioritize one solution (constant gravitational acceleration) over another, reducing learning time. The GNS implementation uses semi-implicit Euler integration to update the next state based on the predicted accelerations. The GNS was trained on 26 CB-Geo MPM simulations of granular flow trajectories with 2500 particles each with 400 timesteps. At each time step, a graph network is created by linking all the material points within an interaction radius of 0.03 m, which results in connecting each node to 128 neighbouring particles. \textcolor{red}{GNS uses the velocities from the previous five steps to predict the acceleration at the next time step. Fig.~\ref{pic19} shows the GNS rollout prediction compared to MPM simulations. GNS successfully replicates the granular flow within a 5\% error in the trajectory estimation for a condition outside the training regime.}

\subsection{The macroscopic modelling with ML models as constitutive models} \label{section5.2}

In addition to the works mentioned above, ML models can also feature the mechanical response of the material at each Gaussian integral point of finite elements in mesh-based numerical methods, such as the FEM. In the following, we will focus on the combination of ML with the most popular FEM and related FEM-DEM multiscale technique, because the idea and workflow of the combination of other methods with ML are similar. In the framework of FEM, the ML model can substitute the phenomenological model or the representative volume element (RVE) embedded in each Gaussian point of the FEM to provide the stress-strain response when solving boundary value problems (BVPs). In the following subsections, the basic framework of the FEM and FEM-DEM method is first introduced concisely, and then the relevant work about the development of the FEM-ML framework will be discussed in detail.

\subsubsection{The framework of the FEM and the coupled FEM-DEM}\label{section5.2.1}

\begin{figure}[ht]
\centering

% 第一排图片
\begin{subfigure}[b]{0.42\textwidth}
  \centering
  \includegraphics[width=\linewidth]{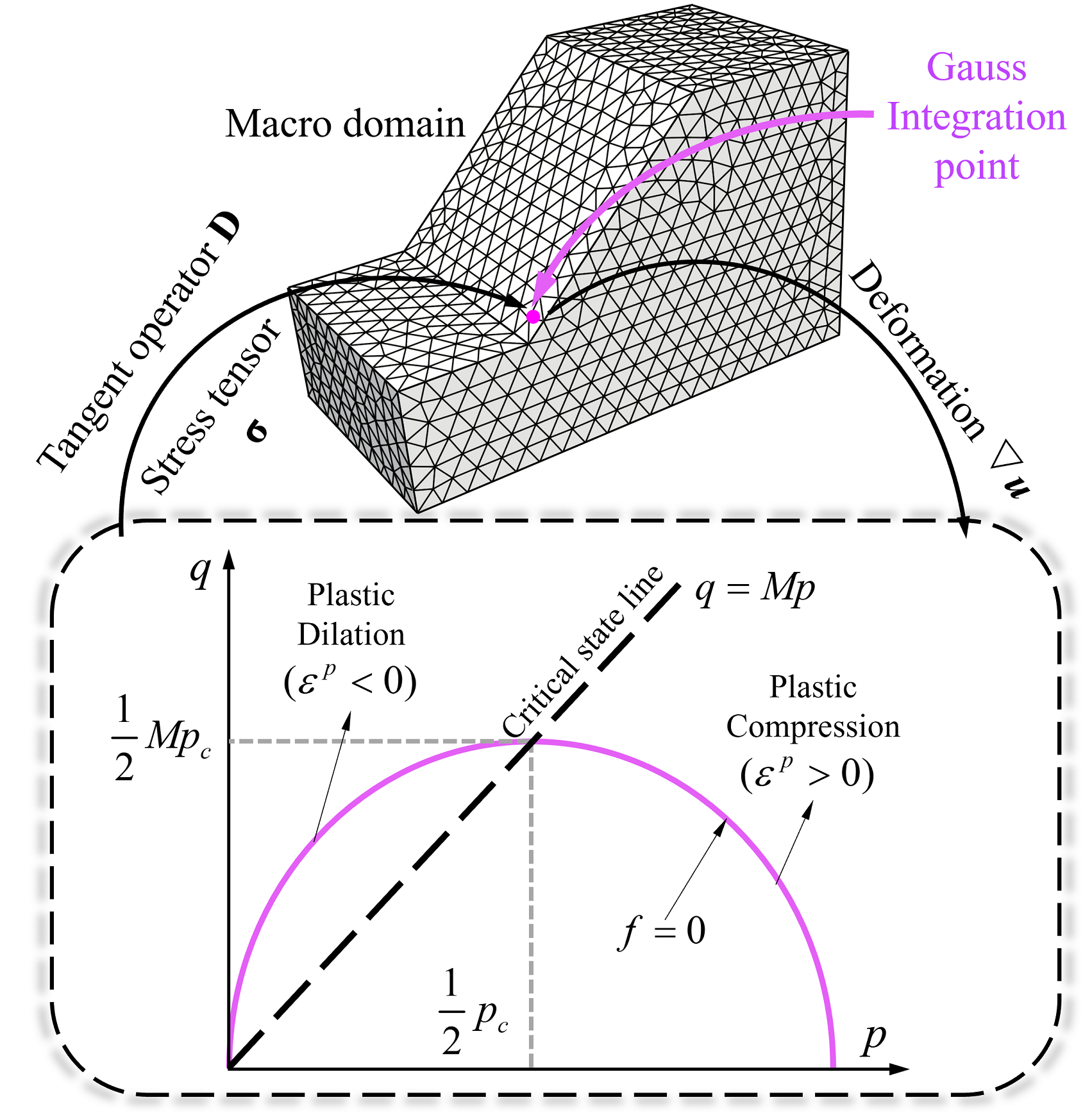}
  \caption{The FEM with critical state-based constitutive model.}
  \label{pic20a}
\end{subfigure}
\hfill
\begin{subfigure}[b]{0.56\textwidth}
  \centering
  \includegraphics[width=\linewidth]{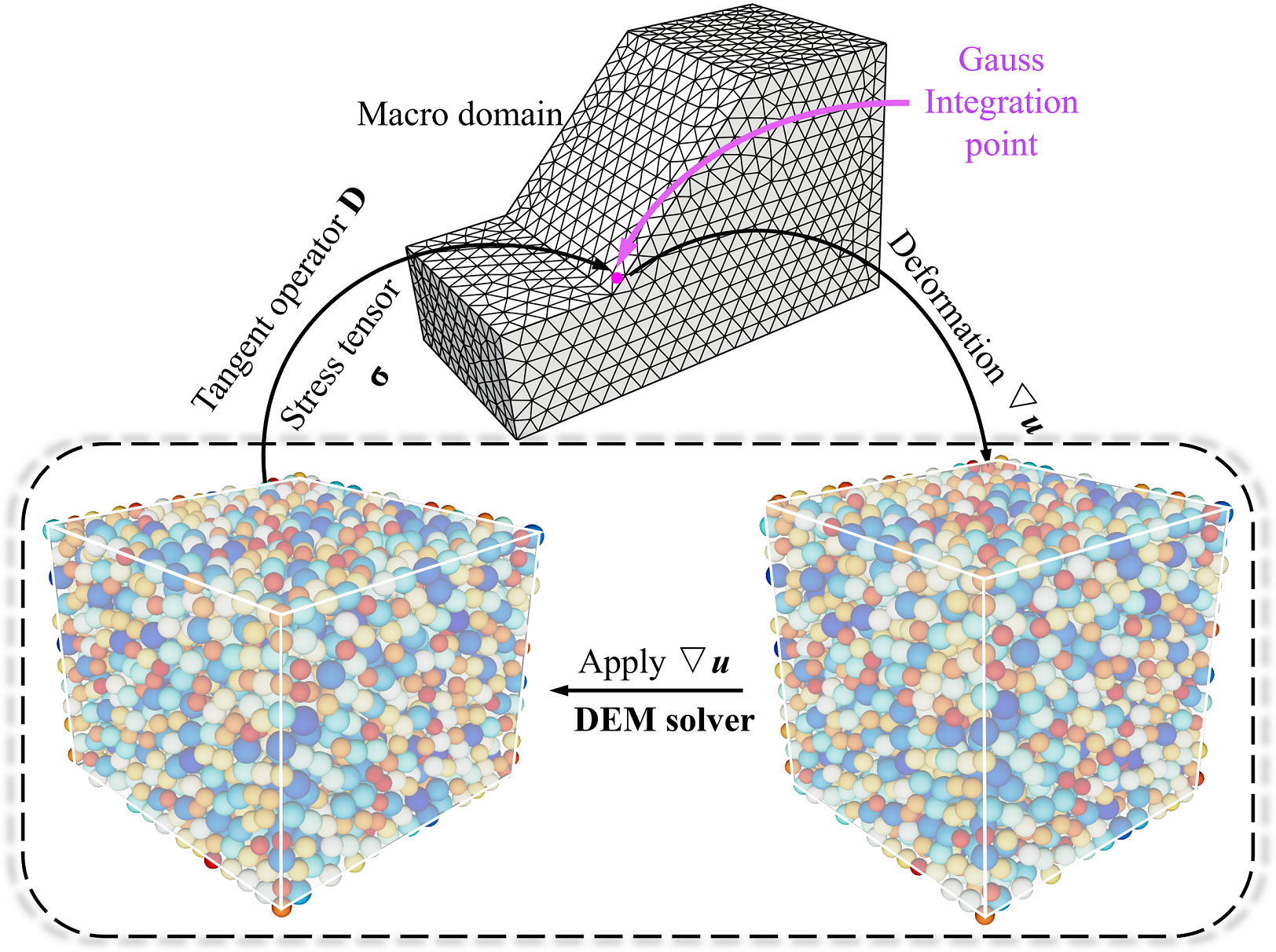}
  \caption{The FEM-DEM multiscale method}
  \label{pic20b}
\end{subfigure}%

\caption{{\color{red}The basic framework of FEM and FEM-DEM.}}
\label{pic20}
\end{figure}

In the FEM, the whole macro domain $\Omega$ is first discretized into a finite number of subdomains (elements) and the phenomenological model (e.g. critical state-based model shown in Fig.~\ref{pic20a}) is embedded into Gaussian points of each FE element to describe the local stress-strain relationship. Without considering the body force, the macroscopic deformation of the geometry domain can be obtained by solving the nodal information of each FE element by the governing equation:
\begin{equation}
\int_{\Omega } \textbf{B}^{T}\mathbf{\sigma}_{t}\, d\Omega =  \int_{\Gamma }\mathbf{N^{T}t}\, d\Gamma = \textbf{F}_{t}
\label{eq:1}
\end{equation}
where $\mathbf{\sigma}_{t}$ and $\textbf{F}_{t}$ represent the stress tensor and external force of the current load step; $\mathbf{t}$ is the boundary traction imposed on the Neumann boundary $\Gamma$ of the $\Omega$; $\mathbf{N}$ represents the shape function; $\textbf{B} = \textbf{L}\textbf{N}$, and $\textbf{L}$ is the differential operator.

In each loading step, as shown in Fig.~\ref{pic20a}, the finite element algorithm fulfil the governing formulation via the following steps: 1) The gradient of nodal displacement $\bigtriangledown \mathbf{u}_{t}$ is calculated and passed into the constitutive model of each Gaussian point to output the nodal stress $\boldsymbol{\sigma}_{t}$ and tangential operator $\textbf{D}_{t}$. 2) Given the $\mathbf{D}_{t}$ (also referred to as the constitutive matrix), $\boldsymbol{\sigma}_{t}$ and boundary condition, the displacement increment $\Delta{\mathbf{u}_{t}}$ is calculated. 3) The nodal displacement is updated according to the obtained $\Delta{\mathbf{u}_{t}}$. However, granular materials normally perform high plasticity. Therefore, the Newton–Raphson method is leveraged to iteratively solve the $\Delta{\mathbf{u}_{t}}$,  ensuring the residual force $\mathbf{R}_{t}$ of the system remains minimized, which can be expressed as:
\begin{equation}
    \mathbf{R}_{t}=\int_{\Gamma }\mathbf{N^{T}t}\, d\Gamma - \int_{\Omega } \textbf{B}^{T}\mathbf{\sigma}_{(t,m-1)}\, d\Omega = \mathbf{K}_{t}\Delta{\mathbf{u}_{(t,m)}}
\end{equation}
where $\Delta{\mathbf{u}_{(t,m)}}$ represents the solved displacement increment at the $m^{th}$ iteration of the current load step, and the global stiffness matrix $\mathbf{K}_{\mathrm{t}}$ is assembled by:
\begin{equation}
    \mathbf{K}_{\mathrm{t}}=\int_{\Omega} \mathbf{{B}^{T}} \mathbf{D}_{t} \mathbf{B} d V
\end{equation}
The iteration is repeatedly executed until the $\mathbf{R}_{t}$ converges to one tiny value.

As shown in Fig.~\ref{pic20b}, in the coupled FEM-DEM algorithm, the local constitutive relationship is provided by the representative volume element (RVE) solved by the DEM. The nodal displacement is calculated by the FEM solver and imposed on each RVE as their boundary conditions (i.e. $\bigtriangledown \mathbf{u}_{t}$). Then, with received boundary conditions, the stress $\boldsymbol{\sigma}_{t}$ and tangential operator $\textbf{D}_{t}$ of each RVE is calculated based on various homogenization methods (e.g. Voigt's hypothesis~\citep{kruyt1998statistical}, Reuss's hypothesis~\citep{chang1994estimates,yimsiri2000micromechanics}). Finally, the solved $\boldsymbol{\sigma}_{t}$ and $\textbf{D}_{t}$ by the DEM are returned to the FEM solver to calculate the displacement increment of each nodal to update the nodal displacement.

Although both the FEM and FEM-DEM can predict the macroscopic characteristics of granular materials, they have their respective limitations. The constitutive models used in the FEM are continuum theory-based, potentially obscuring the discrete nature of granular materials. Furthermore, models utilized in FEM often incorporate multiple variables that lack physical significance but require extensive calibration efforts. Moreover, constitutive models tend to be sophisticated to accommodate specific phenomena and are thus difficult to apply to new data cases. In addition, once the constitutive model is determined in FEM, it remains unadjustable, which restricts the further applicability of FEM to new tests. 

Regarding the FEM-DEM techniques, although they can offset the deficiency of the FEM in capturing the discrete features of materials to some extent, it still suffers from the issue of excessive computational costs.

It is worth noting that to avoid the deficiency of FEM in simulating material large deformation, MPM-DEM is also proposed as a promising way for the investigation of large deformation of granular materials~\citep{liang2019coupled}. To overcome the computational issues of RVE encountered in large deformation, an adaptive RVE model was recently proposed~\citep{wang2021deformation,qu2021adaptive}. Similar to the FEM-DEM framework, the MPM-ML framework can also be developed.

\subsubsection{The FEM-ML framework}\label{section5.2.2}

\begin{figure}[H]
\centering 
\includegraphics[width =0.75
\linewidth,angle=0,clip=true]{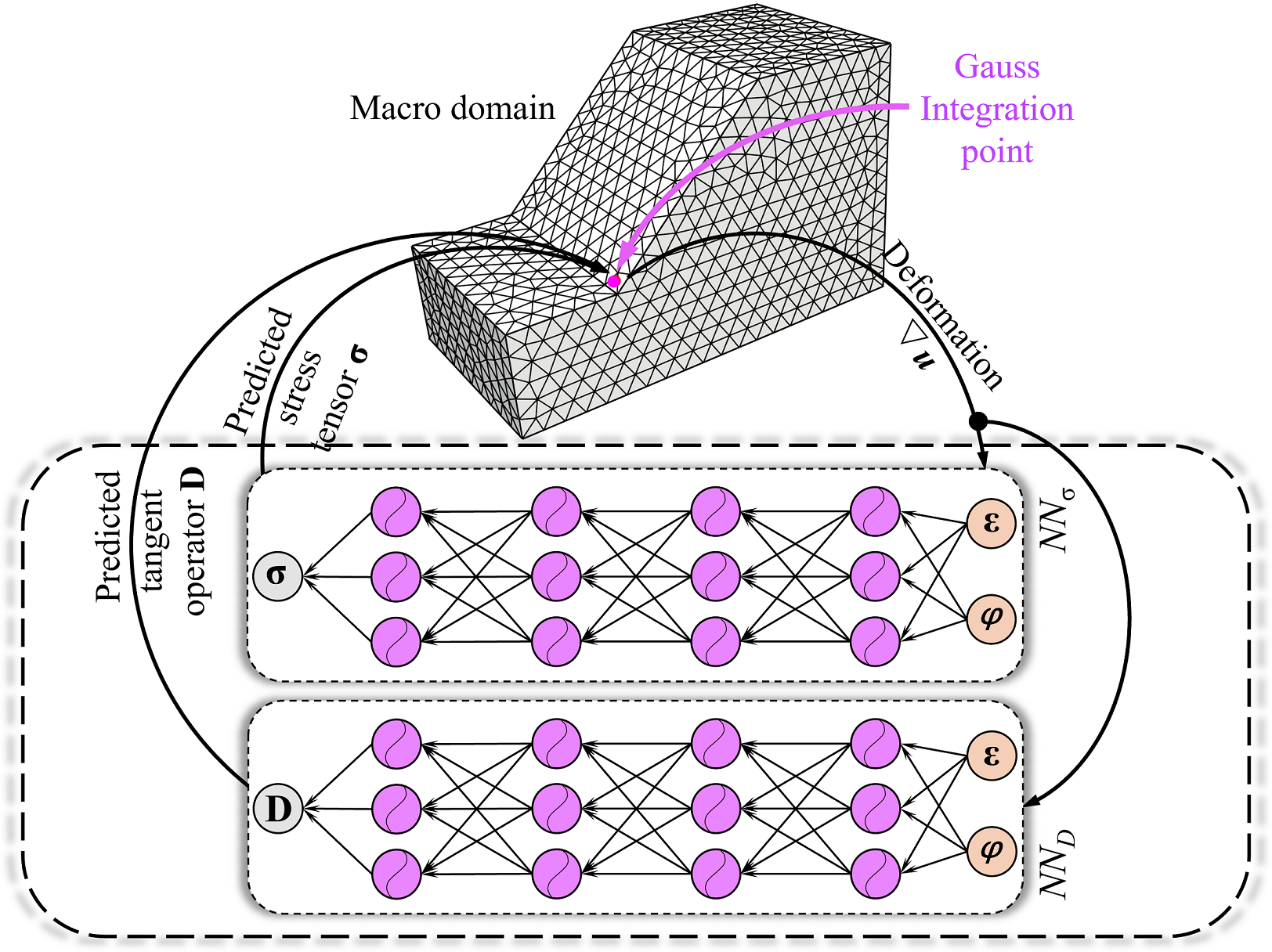}
\caption{ {\color{red}The framework of the FEM-ML algorithm}}
\label{pic21}
\end{figure}

The ML model is featured with ongoing improvement and high computational efficiency in prediction. Thus it is a potential way to circumvent the challenges mentioned above of traditional numerical methods by integrating ML surrogate models into them. As illustrated in Fig.~\ref{pic21}, in FEM or coupled FEM-DEM, replacing constitutive models or RVEs, the trained neural networks offer the necessary tangential operator or stress state according to the received local deformation information of each FE element to accelerate the computational process in BVPs. 

Generally, according to the data used to develop the neural networks, the FEM-ML framework can be categorized into two different types: 1) one is the phenomenological model data-based FEM-ML framework, and 2) the other is the DEM data-based FEM-ML framework, which is demonstrated in the following sections in detail, respectively.

Over the past decades, numerous efforts have been made to represent the material response with ML models in BVPs. The earlier work of constitutive data-based FEM-ML can be traced back to the 1990s, in which the trained ML model functions for the stress calculation in FEM to simulate the material macroscopic behaviour under triaxial~\citep{sidarta1998constitutive}, biaxial, uniaxial cycled compression~\citep{ghaboussi1991knowledge}, and un-uniform loading conditions~\citep{ghaboussi1998autoprogressive}. Subsequently, numerous scholars have carried out related works. For instance, literature~\citep{hashash2004numerical} develops one NN model as an alternative to the constitutive model to accelerate the efficient convergence of the Newton iteration in the FE algorithm; In reference~\citep{jung2006neural}, one rate-dependent NN constitutive model is developed to capture the time-dependent feature of geomaterials in the FEM; Xu et al.~\citep{xu2021learning} compute the tangential matrix with a Cholesky-factored symmetric positive-definite neural network. Based on a dimensionality reduction method (Proper Orthogonal Decomposition), Huang et al.~\citep{huang2020machine} described the intrinsic history-dependency feature of plastic materials with the feedforward neural network and states variables in FEM; Nikolaos N et al.~\citep{vlassis2021sobolev} leveraged the deep learning network to replicate one elastoplastic model, where the plastic flow and yield surface of the model are accurately portrayed through a series of deep network predictions, to obtain the seamless FEM simulation.

Similarly, numerous endeavours have been given to the DEM data-based FEM-ML framework to tackle multiscale simulations. For example, Le et al.~\citep{le2015computational} utilized the neural network to approximate the effective potential, which is used to derive the stress and tangent elastic tensor of materials in the multiscale computation of non-linear elastic materials, to reduce the computational cost. The recurrent neural network is trained with DEM data and integrated into Gaussian points of the FE meshes in the works of Simons et al.~\citep{ghavamian2019accelerating} and Logarzo et al.~\citep{logarzo2021smart} to improve the efficiency of the multiscale scheme; The performance of the Gaussian process and ANN as the agent in the multi-scale BVP was compared in the work of Fuhg et al.~\citep{fuhg2022local}; Replacing RVEs, Guan et al~\citep{guan2023machine} utilized the feedforward neural network to predict the material matrix and local stress to complete the non-linear iteration in the multiscale computation process; Qu et al.~\citep{qu2023deep} developed an active learning-based data-driven modeling framework which can pritorise the most informative data for a trained model and make it be progressively improved via interactive model training and data labelling; Qu et al.~\citep{qu2023data} further developed a transfer learning strategy that can combine the use of well-established phenomenological models, numerical models and physical experiments for data-driven material modeling, thereby reducing the data demands for certain material modeling tasks; {\color{red}Replacing the DEM solver embedded in Gaussian point with one well-trained neural network, Rangel et al.~\citep{rangel2024multiscale} proposed a data-driven FVE-ML multiscale framework to explore the thermomechanical
behavior of granular materials subjected to thermal expansion.}

However, the development of the FEM-ML framework still confronts some problems, which will be discussed from the aspects of the training data and neural networks used. The completeness of the training sample, e.g. fixed time step and the limited stress-strain space of training data, deteriorate computational stability~\citep{zhang2021bilstm,logarzo2021smart} and the robustness of the FEM-ML framework. Furthermore, in the DEM data-based FEM-ML framework, there are no explicit history variables are calculated to feature the loading states in DEM solvers, which performs the history and state-dependent feature of the particle assembly by fabric tensor or energy-based parameters~\citep{eggersmann2019model,karapiperis2021data}. Therefore, parameterizing history states into the macroscopic loading information (e.g. strain and stress) to develop the DEM data-based FEM-ML framework is also a tricky issue, especially when employing the single-step-based neural network to construct the FEM-ML framework. Although multi-timestep deep learning models can avoid this problem~\citep{guan2023finite, guan2023neural}, they are incompatible with the standard FEM computational framework, which operates on a single step.

\subsection{An example: the DEM data-based FEM-ML modelling} \label{section5.3}

A detailed study on time-sequential networks and constitutive data-based FEM-ML framework can be found in the literature~\citep{guan2023finite}. Therefore, in this section, we offer an example of the development of an MLP-based FEM-ML framework with the DEM data where there are no explicit internal variables available via a 2D biaxial compression simulation.

\subsubsection{The process of developing the FEM-ML framework}\label{section5.3.1}

To develop the FEM-ML framework, one biaxial compression FEM-DEM multiscale simulation is performed, and the stress $\boldsymbol{\sigma}^{(t)}$, strain $\boldsymbol{\varepsilon}^{(t)}$, internal variable $\boldsymbol{\varphi}^{(t)}$, as well as tangential operator $\mathbf{D}_t$ of each RVE, are recorded as the training data of ML models. Wherein the absolute accumulated strain increment $\Delta \boldsymbol{\varepsilon}^{(\mathrm{t})}$ is adopted as the history~\citep{huang2020machine} according to the result in Section~\ref{section4.3.2}.

\begin{figure}[H]
\centering 
\includegraphics[width =0.85 \linewidth,angle=0,clip=true]{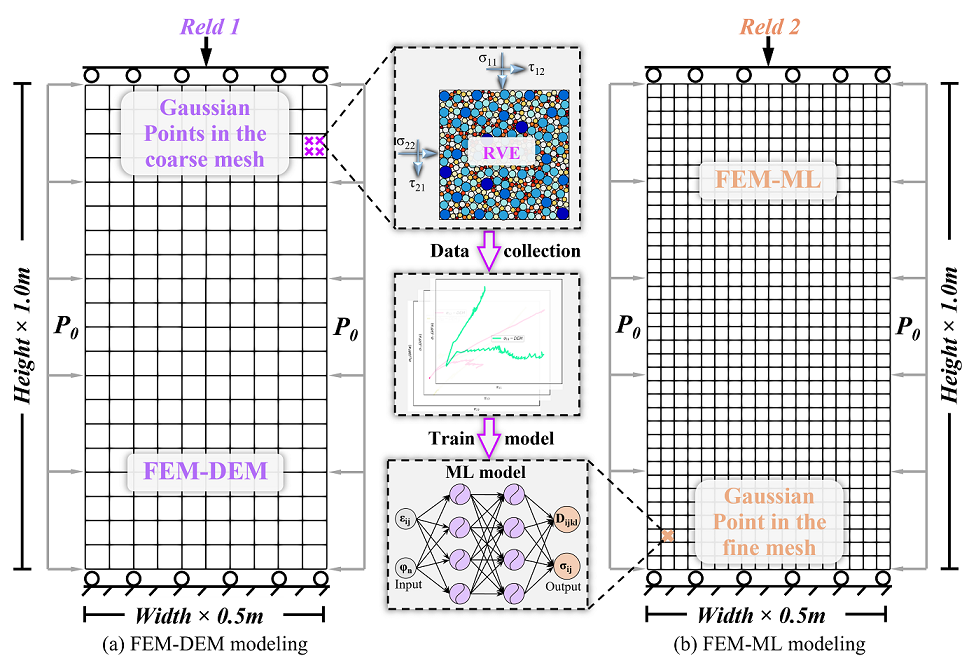}
\caption{An illustration for the DEM data-based FEM-ML framework}
\label{pic22}
\end{figure}

\begin{figure}[H]
\centering 
\includegraphics[width =0.5 \linewidth,angle=0,clip=true]{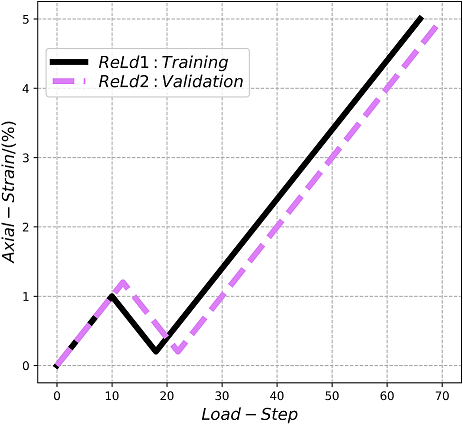}
\caption{Two different reversal loading paths.}
\label{pic23}
\end{figure}

As illustrated in Fig.~\ref{pic22}, during the multiscale modelling process, the bottom boundary of the coarse-meshed geometry is fixed in the vertical and horizontal directions, and both the left and right boundary of the mesh are confined with the constant horizontal pressure ($P_0$ = 100 KPa). A displacement-controlled axial loading is applied to the top surface of the rectangular domain with a fixed velocity (0.001 m/load step) until the axial strain arrives at 0.05 under the reversal loading 1( \textit{Reld1}) path shown in Fig.~\ref{pic23}. The parameters used in each RVE directly refer to the work of Guo et al~\citep{guo2014coupled}. The detailed grain-scale parameters are listed in Table \ref{table5}.

\begin{table}[H]
\centering
\caption{The detailed grain-scale parameters of each RVE in the FEM-DEM scheme}
\label{table5}
\begin{tabular}{cc}
\hline
Parameter                       & Value        \\ \hline
Particle number                 & 450          \\
Particle size range (mm)        & (6 $\sim$12) \\
Young’s modulus (MPa)           & 600          \\
Density (kg/$m^{3}$)             & 2650         \\
Particle frictional coefficient & 0.5          \\
Damping ratio                   & 0.1          \\ \hline
\end{tabular}
\end{table}

Different from continuum-based phenomenological models, whose constitutive function is sufficiently smooth and can be obtained by automatic differentiation (AD) methods~\citep{ghaboussi1998new,hashash2004numerical,jung2006neural, abadi2016tensorflow,ghavamian2019accelerating}, the stress-strain curve of granular materials has dramatic disturbance, and thus the $\mathrm{AD}$ method cannot provide the proper tangential operator during the Newton-Raphson iteration in FEM. Alternatively, the method proposed in the work of Guan et al.~\citep{guan2023machine} is resorted in this case, i.e. using the neural network to provide the material matrix. Consequently, two MLP models $\left(N N_\sigma\right.$ and $\left.N N_{\mathrm{D}}\right)$ are trained to provide the necessary stress $\boldsymbol{\sigma}^{(\mathrm{t})}$ and material matrix $\mathbf{D}_t$ for the $\mathrm{FE}$ algorithm, which can be expressed as:
\begin{equation}
    \sigma^{(t)}=N N_{\boldsymbol{\sigma}}\left(\boldsymbol{\varepsilon}^{(t)}, \Delta \boldsymbol{\varepsilon}^{(t)}, \mathbf{W}, \mathbf{b}\right)
\end{equation}
\begin{equation}
    \mathrm{D}_t=N N_D\left(\boldsymbol{\varepsilon}^{(t)}, \Delta \boldsymbol{\varepsilon}^{(t)}, \mathbf{W}, \mathbf{b}\right)
\end{equation}

As demonstrated in Fig.~\ref{pic22}, the trained networks are embedded into each Gaussian point of the fine mesh to construct the FEM-ML framework, which is then leveraged to solve the same BVP of biaxial compression but under the \textit{Reld2} path showcased in Fig.~\ref{pic23}.

\subsubsection{The prediction results with the developed FEM-ML scheme}\label{section5.3.2}

\begin{figure}[H]
\centering

% 第一排图片
\begin{subfigure}{.5\textwidth}
  \centering
  \includegraphics[width=\linewidth]{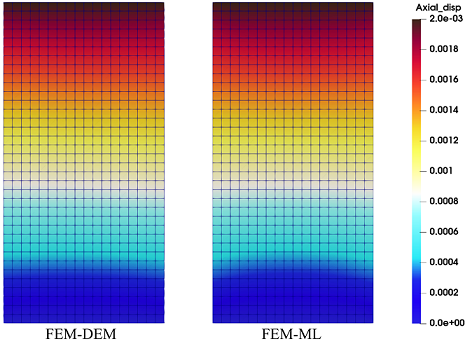}
  \caption{Load step=2}
  \label{pic24a}
\end{subfigure}%
\begin{subfigure}{.5\textwidth}
  \centering
  \includegraphics[width=\linewidth]{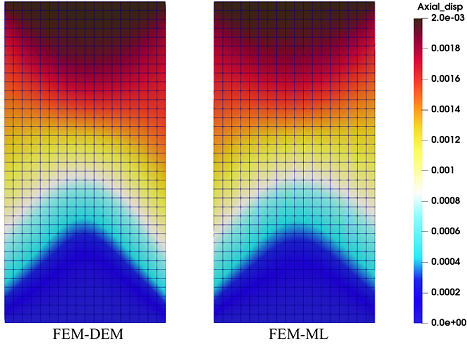}
  \caption{Load step=22}
  \label{pic24b}
\end{subfigure}

\vspace{1em} % 添加一些垂直空间

% 第二排图片
\begin{subfigure}{.5\textwidth}
  \centering
  \includegraphics[width=\linewidth]{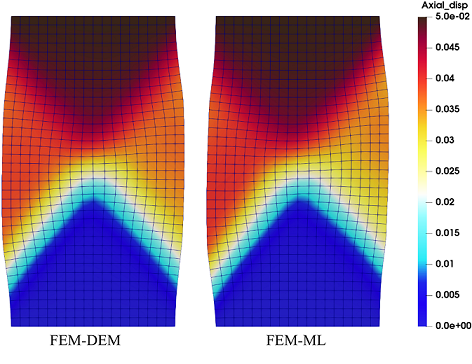}
  \caption{Load step=70}
  \label{pic24c}
\end{subfigure}%
\begin{subfigure}{.47\textwidth}
  \centering
  \includegraphics[width=\linewidth]{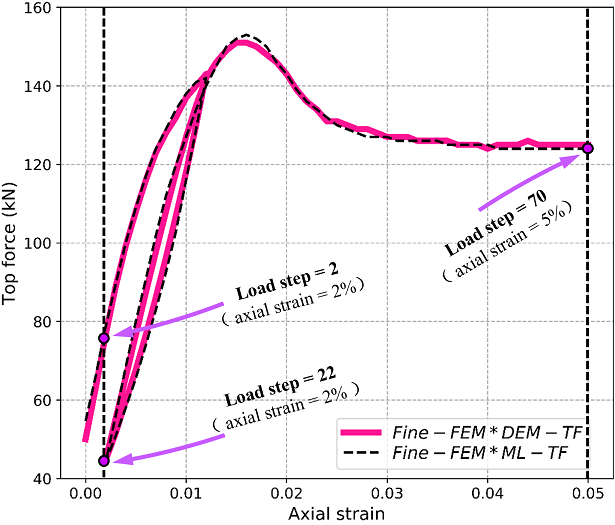}
  \caption{The top force}
  \label{pic24d}
\end{subfigure}

\caption{The solved axial displacement field under the $Reld2$ path in fine meshes~\citep{wang2024multi}}
\label{pic24}
\end{figure}

\begin{figure}[H]
\centering

% 第一排图片
\begin{subfigure}{.5\textwidth}
  \centering
  \includegraphics[width=\linewidth]{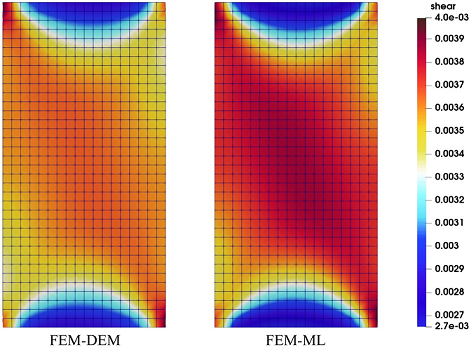}
  \caption{Load step=2}
  \label{pic25a}
\end{subfigure}%
\begin{subfigure}{.5\textwidth}
  \centering
  \includegraphics[width=\linewidth]{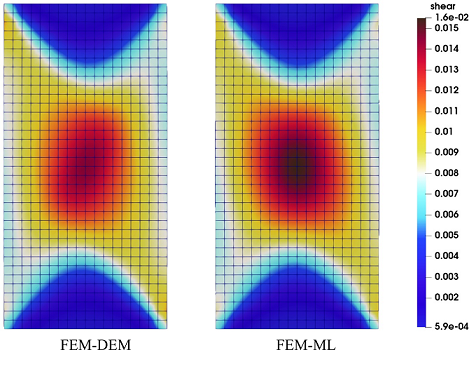}
  \caption{Load step=22}
  \label{pic25b}
\end{subfigure}

\vspace{1em} % 添加一些垂直空间

% 第二排图片
\begin{subfigure}{.5\textwidth}
  \centering
  \includegraphics[width=\linewidth]{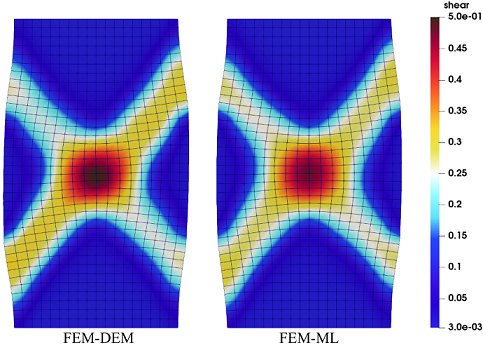}
  \caption{Load step=70}
  \label{pic25c}
\end{subfigure}%
\begin{subfigure}{.47\textwidth}
  \centering
  \includegraphics[width=\linewidth]{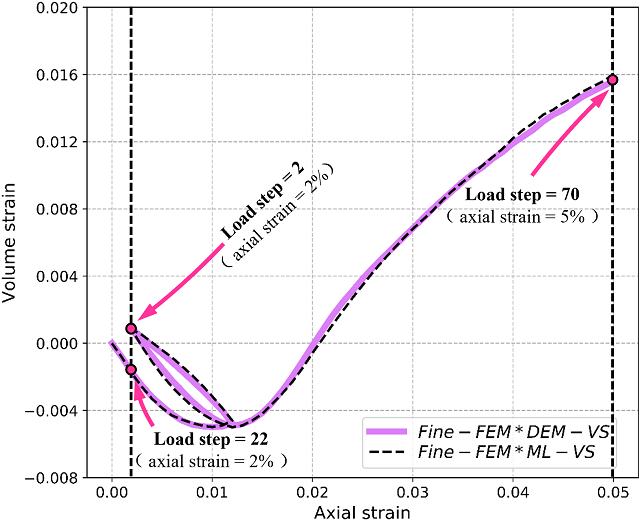}
  \caption{The top force}
  \label{pic25d}
\end{subfigure}

\caption{ The solved shear strain field under the $Reld2$ path in fine meshes~\citep{wang2024multi}}
\label{pic25}
\end{figure}

As one controlled experiment, an additional FEM-DEM simulation is conducted with identical boundary conditions and \textit{Reld2} path within the finely-meshed sample to validate the performance of the developed FEM-ML scheme. Therefore, a total of three biaxial compression simulations are implemented. The detailed experimental parameters of each simulation are listed in Table \ref{table6}. The solved shear strain and axial displacement fields by both the FEM-ML approach and the corresponding FEM-DEM scheme are compared and demonstrated in Fig.~\ref{pic24} and Fig.~\ref{pic25}.

\begin{table}[H]
\caption{The experimental parameters of the three biaxial compression modelling}
\label{table6}
\scalebox{0.8}{
\begin{tabular}{cccccccccc}
\hline
No. & Mesh   & Method  & \begin{tabular}[c]{@{}c@{}}Loading\\ path\end{tabular} & \begin{tabular}[c]{@{}c@{}}Data for\\ using\end{tabular} & \begin{tabular}[c]{@{}c@{}}Total\\ load steps\end{tabular} & \begin{tabular}[c]{@{}c@{}}Time\\ consuming\end{tabular} & \begin{tabular}[c]{@{}c@{}}Unloading \\ step\end{tabular} & \begin{tabular}[c]{@{}c@{}}Reloading\\ step\end{tabular} & \begin{tabular}[c]{@{}c@{}}The number\\  of RVEs\end{tabular} \\ \hline
1   & Coarse & FEM-DEM & Reld1                                                  & Training                                                 & 66                                                         & 2.12 hours                                               & 10                                                        & 18                                                       & 800                                                           \\
2   & Fine   & FEM-DEM & Reld2                                                  & Validation                                               & 70                                                         & 9.65 hours                                               & 12                                                        & 22                                                       & 2592                                                          \\
3   & Fine   & FEM-ML  & Reld2                                                  & Testing                                                  & 70                                                         & 0.016 hours                                              & 12                                                        & 22                                                       & 2592                                                          \\ \hline
\end{tabular}
}
\end{table}

Fig.~\ref{pic24} demonstrates the solved axial displacement fields and top force (TF) of the fine-meshed sample by both the FEM-ML and FEM-DEM approaches. Where the axial displacement at some special loading moments (i.e. the $2^{\text {nd }}, 22^{\text {nd }}$, and $70^{\text {th }}$ load steps) are respectively plotted in Figs.~\ref{pic24a},~\ref{pic24b}, and~\ref{pic24c}; Fig.~\ref{pic24d} illustrates the change of the top force of the sample during the simulation process. Similarly, the shear strain fields at the $2^{\text {nd }}, 22^{\text {nd }}$, and $70^{\text {th }}$ load steps along with the volume strain (VS) of the geometry domain in the FEM-DEM and FEM-ML modelling are showcased in Fig.~\ref{pic25}.

The results showcased in Fig.~\ref{pic24}-\ref{pic25} demonstrate that the FEM-ML scheme can still accurately model the deformation feature of granular materials in the same BVP, although the mesh density is increased and the loading path is changed, which proves that the well-developed FEM-ML approach possesses the extrapolation capability to some extent.

\subsection{Summary}\label{section5.4}

{\color{red}This section focuses on recent developments in coupling ML techniques with continuum-based numerical methods, which are categorized into two main areas: the integration of ML models in 1) meshless methods, such as SPH and the MPM, and 2) mesh-based methods, including the FEM and multiscale FEM-DEM frameworks.}

\begin{table}[H]
\centering
\caption{{\color{red}The overview of advantages and limitations of ML-aided macroscopic simulation methods}}
\label{summary5}
\scalebox{0.8}{
\begin{tabular}{c|c|l|l}
\hline
\begin{tabular}[c]{@{}c@{}}ML-aided\\ macroscopic\\ simulation\end{tabular}
& Types & \multicolumn{1}{c|}{Advantages} & \multicolumn{1}{c}{Disadvantages}\\

\hline
\begin{tabular}[c]{@{}c@{}}Geometry information\\
network-based\\ML simulator\end{tabular}&GNS 
& \begin{tabular}[c]{@{}l@{}}
1). Strong generalization capability\\ across meshless numerical methods\\ 
2). Low computational complexity\\ 
3). High computational efficiency
\end{tabular}

& \begin{tabular}[c]{@{}l@{}}
1). Requirement for large training memory\\ 
2). Error accumulation problem\\ 
3). Weaker suitability to systems where\\interpolation points are added or removed
\end{tabular}\\
\hline
\multirow{2}{*}{\begin{tabular}[c]{@{}c@{}}Data-driven\\FEM-ML multiscale\\ methods
\end{tabular}}

& \begin{tabular}[c]{@{}c@{}} Phenomenological \\ model data-based
\end{tabular}
& \multirow{2}{*}{\begin{tabular}[c]{@{}l@{}}
1). Higher computational efficiency\\ 
2). Potential for ongoing enhancement\end{tabular}} 
& \multirow{2}{*}{\begin{tabular}[c]{@{}l@{}}
1). Requirement for artificially\\ added internal variables\\
2). Limited generalization ability\end{tabular}}\\ \cline{2-2}
& \begin{tabular}[c]{@{}c@{}}DEM\\data-based\end{tabular}& & \\
\hline
\end{tabular}}
\end{table}

{\color{red}Table~\ref{summary5} presents a concise summary of the advantages and limitations of each aspect. In the context of continuum-based numerical methods, the extensive spatial interactions among interpolation particles underscore the suitability of geometry information-based neural networks. These networks facilitate permutation-invariant learning, which enables them to efficiently capture complex spatial relationships among particles and bypass complicated physical equations. Consequently, compared to traditional continuum-based particle methods, the geometry information network-based simulator significantly reduces the computational complexity when simulating the deformation of granular materials, leading to a remarkable improvement in overall simulation efficiency.

On the other hand, compared to the traditional FEM, especially the FEM-DEM approach, the superior computational efficiency of the ML-based constitutive model mentioned in Section~\ref{section4.4} can significantly reduce the computational cost and time of the FEM-ML multiscale method. Additionally, due to the adaptive nature of ML surrogate models, the performance of the FEM-ML scheme can be continuously enhanced given sufficient training data, enabling it to handle increasingly complex tasks more effectively.

However, it is important to note that training geometry information-based networks typically requires significant computational memory due to the large number of particles in a single simulation case. Furthermore, when using Graph Neural Networks (GNNs) to predict system rollouts over extended periods, error accumulation becomes a challenge. This issue arises because the dynamic information at each time step is updated based on the particle positions from the preceding time step, leading to the propagation of errors throughout the simulation.

In the development of the FEM-ML scheme, both single-step and multi-step surrogate models serve as alternatives to phenomenological models or RVEs embedded at Gaussian points within the FEM framework. While time-sequence neural networks can naturally capture the history-dependent constitutive behavior of granular materials, they are sensitive to changes in loading intervals, and their multi-step nature is not aligned with the conventional FEM algorithms. This discrepancy requires additional efforts to integrate time-sequence networks into a FEM solver. In contrast, single-step neural networks are inherently compatible with the FEM algorithms and robust to variations in loading intervals. However, they still face the challenge of identifying appropriate internal variables to capture different loading histories adequately.

Additionally, the generalization capability of current FEM-ML strategies is often limited by the completeness of training data. For instance, the simulation result obtained in Section~\ref{section5.3} demonstrates the developed FEM-ML approach has a certain extrapolation capability. Although the mesh density of the solving domain increases and a different but similar loading path is applied, the FEM-ML scheme can still acquire a comparable solution to the corresponding FEM-DEM simulation. However, when encountering a new situation where the geometry domain and boundary conditions vary considerably, the performance of the current FEM-ML solver will deteriorate ~\citep{wang2024multi}. This limitation makes many FEM-ML approaches case-specific and restricts their applicability to more general scenarios.}

{\color{red}
Based on the summary presented above, several promising directions for future research in ML-aided macroscopic simulation methods can be identified. These avenues include employing active learning to identify informative data for training trustworthy ML models, Bayesian or Gaussian processes for uncertainty quantification, and the development of hybrid computational frameworks that integrate neural networks with traditional numerical methods, which are detailed as follows:

1). Application of active learning to mitigate error accumulation: One of the key challenges in using ML-based simulators, particularly GNS, is the accumulation of errors during long-term system rollouts. Active learning could be employed to dynamically select new data points that the model finds most uncertain, enabling iterative retraining of the model to correct errors. This approach would improve the robustness of ML models over extended simulations and ensure more reliable results, especially in complex, multi-step scenarios.

2). Incorporating Bayesian or Gaussian processes for reliability assessment: To improve the reliability of ML-based constitutive models, future research could incorporate Bayesian networks or Gaussian processes, which provide probabilistic estimates along with predictions. These approaches enable uncertainty quantification in model outputs, allowing researchers to assess the confidence in predictions. Incorporating such probabilistic models would enhance the trustworthiness of simulations by identifying areas where predictions may be less reliable and require further retraining efforts.

3). Development of hybrid computational schemes: Utilizing the insights from Bayesian or Gaussian processes, future research could focus on developing hybrid computational schemes that integrate the strengths of ML models with traditional numerical methods. In such frameworks, ML models could be applied in regions where their predictions have the highest certainty, while traditional methods could be used in areas where ML models exhibit lower reliability. This approach would improve computational efficiency while maintaining high accuracy, offering a robust solution for complex simulations with high efficiency and precision.}

\section{Discussion}\label{section6}

The weak extrapolation capability of the ML models is one primary reason that limits their further application in modelling granular materials. Research efforts have been focused on developing advanced techniques that specifically address the relevant issue, such as the proposal of Physics-informed neural networks (PINNs) and Fourier neural operators(FNOs).

The idea behind PINNs~\citep{raissi2019physics} is to enforce the neural network to satisfy some known physical laws or governing equations by introducing a loss function that incorporates these physical constraints. The total loss function for a PINN is typically a combination of a data loss term $L_{\text {data }}(\theta)$ and a physics loss term $L_{\text {physics }}(\theta)$, which can be expressed as: $L(\theta)=L_{\text {data }}(\theta)+\lambda L_{\text {physics }}(\theta)$. Here, $\lambda$ is a hyperparameter that balances the importance of data fidelity and physics consistency. $\theta$ is a set of parameters (weights and biases). The data loss term can be a standard mean squared error (MSE) that compares predictions with observed data. The physics loss term enforces that the network's predictions satisfy the underlying physics, expressed as a partial differential equation (PDE) or other constraints using a differential operator representing the governing equations, which can be computed using automatic differentiation. For example, in the work of Eghbalian et al.~\citep{eghbalian2023physics}, plasticity concepts like additive strain decomposition and hypoelasticity are incorporated into the network design to capture the pressure-dependent yielding and dilatancy behaviours of sand. This physics-informed structure enables more efficient training with improved generalization compared to standard neural networks, as evidenced by the PINN's stable predictions across diverse loading paths unseen in the training data. Although the PINN can improve the robustness of the ML model to a certain extent, they are still largely restricted to the boundary conditions on which they were trained and are not generalizable beyond the training regime. {\color{red} Additionally, because many granular media-related phenomena are not easily simplified using PDEs or Ordinary Differential Equations (ODEs), the application of PINNs in simulating granular media remains limited. Several examples in this field include using PINNs or neural operator for consolidation analysis in soils ~\citep{zhang2022physics,mandl2024separable} and for modelling particle aggregation and breakage processes in chemical engineering~\citep{chen2021physics}. Both examples rely on the prerequisite that PDEs have been well-established to characterize these problems.}

On the other hand, operator learning, particularly through the lens of FNOs~\citep{roberts2021learning,li2023fourier}, offers a transformative approach to modelling in scientific computing. Traditional deep learning models are designed to learn mappings between finite-dimensional vectors, but operator learning seeks to understand the mappings between entire function spaces. This is achieved by training on pairs of input-output functions, rather than discrete data points. The incorporation of spectral methods allows the model to operate in infinite-dimensional function spaces without being constrained by mesh resolutions. By representing functions in the Fourier space, FNOs can efficiently learn operators by applying Fourier transforms and diagonal matrix multiplications within their layers. This not only reduces the need for excessive parametrization but also ensures that the learned operator behaves smoothly across the function space. Furthermore, the inherent properties of the Fourier space, such as smoothness and regularization, {\color{red}enable FNOs to efficiently capture global patterns and generalize well within the scope of the training data, making them a promising tool for a wide range of scientific applications.} 

{\color{red}
Another significant advancement in operator learning is the Deep Operator Network (DeepONet; \citep{lu2021learning,goswami2023physics,mandl2024separable}). This architecture is grounded in the Universal Approximation Theorem for Operators, which extends the classical neural network theory to function spaces. The theorem posits that certain neural network architectures can approximate a wide class of nonlinear operators between function spaces with arbitrary precision.
DeepONet implements this theorem through a novel two-part structure: a branch network and a trunk network. The branch network processes input functions, while the trunk network handles the evaluation points of output functions. DeepONet approximates operators by representing them as a sum of basis functions multiplied by coefficients, implemented through a dot product operation between the outputs of these networks. The trunk network learns the basis functions, while the branch network computes the coefficients based on the input function.
This architecture allows DeepONet to solve multi-scale problems~\citep{liu2021multiscale} and complex non-linear systems~\citep{mandl2024separable}.

While FNOs and DeepONet operate in general Banach spaces, the Basis-to-Basis (B2B) approach~\citep{ingebrand2024b2b} leverages the geometric properties of Hilbert spaces to generate more interpretable and generalizable operators. By exploiting the inner product structure, spectral theory, and optimization geometry of Hilbert spaces, B2B learns basis functions for both input and output spaces, and then maps between their coefficients. This structure allows B2B to handle variable input locations, interpolate effectively, and potentially extrapolate beyond the training domain, while offering improved interpretability and analytical tractability.

Although relatively unexplored in the space of granular media, operator learning approaches like FNO, DeepONet, and B2B have the potential to revolutionize and accelerate numerical simulations in this field. These methods offer promising avenues for modeling complex granular systems, predicting their behavior under various conditions, and potentially uncovering new insights into the fundamental physics governing granular media.}

\section{Conclusion} \label{conclusion}

This work reviewed the major advancement in ML-aided modelling of granular media, specifically including the application of the ML method in microscopic grain scale computation, the development of the data-driven constitutive model for granular materials, and the ML-aided macroscopic simulation of granular media. 

On the microscopic scale, we mainly focus on the recent development of two different types of grain information-based ML models. The first one is the contact information-based ML model, and the other is the grain-level kinematic features-based ML model. The high computation efficiency of the ML algorithm makes these models promising ways to accelerate the computational process of particle/grain-based numerical techniques, and the black-box characteristics make these ML models very user-friendly. However, it is essential to recognize that the development of such grain contact information-based ML models is also subjected to certain challenges. The variations of particle shape, diverse contact situations, and the absence of consistent contact theories make it nearly impossible to generate a high-quality and comprehensive training set that covers all contact conditions. Consequently, this limitation hinders the generalization of the ML-based contact surrogate models in practical engineering applications. On the other hand, it has been observed that a majority of existing kinematic features-based ML models ignored some essential physical motion information of grains, e.g. particle rotation, and angular velocity. These two main issues are the primary obstacles impeding the continued advancement of particle information-based ML models at the grain level.

Both time-sequential and single-step-based neural networks can be leveraged to derive the constitutive relationship of granular materials using various types of training data. Wherein the distinctive architecture of these time-sequence neural networks inherently enables them to capture the history-dependent stress-strain response of granular materials. For single-step-based networks, to achieve comparable performance to time-sequence neural networks, the incorporation of artificially introduced internal/historical variables is necessary to identify loading states. Since the ML-based constitutive models directly acquire the constitutive laws from generated stress-strain data, they can naturally circumvent assumption and intricate mathematical formulation and evolve continuously with the expansion of the training dataset, making them naturally overcome the problems confronted by the traditional continuum-theory-based phenomenological constitutive models. However, the progress of ML-based constitutive models is primarily hindered by their inherently weaker extrapolation ability of neural networks. This limitation implies that the training set for a general ML-based constitutive model has to encompass diverse stress-strain paths and various material types to the greatest extent, which is challenging to achieve. Therefore, most existing ML constitutive models can only accurately replicate the stress-strain response of specific granular materials under certain loading conditions.

Utilizing various neural networks, data-driven simulators or ML-aided numerical solvers have been developed to accelerate the macroscopic simulation process of granular materials. Due to their high computational efficiency and black
-box characteristics, ML models bypass the most time-consuming process in traditional physical solving processes and directly provide essential state parameters for each solving iteration in the numerical algorithm, thereby accelerating the overall computation process. However, similar to the ML-based constitutive model and grain information-based ML simulations, the ML-assisted macroscopic simulation methods also suffer from the issue of weaker extrapolation capability, and most of the current works are case-by-case. Additionally, it is essential to consider the avoidance of the error accumulation problem, particularly when employing the ML model in explicit dynamic numerical methods to predict system rollouts over an extended duration. Any prediction error from previous time steps can impact subsequent predictions, potentially leading to the breakdown of the ML system.

\bibliographystyle{elsarticle-harv1} 
\biboptions{comma,numbers}
\bibliography{main}

\end{document}